\documentclass[3p]{elsarticle}

 \usepackage{graphics}
 \usepackage{graphicx}
 \usepackage{epsfig}

\usepackage{amssymb}
\usepackage{amsthm}
\usepackage{amsmath}

\usepackage{empheq}
\usepackage{epstopdf}
\usepackage{changepage}
\usepackage{rotating}
\usepackage{adjustbox}
\usepackage{color}
\usepackage{xcolor}
\usepackage{dsfont}
\usepackage{float}
\usepackage{lineno}
\usepackage{resizegather}
\usepackage{multicol}
\usepackage{booktabs} % mejora la calidad de las tablas y proporciona nuevas opciones
\usepackage{colortbl} % coloreado en tablas
\usepackage{multirow} % tablas con celdas de varias lineas
\usepackage{longtable,tabu} % tablas
\usepackage{hhline}   % mejora las lineas horizontales en las tablas
\usepackage{anyfontsize} % tamanos de fuente personalizados
\usepackage{everysel}
\usepackage{psfrag}
\usepackage{bbm}
\usepackage{bm}
\usepackage{pgf,tikz}
\usetikzlibrary{arrows}

\newtheorem{teorema}{Theorem}

\newtheorem{remark}[teorema]{Remark}

\allowdisplaybreaks 

\newcommand{\R}{\mathbb{R}}

\newcommand{\p}{\mathbb{\wp}}
\newcommand{\np}{p}

\newcommand{\nextud}{\vec{n}_{ij}}
\newcommand{\nstd}{\vec{n}_{j}}
\newcommand{\etah}{ \hat{\bm{p}}}

\newcommand{\nv}{\vec{n}}
\newcommand{\B}{\mathcal{B}}
\newcommand{\TF}{\mathbf{F}}
\newcommand{\TFe}{\mathbf{F}^{\theta}}
\newcommand{\TL}{\mathbf{L}}
\newcommand{\TLe}{\mathbf{L}^{\theta}}

\newcommand{\st}{ {st}}

\newcommand{\Gg}{ \bm{\mathcal{G}}}
\newcommand{\Mphi}{\bar{\bm{M}}}
\newcommand{\Mpsi}{\hat{\bm{M}}}
\newcommand{\idm}{\mathbf{I}}	
\newcommand{\vv}{{\mathbf{v}}}	
 
\newcommand{\rhov}{\rho {\mathbf{v}}}
\newcommand{\rhoEE}{\rho E}      	
\newcommand{\rhoe}{\rho {e}}
\newcommand{\rhok}{\rho {k}}

\newcommand{\xx}{ \mathbf{x}}
\newcommand{\TT}{\mathbf{T}}
\newcommand{\QQ}{\mathbf{R}}

\newcommand{\Ni}{ N_e}
\newcommand{\Nj}{ N_d}

\newcommand{\dx}{ d\xx}

\newcommand{\D}{\bm{\mathcal{D}}}

\newcommand{\MEPS}{{\bm{\mathcal{E}}}}
\newcommand{\LL}{ \bm{\mathcal{M}}}

\newcommand{\Q}{\bm{\mathcal{Q}}}
\newcommand{\RM}{\bm{\mathcal{R}}}
\newcommand{\LM}{\bm{\mathcal{L}}}

\newcommand{\MM}{\mathcal{\bm{M}}}

\newcommand{\ie}{ e}
\newcommand{\pph}{ \bar{\bm{p}}}

\newcommand{\rhoeh}{ \overline{\bm{\rho \ie}}}
\newcommand{\rhokh}{ \overline{\bm{\rho k}}}
\newcommand{\rhoEEh}{ \overline{\bm{\rho E}}}
\newcommand{\rhovh}{ \widehat{\bm{\rho \mathbf{v}}}}

\newcommand{\hh}{ \hat{\mathbf{H}}}

\newcommand{\diff}[2]{\frac{\partial {#1} }{\partial {#2} } }
\newcommand{\tphi}{\phi}
\newcommand{\tpsi}{\psi}
\newcommand{\tbphi}{\mathbf{\phi}}
\newcommand{\tbpsi}{\mathbf{\psi}}
\newcommand{\TTst}{\mathbf{T}}
\newcommand{\TTlst}{\TT_{\ell(j),j}}
\newcommand{\TTrst}{\TT_{r(j),j}}
\newcommand{\QQst}{\mathbf{R}}
\newcommand{\Nphist}{N_{\phi}}
\newcommand{\Npsist}{N_{\psi}}
\newcommand{\Gammast}{\Gamma}
\newcommand{\dxt}{d\xx}
\newcommand{\dSt}{dS}
\newcommand{\bbtheta}{\bar{\boldsymbol{\theta}}}

\newcommand{\btheta}{\bar{\theta}}

\newcommand{\bbrho}{\bar{\boldsymbol{\rho}}}
\newcommand{\hbrho}{\hat{\boldsymbol{\rho}}}
\newcommand{\brho}{\bar{\rho}}
\newcommand{\hrho}{\hat{\rho}}
\newcommand{\bbpi}{\bar{\mathbf{p}}}
\newcommand{\hbpi}{\hat{\mathbf{p}}}
\newcommand{\bpi}{\bar{p}}

\newcommand{\hbv}{\hat{\mathbf{v}}}
\newcommand{\bbv}{\bar{\mathbf{v}}}

\newcommand{\bov}{\mathbf{v}}
\newcommand{\hbrv}{\widehat{\boldsymbol{\rho} \mathbf{v}}}
\newcommand{\bbrv}{\overline{\boldsymbol{\rho} \mathbf{v}}}
\newcommand{\hrv}{\widehat{ \rho \mathbf{v}}}
\newcommand{\brv}{\overline{\rho \mathbf{v}}}

\newcommand{\bbrE}{\overline{\boldsymbol{\rho} \bm{E}}}

\newcommand{\brE}{\overline{\rho E}}

\newcommand{\bsigma}{\boldsymbol{\sigma}}

\newcommand{\Fte}[2]{ \bar{\mathbf{F}}_{t}^{\theta} \left({#1},{#2}\right) }
\definecolor{ttzzqq}{rgb}{0.2,0.6,0}
\definecolor{qqttcc}{rgb}{0,0.2,0.8}
\definecolor{qqttzz}{rgb}{0,0.2,0.6}
\definecolor{ffqqqq}{rgb}{1,0,0}
\definecolor{qqwuqq}{rgb}{0,0.39,0}
\definecolor{zzttqq}{rgb}{0.6,0.2,0}
\definecolor{qqqqff}{rgb}{0,0,1}
\definecolor{ttttqq}{rgb}{0.2,0.2,0}
\definecolor{qqwwtt}{rgb}{0,0.4,0.2}
\definecolor{ubqqys}{rgb}{0.29,0,0.51}
\definecolor{wwttqq}{rgb}{0.4,0.2,0}
\definecolor{uuuuuu}{rgb}{0.27,0.27,0.27}
\definecolor{qqzzff}{rgb}{0,0.6,1}
\definecolor{xdxdff}{rgb}{0.49,0.49,1}
\definecolor{ccwwqq}{rgb}{0.8,0.4,0}
\definecolor{ttqqqq}{rgb}{0.2,0,0}
\definecolor{qqzzcc}{rgb}{0,0.6,0.8}

\definecolor{cr1}{RGB}{0,0,0}
\definecolor{cr2}{RGB}{0,0,0}
\definecolor{cr12}{RGB}{0,0,0}

\journal{Computers \& fluids}
\date{}

\begin{document}

\begin{frontmatter}

\title{Efficient high order accurate staggered semi-implicit discontinuous Galerkin methods for natural convection problems}

\author[LAM]{S. Busto}
\ead{saray.busto@unitn.it}

\author[LAM]{M. Tavelli}
\ead{m.tavelli@unitn.it}

\author[Fer]{W. Boscheri}
\ead{walter.boscheri@unife.it}

\author[LAM]{M. Dumbser\corref{cor1}}
\ead{michael.dumbser@unitn.it}

\cortext[cor1]{Corresponding author}

\address[LAM]{Laboratory of Applied Mathematics, DICAM, University of Trento, via Sommarive 14, IT-38050 Trento, Italy}

\address[Fer]{Department of Mathematics and Computer Science, University of Ferrara, via Machiavelli 30, IT-44121 Ferrara, Italy}

% % % % % % % % % % % % % % % % % % % % % % % % % % % % % %
%                   Abstract                              %
% % % % % % % % % % % % % % % % % % % % % % % % % % % % % %
\begin{abstract}
In this article we propose a new family of high order 
staggered semi-implicit discontinuous Galerkin (DG) methods for 
the simulation of natural convection problems. Assuming small temperature 
fluctuations, the Boussinesq approximation is valid and in this
case the flow can simply be modeled by the incompressible Navier-Stokes 
equations coupled with a transport equation for the temperature and a buoyancy source term
in the momentum equation. 
Our numerical scheme is developed starting from the work presented in \cite{TD14,TD15,TD16}, 
in which the spatial domain is discretized using a face-based staggered 
unstructured mesh. The pressure and temperature variables are defined on 
the primal simplex elements, while the velocity is assigned to the dual grid. 
For the computation of the advection and diffusion terms, two different 
algorithms are presented: i) a purely Eulerian upwind-type scheme and 
ii) {\color{cr1} an Eulerian-Lagrangian} approach. The first methodology leads to a conservative scheme whose 
major drawback is the time step restriction imposed by the CFL stability condition 
due to the explicit discretization of the convective terms. 
On the contrary, computational efficiency can be 
notably improved relying on {\color{cr1} an Eulerian-Lagrangian} approach in which the Lagrangian trajectories of the flow are  tracked back. This method leads to an unconditionally stable scheme if the diffusive terms 
are discretized implicitly. Once the advection and diffusion contributions have been computed, the pressure Poisson 
equation is solved and the velocity is updated. 
As a second model for the computation of buoyancy-driven flows, in this paper we 
also consider the full compressible Navier-Stokes equations. The staggered semi-implicit DG method 
first proposed in \cite{TD17} for all Mach number flows is properly extended to account for the 
gravity source terms arising in the momentum and energy conservation laws.  
In order to assess the validity and the robustness of our novel class of staggered semi-implicit 
DG schemes, several classical benchmark problems are considered, showing in all cases a good 
agreement with available numerical reference data. Furthermore, a detailed comparison between the 
incompressible and the compressible solver is presented. Finally, advantages and disadvantages of the Eulerian and the {\color{cr1} Eulerian-Lagrangian} methods for the discretization of the nonlinear convective terms 
are carefully studied.
\end{abstract}

% % % % % % % % % % % % % % % % % % % % % % % % % % % % % %
%                   Keywords                              %
% % % % % % % % % % % % % % % % % % % % % % % % % % % % % %

\end{frontmatter}
% \linenumbers

%                 Introduction                            %
% % % % % % % % % % % % % % % % % % % % % % % % % % % % % %
\section{Introduction}

Natural convection problems play an important role in computational fluid dynamics. 
They appear in numerous engineering applications and natural phenomena ranging from the design 
of cooling devices in industrial processes, electronics, building isolation or solar energy collectors, to the simulation of atmospheric flows. In the last decades, the scientific community has put a lot of efforts into the study of these phenomena, see e.g. \cite{GR08,BB11,BZG14,DRB17,GCD18,Yi18} for a non-exhaustive
overview.
Nowadays, the main challenge is to develop efficient high order numerical methods which are 
able to capture even small scale structures of the flow, avoiding the use of RANS 
turbulence models (see \cite{MNG15,MS18}). In this paper, we propose a novel 
family of high order accurate staggered semi-implicit discontinuous Galerkin (DG) methods, 
which extends the works presented in \cite{TD14,TD16,TD17} appropriately to deal also with 
gravity driven flows.  

Depending on the magnitude of the temperature perturbation and on the importance of density changes, natural convection problems are usually divided into two main groups.
If the Mach number and the temperature fluctuations are small, the \textit{incompressible} Navier-Stokes 
equations under the usual Boussinesq assumption can be applied. Otherwise, the full \textit{compressible} Navier-Stokes equations have to be employed. In the following, we will mainly focus on the first case. However, in this paper also the compressible model will be considered, thus allowing for a direct comparison of the results obtained using the two different systems of governing partial differential equations. 
Therefore, we will be able to further validate the applicability of the Boussinesq approach for the flow regimes we are interested in.  

In the literature there are numerous approaches that have been proposed for the solution 
of the Navier-Stokes equations, such as finite difference methods \cite{HW65,PS72,Pat80,Kan86} 
or continuous finite element schemes \cite{TH73,BH82,HMM86,For81,Ver91,HR82,HR88}. 
Nevertheless, the construction of high order numerical methods, and especially of high 
order discontinuous Galerkin (DG) finite element schemes, is still a very active research field, 
which has started with the pioneering works of Bassi and Rebay \cite{BR97},
and Baumann and Oden \cite{BO99,BO99b}. Later, several high order DG methods 
for the incompressible and compressible Navier-Stokes equations have been proposed, see for 
example \cite{BCPR06,HH06,BCPR07,Gas07,SFE07,HH08,Dum10,FW11,NPC11,RC12,RCV13,CAB13,KKO13}. We also would like to mention recent works on 
semi-implicit DG schemes that can be found in \cite{TBR13,GR10,Dol08,DF04,DFH07},  
to which our approach is indirectly related. 

The algorithm proposed in this article makes use of the novel family of 
staggered semi-implicit DG schemes that has been introduced in \cite{TD14,TD15,TD16} for the
incompressible Navier-Stokes equations in two and three space dimensions and which was later also 
extended to the full compressible regime in \cite{TD17}, following the work outlined in \cite{PM05,TV12,DC16}.
These arbitrary high order accurate DG schemes are constructed on \textit{staggered unstructured meshes}. 
The pressure, the density and the energy are defined on the triangular or tetrahedral 
\textit{primal grid}, whereas the velocity is computed on a \textit{face-based staggered dual mesh}.
While the use of staggered grids is a very common practice in the finite difference 
and finite volume framework (see e.g. \cite{HW65,PS72,CasulliCompressible}), its use is not so widespread in the context of 
high order DG schemes. The first staggered DG methods, which adopted a \textit{vertex-based} 
dual grid, have been proposed in  \cite{LSTZ07,LSTZ08}. Other recent high order 
staggered DG algorithms that rely on an \textit{edge-based} dual grid have been {\color{cr1} advanced} 
in \cite{CE06,CL12}. For high order staggered semi-implicit discontinuous Galerkin schemes on 
uniform and adaptive Cartesian meshes, see \cite{FambriDumbser,AMRDGSI}.   

Focusing on the incompressible model, we propose two different approaches for the 
computation of the nonlinear convective terms. On the one hand, we consider the methodology already 
introduced in \cite{TD16}. There, the convective subsystem for the velocity is solved considering 
the Rusanov flux function for an explicit upwind-type discretization of the nonlinear convective 
terms. Instead, the viscous terms are discretized implicitly, making again use of the dual mesh 
in order to obtain the discrete gradients, without needing any additional numerical flux function
for the viscous terms. One of the major 
drawbacks of this approach is its high computational cost coming from the small 
time step dictated by the CFL stability condition due to the explicit discretization of the convective terms. Moreover, to avoid spurious oscillations, a limiter 
should be used (see \cite{TD17}). As an alternative option, which is at the same time able to deal with 
large gradients and substantially reduces the computational cost, we propose the use of an \textit{{\color{cr1} Eulerian-Lagrangian}} approach, 
recently forwarded in \cite{TB19} also in the context of high order {\color{cr12} in space}
staggered DG schemes. The trajectory of the flow particles is
followed backward in time by integrating the associated trajectory equations at 
the aid of a high order Taylor series expansion, where time derivatives are 
replaced by spatial derivatives using the Cauchy-Kovalevskaya procedure, similar to the ADER approach
of Toro and Titarev \cite{toro4,titarevtoro,Toro:2006a}. 
The high order spatial discretization of the DG scheme is then employed to 
obtain a  high order time integration for each point needed to
solve numerically the advection part of the governing equations.
For further information on efficient {\color{cr1} semi-Lagrangian and Eulerian-Lagrangian} schemes we refer the reader to
\cite{Wel55,Wii59,Rob81,BD82,Cas88,CS11,BDR13,BPR19,Bon00,RBS06,BFR18}.

The use of the Boussinesq assumption yields the coupling of the 
incompressible Navier-Stokes equations with an additional 
conservation equation for the temperature. 
The computation of the related advection and diffusion terms is performed 
similarly to what is done for the velocity in the momentum equation. Nevertheless, let us remark 
that the temperature is defined on the primal mesh so that interpolation 
from one mesh to the other is avoided in the fully Eulerian approach. 
Once the new temperature is known, the gravity source term in the momentum equation can be evaluated.
Finally, the pressure Poisson equation is solved and the velocity at the new time step is computed.

Regarding the compressible Navier-Stokes equations, 
we extend the numerical scheme introduced in \cite{TD17} to consider 
the additional gravity terms. To this end, two new terms are included 
in the pressure system which has been obtained by formal substitution 
of the discrete momentum equation into the discrete energy equation. 
The first gravity term, coming from the momentum equation, is computed 
jointly with the convective and viscous terms of the momentum equation 
at the beginning of each time step. Meanwhile, the gravity term embedded in the energy equation is computed at each Picard iteration 
using the updated values of the linear momentum density.

The rest of the paper is organized as follows. In Section 
\ref{sec:governing_equations} we recall the incompressible and compressible 
Navier-Stokes equations. For the incompressible model, the Boussinesq 
assumption is made to account for fluid flow with small temperature variations 
under gravity effects. 
Concerning the compressible model, we consider the full   
Navier-Stokes equations including the conservation law for the total energy density
and assuming here the equation of state for an ideal gas. 
Section \ref{sec:numerical_method} is devoted to the description of the 
semi-implicit staggered DG method used to solve the incompressible 
model in two and three space dimensions. We start by recalling some basic 
definitions about the usage of staggered meshes and the polynomial spaces which are employed. 
Next, we derive the numerical method considering two different frameworks 
for the discretization of convective and diffusive terms, namely an Eulerian and 
{\color{cr1} an Eulerian-Lagrangian} approach. The extension of the algorithm to the compressible 
case is described in Section \ref{sec:numerical_method_cns}. 
Several benchmarks are presented in Section \ref{sec:num_res}, aiming at 
assessing the validity, efficiency and the robustness of our novel numerical schemes. 
The main pros and cons 
of the Eulerian and the {\color{cr1} Eulerian-Lagrangian} approaches are analyzed as well. 
Finally, we compare the results obtained with the incompressible solver against 
those computed with the compressible solver in the low Mach number regime.

% % % % % % % % % % % % % % % % % % % % % % % % % % % % % %
%               Mathematical model                        %
% % % % % % % % % % % % % % % % % % % % % % % % % % % % % %
\section{Governing equations}\label{sec:governing_equations}

As already mentioned, natural convection problems may be 
studied using two different models: the incompressible and the compressible Navier-Stokes equations, both including proper gravitational terms. The choice of the model usually 
depends on specific features of the flow under consideration, like 
the magnitude of the temperature fluctuations or the importance of 
capturing density variations. Here, we are mainly interested 
in small temperature changes, so that we will firt focus on the incompressible case. 
Later, even the full compressible model will be studied and numerical results will be compared considering both approaches.

% % % % % % % % % % % % % % % % % % % % % % % % % % % % % %
% % % % % % % % % % % % % % % % % % % % % % % % % % % % % %
\subsection{Incompressible Navier-Stokes equations}
Let us consider the laminar flow of a single phase Newtonian fluid without 
neither radiation nor chemical reactions. Let furthermore $\beta$ be the thermal expansion coefficient of the fluid,
$\theta_{0}$ the reference temperature of the flow and 
$\theta_{\mathrm{b}}-\theta_{0}$ the maximum temperature fluctuation. Under the assumption
\begin{equation}
\beta (\theta_{\mathrm{b}}-\theta_{0})\ll 1,
\end{equation}
the Boussinesq approximation for buoyancy-driven flows holds. 
Therefore, natural convection problems with small 
temperature gradients may be analyzed by solving the 
system of incompressible Navier-Stokes equations 
coupled with an energy conservation equation through 
an additional source term in the momentum equation. The governing 
PDE system reads as follows:
\begin{eqnarray}
\nabla \cdot \mathbf{v}=0 \label{eq:CS_2},\\
\frac{\partial \mathbf{v}}{\partial t}+\nabla \cdot \mathbf{F}^{ \mathbf{v}}_c + \nabla p= \nabla \cdot \left( \nu \nabla \mathbf{v} \right) +\left(  1- \beta \delta\theta\right) \mathbf{g}  \label{eq:CS_2_2_0}, \\
\frac{\partial{\theta}}{\partial{t}}+\nabla \cdot \TFe_c= \nabla \cdot \left( \alpha \nabla \theta\right),
\label{eq:cons_eq_temperature}
\end{eqnarray}
where $\mathbf{v}$ is the velocity field; $p=P/\rho$ 
indicates the normalized fluid pressure; $P$ is the 
physical pressure and $\rho$ is the fluid density, which according to the Boussinesq
approximation is assumed to be constant everywhere apart from the gravity source term
in the momentum equation; 
$\nu=\mu / \rho$ is the kinematic viscosity  coefficient; 
$\mathbf{F}^{ \mathbf{v}}_c=\mathbf{v} \otimes \mathbf{v}$ is the flux 
tensor of the nonlinear convective terms; 
$\delta \theta := \theta-\theta_{0}$ is the temperature difference;
$\mathbf{g}$ is the gravity acceleration; $\theta$ is the temperature; $\mathbf{F}_c^{\theta}= \mathbf{v}\theta$ is the flux 
tensor of the nonlinear convective terms of the energy equation; 
$\alpha = \kappa / (\rho c_P)$ is the thermal diffusivity, which depends on the thermal conductivity $\kappa$, the density $\rho$ and the heat capacity $c_P$. Let us also introduce the following notation:
\begin{eqnarray}
\TF^{ \mathbf{v}}:=\TF_c^{ \mathbf{v}}-  \nu \nabla \mathbf{v}, \qquad \TL^{ \mathbf{v}}\left(\mathbf{v}\right) := \frac{\partial }{\partial t}\mathbf{v}+\nabla \cdot \TF^{ \mathbf{v}}, \\
\TFe:=\TFe_c-\nabla \cdot \alpha \nabla \theta, \qquad 
\TLe\left(\theta,\mathbf{v}\right) := \frac{\partial }{\partial t}\theta + \nabla \cdot \TFe.
\end{eqnarray}

% % % % % % % % % % % % % % % % % % % % % % % % % % % % % %
% % % % % % % % % % % % % % % % % % % % % % % % % % % % % %
\subsection{Compressible Navier-Stokes equations}
Large temperature fluctuations may produce substantial changes in 
the density of a fluid. As a consequence, the incompressible model 
\eqref{eq:CS_2}-\eqref{eq:cons_eq_temperature} is no longer valid 
for large temperature fluctuations.  
In this case, one must use the full compressible 
Navier-Stokes equations under gravitational effects, that read
\begin{eqnarray}
\frac{\partial \rho}{\partial t}+ \nabla \cdot \left( \rho\mathbf{v}\right) =0 \label{eq:CNS_mass},\\
\frac{\partial \rho\mathbf{v}}{\partial t}+\nabla \cdot \mathbf{F}^{\rho\mathbf{v}}_c + \nabla P= \nabla \cdot \boldsymbol{\sigma} + \rho \mathbf{g}  \label{eq:CNS_momentum}, \\
\frac{\partial{\rho E}}{\partial{t}}+\nabla \cdot \mathbf{F}^{\rho E}_c= \nabla \cdot \left( \boldsymbol{\sigma}\mathbf{v} + \kappa \nabla \theta\right)+ \rho\mathbf{g}\cdot\mathbf{v},
\label{eq:CNS_energy}
\end{eqnarray}
where  $\mathbf{F}^{\rho\mathbf{v}}_c = \rho\mathbf{v} \otimes \mathbf{v}$ 
is the convective flux for the momentum equation; 
$\boldsymbol{\sigma}= \mu (\nabla \vv +\nabla \vv^\top)-\frac{2}{3} \mu \left( \nabla \cdot \vv \right)\idm $ 
is the viscous stress tensor;  
$\rhoEE=\rhoe+\rhok$ is the total energy density; $\rhok = \frac{1}{2}\rho \vv^2$ 
is the kinetic energy density; $e=e(P,\rho)$ represents the specific internal 
energy per unit mass and is given by the equation of state (EOS) as a function 
of the pressure $P$ and the density $\rho$; $H = e+\frac{P}{\rho}$ denotes 
the specific enthalpy; $\mathbf{F}^{\rho E}_c= \mathbf{v} \left( \rho E + P\right)= \rho \mathbf{v} k + \rho\mathbf{v} H$ 
is the convective flux for the energy conservation equation; $\kappa$ is the thermal conductivity coefficient. Let us also define the following operators:
\begin{equation}
\mathbf{F}^{\rho\mathbf{v}} := \mathbf{F}^{\rho\mathbf{v}}_c - \boldsymbol{\sigma},  \qquad \TL^{\rho\mathbf{v}}\left(\rho\mathbf{v}\right) := \frac{\partial }{\partial t}\mathbf{v}+\nabla \cdot \TF^{\rho\mathbf{v}}.
\end{equation}

In our approach we assume that we are dealing with an ideal gas, so that 
the thermal and caloric equation of state (EOS) that are needed to close the above system are given by
\begin{equation}
P= \rho R \theta, \qquad e=c_{v} \theta, \qquad \textnormal{hence} \qquad P = (\gamma-1) \rho e, 
\end{equation}
with $R = c_{P}+c_{v}$ denoting the specific gas constant; $c_{P}$ and 
$c_{v}$ representing the heat capacities at constant pressure and at constant volume, respectively, 
and $\gamma = c_P / c_v$ being the usual ratio of specific heats. 

% % % % % % % % % % % % % % % % % % % % % % % % % % % % % %
% % % % % % % % % % % % % % % % % % % % % % % % % % % % % %
%               Numerical method                          %
% % % % % % % % % % % % % % % % % % % % % % % % % % % % % %
% % % % % % % % % % % % % % % % % % % % % % % % % % % % % %
\section{Numerical method for the incompressible model}\label{sec:numerical_method}
System \eqref{eq:CS_2}-\eqref{eq:cons_eq_temperature} will be solved starting by the
staggered semi-implicit discontinuous Galerkin scheme detailed in \cite{TD14,TD15,TD16}. Here, we recall the main ingredients of the algorithm, while for an exhaustive description the reader is referred to the aforementioned references.

% % % % % % % % % % % % % % % % % % % % % % % % % % % % % %
% % % % % % % % % % % % % % % % % % % % % % % % % % % % % %
\subsection{Staggered unstructured mesh}\label{sec:mesh}
The computational domain is discretized using a face-based
staggered unstructured meshes, as adopted in \cite{TD14,TD15,TD16,TD17}.
In what follows, we briefly summarize the grid construction
and the main notation for the two dimensional triangular grid.
After that, the primal and dual spatial elements are extended
to the three dimensional case.

The spatial computational domain 
$\Omega \subset \mathbb{R}^2$ is covered with a set of $\Ni$ 
non-overlapping triangular elements $\TT_i$ with $i=1 \ldots \Ni$. 
By denoting with $\Nj$ the total number of edges, the $j-$th edge 
will be called $\Gamma_j$. $\B(\Omega)$ refers to the set of indices 
$j$ corresponding to boundary edges.
The three edges of each triangle $\TT_i$ constitute the set $S_i$ 
defined by $S_i=\{j \in [1,\Nj] \,\, | \,\, \Gamma_j \mbox{ is an edge of }\TT_i \}$. 
For every $j\in [1\ldots \Nj]-\B(\Omega)$ there exist two triangles 
$i_1$ and $i_2$ that share $\Gamma_j$. We assign arbitrarily a left 
and a right triangle called $\ell(j)$ and $r(j)$, respectively. 
The standard positive direction is assumed to be from left to right. 
Let $\nv_{j}$ denote the unit normal vector defined on the edge $j$ 
and oriented with respect to the positive direction. 
For every triangular element $i$ and edge $j \in S_i$,
the neighbor triangle of element $\TT_i$ that share the edge $\Gamma_j$ 
is denoted by $\p(i,j)$.

\begin{figure}[ht]
	\begin{center}
	    \begin{tikzpicture}[line cap=round,line join=round,>=triangle 45,x=0.6373937677053826cm,y=0.6177884615384613cm]
	    \clip(2.11,-8.53) rectangle (16.23,3.95);
	    \fill[color=zzttqq,fill=zzttqq,fill opacity=0.1] (5.19,-3.03) -- (9,3) -- (13,-5) -- cycle;
	    \fill[color=qqwuqq,fill=qqwuqq,fill opacity=0.05] (9,3) -- (14.49,2.37) -- (13,-5) -- cycle;
	    \fill[color=qqwuqq,fill=qqwuqq,fill opacity=0.05] (9,3) -- (4.13,2.43) -- (5.19,-3.03) -- cycle;
	    \fill[color=qqwuqq,fill=qqwuqq,fill opacity=0.05] (5.19,-3.03) -- (4.77,-7.83) -- (13,-5) -- cycle;
	    \fill[color=zzttqq,fill=zzttqq,fill opacity=0.1] (9.27,-1.79) -- (9,3) -- (5.67,0.43) -- (5.19,-3.03) -- cycle;
	    \fill[color=zzttqq,fill=zzttqq,fill opacity=0.1] (9.27,-1.79) -- (9,3) -- (12.15,1.37) -- (13,-5) -- cycle;
	    \fill[color=zzttqq,fill=zzttqq,fill opacity=0.1] (5.19,-3.03) -- (7.47,-5.49) -- (13,-5) -- (9.27,-1.79) -- cycle;
	    \fill[color=qqttzz,fill=qqttzz,fill opacity=0.1] (13,-5) -- (5.19,-3.03) -- (9.27,-1.79) -- cycle;
	    \draw [color=zzttqq] (5.19,-3.03)-- (9,3);
	    \draw [color=zzttqq] (9,3)-- (13,-5);
	    \draw [color=zzttqq] (13,-5)-- (5.19,-3.03);
	    \draw [color=qqwuqq] (9,3)-- (14.49,2.37);
	    \draw [color=qqwuqq] (14.49,2.37)-- (13,-5);
	    \draw [color=qqwuqq] (13,-5)-- (9,3);
	    \draw [color=qqwuqq] (9,3)-- (4.13,2.43);
	    \draw [color=qqwuqq] (4.13,2.43)-- (5.19,-3.03);
	    \draw [color=qqwuqq] (5.19,-3.03)-- (9,3);
	    \draw [color=qqwuqq] (5.19,-3.03)-- (4.77,-7.83);
	    \draw [color=qqwuqq] (4.77,-7.83)-- (13,-5);
	    \draw [color=qqwuqq] (13,-5)-- (5.19,-3.03);
	    \draw (9.51,-1.19) node[anchor=north west] {$i$};
	    \draw (12.1,1.7) node[anchor=north west] {$\mathbb{\wp}(i,j_1)$};
	    \draw (4.3,1.6) node[anchor=north west] {$\mathbb{\wp}(i,j_2)$};
	    \draw (7,-5.5) node[anchor=north west] {$\mathbb{\wp}(i,j_3)$};
%	    \draw (11.25,-0.47) node[anchor=north west] {$j_1$};
	    \draw (6.64,1.43) node[anchor=north west] {$\Gamma_{j_2}$};
	    \draw (8.63,-3.93) node[anchor=north west] {$\Gamma_{j_3}$};
	    \draw (4.39,-2.79) node[anchor=north west] {$n_1$};
	    \draw (13.07,-4.83) node[anchor=north west] {$n_2$};
	    \draw (9.05,4.03) node[anchor=north west] {$n_3$};
	    \draw (7.71,0.35) node[anchor=north west] {$\TT_i$};
	    \draw (11.5,2.6) node[anchor=north west] {$\TT_{\mathbb{\wp}(i,j_1)}$};
	    \draw [color=zzttqq] (9.27,-1.79)-- (9,3);
	    \draw [color=zzttqq] (9,3)-- (5.67,0.43);
	    \draw [color=zzttqq] (5.67,0.43)-- (5.19,-3.03);
	    \draw [color=zzttqq] (5.19,-3.03)-- (9.27,-1.79);
	    \draw [color=zzttqq] (9.27,-1.79)-- (9,3);
	    \draw [color=zzttqq] (9,3)-- (12.15,1.37);
	    \draw [color=zzttqq] (12.15,1.37)-- (13,-5);
	    \draw [color=zzttqq] (13,-5)-- (9.27,-1.79);
	    \draw [color=zzttqq] (5.19,-3.03)-- (7.47,-5.49);
	    \draw [color=zzttqq] (7.47,-5.49)-- (13,-5);
	    \draw [color=zzttqq] (13,-5)-- (9.27,-1.79);
	    \draw [color=zzttqq] (9.27,-1.79)-- (5.19,-3.03);
	    \draw (10.49,1.15) node[anchor=north west] {$\QQ_{j_1}$};
	    \draw [color=ffqqqq](10.07,-0.83) node[anchor=north west] {$\Gamma_{j_1}$};
	    \draw [line width=1.6pt,color=ffqqqq] (9,3)-- (13,-5);
	    \draw [color=qqttzz] (13,-5)-- (5.19,-3.03);
	    \draw [color=qqttzz] (5.19,-3.03)-- (9.27,-1.79);
	    \draw [color=qqttzz] (9.27,-1.79)-- (13,-5);
	    \draw [color=qqttzz](8.35,-2.17) node[anchor=north west] {$\TT_{i,j_3}$};
	    \draw (5.19,-3.03)-- (5.67,0.43);
	    \draw (5.67,0.43)-- (9,3);
	    \draw (9,3)-- (9.27,-1.79);
	    \draw (9.27,-1.79)-- (5.19,-3.03);
	    \draw (9,3)-- (12.15,1.37);
	    \draw (12.15,1.37)-- (13,-5);
	    \draw (13,-5)-- (9.27,-1.79);
	    \draw (13,-5)-- (7.47,-5.49);
	    \draw (7.47,-5.49)-- (5.19,-3.03);
	    \begin{scriptsize}
	    \fill [color=qqqqff] (5.19,-3.03) circle (1.5pt);
	    \fill [color=qqqqff] (9,3) circle (1.5pt);
	    \fill [color=qqqqff] (13,-5) circle (1.5pt);
	    \fill [color=qqqqff] (9.27,-1.79) circle (1.5pt);
	    \fill [color=qqqqff] (14.49,2.37) circle (1.5pt);
	    \fill [color=qqqqff] (4.13,2.43) circle (1.5pt);
	    \fill [color=qqqqff] (4.77,-7.83) circle (1.5pt);
	    \fill [color=qqqqff] (12.15,1.37) circle (1.5pt);
	    \fill [color=qqqqff] (5.67,0.43) circle (1.5pt);
	    \fill [color=qqqqff] (7.47,-5.49) circle (1.5pt);
	    \end{scriptsize}
	    \end{tikzpicture}
		\caption{Triangular mesh element with its three neighbors and the associated staggered edge-based dual control volumes, together with the notation
			used throughout the paper.}
		\label{fig.1}
	\end{center}
\end{figure}
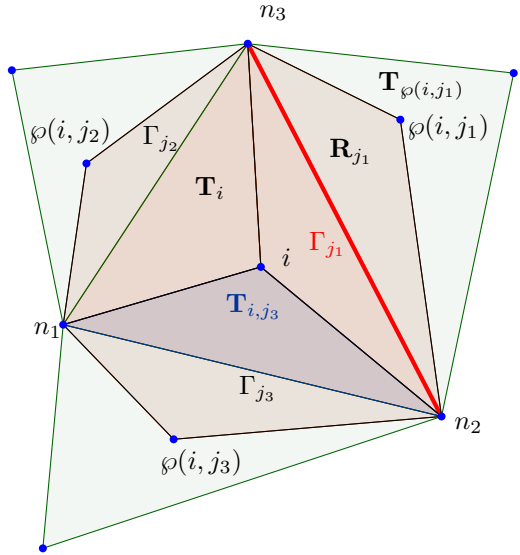

For every $j\in [1, \Nj]-\B(\Omega)$ the quadrilateral element 
associated to $j$ is called $\QQ_j$ and it is defined, in general, 
by the two centers of gravity of $\ell(j)$ and $r(j)$ and the two 
terminal nodes of $\Gamma_j$, see also \cite{BDDV98,THD09,TD14sw,CL12}.
We denote by $\TT_{i,j}=\QQ_j \cap \TT_i$ the intersection element for 
every $i$ and $j \in S_i$. Figure $\ref{fig.1}$ summarizes the notation, the primal triangular mesh and the dual quadrilateral grid. 
According to \cite{TD15}, we will call the \textit{main grid}, or \textit{primal grid}, the mesh of triangular elements $\{\TT_i \}_{i \in [1, \Ni]}$, whereas the quadrilateral grid $\{\QQ_j \}_{j \in [1, N_d]}$ is addressed as the \textit{dual grid}. 

The definitions given above are then readily extended to three space dimensions within a domain $\Omega \subset \mathbb{R}^3$. 
\begin{figure}[h!t]
	\begin{center}
		\begin{tikzpicture}[line cap=round,line join=round,>=triangle 45,x=0.8cm,y=0.8cm]
		\clip(3.26,-5.21) rectangle (8.37,1.57);
		\fill[color=zzttqq,fill=zzttqq,fill opacity=0.1] (4,-4) -- (8,-4) -- (4,-2) -- cycle;
		\fill[color=zzttqq,fill=zzttqq,fill opacity=0.1] (4,-2) -- (6,1) -- (8,-4) -- cycle;
		\fill[color=zzttqq,fill=zzttqq,fill opacity=0.1] (4,-2) -- (6,1) -- (4,-4) -- cycle;
		\fill[color=zzttqq,fill=zzttqq,fill opacity=0.1] (6,1) -- (4,-4) -- (8,-4) -- cycle;
		\fill[color=zzttqq,fill=zzttqq,fill opacity=0.1] (16,-4) -- (20,-4) -- (16,-2) -- cycle;
		\fill[color=zzttqq,fill=zzttqq,fill opacity=0.1] (16,-2) -- (18,1) -- (20,-4) -- cycle;
		\fill[color=zzttqq,fill=zzttqq,fill opacity=0.1] (18,1) -- (16,-4) -- (20,-4) -- cycle;
		\fill[color=zzttqq,fill=zzttqq,fill opacity=0.1] (16,-2) -- (12.56,-0.2) -- (16,-4) -- cycle;
		\fill[color=qqqqff,fill=qqqqff,fill opacity=0.1] (12.56,-0.2) -- (15.16,-1.08) -- (16,-4) -- cycle;
		\fill[color=qqqqff,fill=qqqqff,fill opacity=0.1] (15.16,-1.08) -- (16,-2) -- (12.56,-0.2) -- cycle;
		\fill[color=qqqqff,fill=qqqqff,fill opacity=0.1] (15.16,-1.08) -- (16,-4) -- (16,-2) -- cycle;
		\fill[color=qqqqff,fill=qqqqff,fill opacity=0.1] (12.56,-0.2) -- (13.78,-2.9) -- (16,-4) -- cycle;
		\fill[color=qqqqff,fill=qqqqff,fill opacity=0.1] (13.78,-2.9) -- (16,-2) -- (12.56,-0.2) -- cycle;
		\fill[color=qqqqff,fill=qqqqff,fill opacity=0.1] (13.78,-2.9) -- (16,-4) -- (16,-2) -- cycle;
		\fill[color=zzttqq,fill=zzttqq,fill opacity=0.15] (26,-6) -- (28,-3) -- (30,-8) -- cycle;
		\draw [color=zzttqq] (4,-4)-- (8,-4);
		\draw [color=zzttqq] (4,-2)-- (4,-4);
		\draw [color=zzttqq] (4,-2)-- (6,1);
		\draw [color=zzttqq] (6,1)-- (8,-4);
		\draw [dash pattern=on 1pt off 1pt,color=zzttqq] (8,-4)-- (4,-2);
		\draw [color=zzttqq] (4,-2)-- (6,1);
		\draw [color=zzttqq] (6,1)-- (4,-4);
		\draw [color=zzttqq] (4,-4)-- (4,-2);
		\draw [color=zzttqq] (6,1)-- (4,-4);
		\draw [color=zzttqq] (4,-4)-- (8,-4);
		\draw [color=zzttqq] (8,-4)-- (6,1);
		\draw [color=zzttqq] (16,-4)-- (20,-4);
		\draw [color=zzttqq] (16,-2)-- (16,-4);
		\draw [color=zzttqq] (16,-2)-- (18,1);
		\draw [color=zzttqq] (18,1)-- (20,-4);
		\draw [dash pattern=on 1pt off 1pt,color=zzttqq] (20,-4)-- (16,-2);
		\draw [color=zzttqq] (18,1)-- (16,-4);
		\draw [color=zzttqq] (16,-4)-- (20,-4);
		\draw [color=zzttqq] (20,-4)-- (18,1);
		\draw [color=zzttqq] (16,-2)-- (16.67,-2.33);
		\draw [shift={(16,-3)},dotted]  plot[domain=1.11:2.46,variable=\t]({1*4.47*cos(\t r)+0*4.47*sin(\t r)},{0*4.47*cos(\t r)+1*4.47*sin(\t r)});
		\draw [color=zzttqq] (12.56,-0.2)-- (16,-4);
		\draw [color=zzttqq] (16,-4)-- (16,-2);
		\draw [shift={(16,-3)},dotted]  plot[domain=0.66:1.98,variable=\t]({1*2.12*cos(\t r)+0*2.12*sin(\t r)},{0*2.12*cos(\t r)+1*2.12*sin(\t r)});
		\draw [color=qqqqff] (12.56,-0.2)-- (15.16,-1.08);
		\draw [color=qqqqff] (15.16,-1.08)-- (16,-4);
		\draw [color=qqqqff] (16,-4)-- (12.56,-0.2);
		\draw [color=qqqqff] (15.16,-1.08)-- (16,-2);
		\draw [color=qqqqff] (12.56,-0.2)-- (15.16,-1.08);
		\draw [color=qqqqff] (15.16,-1.08)-- (16,-4);
		\draw [color=qqqqff] (16,-4)-- (16,-2);
		\draw [color=qqqqff] (16,-2)-- (15.16,-1.08);
		\draw [color=qqqqff] (12.56,-0.2)-- (13.78,-2.9);
		\draw [color=qqqqff] (13.78,-2.9)-- (16,-4);
		\draw [color=qqqqff] (16,-4)-- (12.56,-0.2);
		\draw [dash pattern=on 1pt off 1pt,color=qqqqff] (16,-2)-- (12.56,-0.2);
		\draw [color=qqqqff] (12.56,-0.2)-- (13.78,-2.9);
		\draw [color=qqqqff] (13.78,-2.9)-- (16,-4);
		\draw [color=qqqqff] (16,-4)-- (16,-2);
		\draw [dash pattern=on 1pt off 1pt,color=qqqqff] (16,-2)-- (13.78,-2.9);
		\draw [color=zzttqq] (26,-8)-- (30,-8);
		\draw [color=zzttqq] (26,-6)-- (26,-8);
		\draw [color=zzttqq] (26,-6)-- (28,-3);
		\draw [color=zzttqq] (28,-3)-- (30,-8);
		\draw [color=zzttqq] (30,-8)-- (26,-6);
		\draw [color=zzttqq] (26,-6)-- (28,-3);
		\draw [color=zzttqq] (28,-3)-- (26,-8);
		\draw [color=zzttqq] (26,-8)-- (26,-6);
		\draw [color=zzttqq] (28,-3)-- (26,-8);
		\draw [color=zzttqq] (26,-8)-- (30,-8);
		\draw [color=zzttqq] (30,-8)-- (28,-3);
		\draw [dash pattern=on 1pt off 1pt,color=qqqqff] (30,-4)-- (26,-6);
		\draw [color=qqqqff] (30,-8)-- (30,-4);
		\draw [color=qqqqff] (28,-3)-- (30,-4);
		\draw [color=qqqqff] (30,-4)-- (30,-8);
		\draw [color=qqqqff] (30,-8)-- (28,-3);
		\draw [color=qqqqff] (26,-6)-- (28,-3);
		\draw [color=qqqqff] (28,-3)-- (30,-4);
		\draw [dash pattern=on 1pt off 1pt,color=ffqqqq] (28.46,-5.11)-- (27.23,-5.8);
		\draw [color=qqqqff] (28.67,-4.68)-- (30,-4);
		\draw [color=ffqqqq] (27.23,-5.8)-- (27.86,-5.45);
		\draw (5.56,-1.25) node[anchor=north west] {$\TT_i$};
		\draw (6.85,0.18) node[anchor=north west] {$\Gamma_{j_1}$};
		\draw (6.8,-2.49) node[anchor=north west] {$\Gamma_{j_2}$};
		\draw (4.02,-0.04) node[anchor=north west] {$\Gamma_{j_3}$};
		\draw (4.63,-4.18) node[anchor=north west] {$\Gamma_{j_4}$};
		\draw (14.4,0.1) node[anchor=north west] {$\QQ_j$};
		\draw (16.02,-0.08) node[anchor=north west] {$i$};
		\draw (17.82,-1.74) node[anchor=north west] {$\TT_i$};
		\draw (13.21,-3.03) node[anchor=north west] {$\mathbb{\wp}(i,j)$};
		\draw [->,dash pattern=on 1pt off 1pt] (28.02,-5.62) -- (28.66,-5.29);
		\draw (27.13,-5.78) node[anchor=north west] {$\ell(j)$};
		\draw (28.99,-4.46) node[anchor=north west] {$r(j)$};
		\draw (28.36,-5.41) node[anchor=north west] {$\vec{n_j}$};
		\draw (30.3,-7.62) node[anchor=north west] {$\Gamma_j$};
		\begin{scriptsize}
		\fill [color=uuuuuu] (4,-4) circle (1.0pt);
		\fill [color=uuuuuu] (8,-4) circle (1.0pt);
		\fill [color=uuuuuu] (4,-2) circle (1.0pt);
		\fill [color=uuuuuu] (6,1) circle (1.0pt);
		\fill [color=uuuuuu] (16,-4) circle (1.0pt);
		\fill [color=uuuuuu] (20,-4) circle (1.0pt);
		\fill [color=uuuuuu] (16,-2) circle (1.0pt);
		\fill [color=uuuuuu] (18,1) circle (1.0pt);
		\fill [color=uuuuuu] (12.56,-0.2) circle (1.0pt);
		\draw [color=ffqqqq] (17.68,-1.7) circle (1.5pt);
		\draw [color=ffqqqq] (15.16,-1.08) circle (1.5pt);
		\draw [color=ffqqqq] (13.78,-2.9) circle (1.5pt);
		\fill [color=uuuuuu] (26,-8) circle (1.0pt);
		\fill [color=uuuuuu] (30,-8) circle (1.0pt);
		\fill [color=uuuuuu] (26,-6) circle (1.0pt);
		\fill [color=uuuuuu] (28,-3) circle (1.0pt);
		\fill [color=qqqqff] (30,-4) circle (1.5pt);
		\draw[color=qqqqff] (0.19,5.94) node {$L$};
		\draw [color=ffqqqq] (28.46,-5.11) circle (1.5pt);
		\draw [color=ffqqqq] (27.23,-5.8) circle (1.5pt);
		\end{scriptsize}
		\end{tikzpicture}
		\begin{tikzpicture}[line cap=round,line join=round,>=triangle 45,x=0.8cm,y=0.8cm]
		\clip(11.74,-4.85) rectangle (20.91,2.12);
		\fill[color=zzttqq,fill=zzttqq,fill opacity=0.1] (4,-4) -- (8,-4) -- (4,-2) -- cycle;
		\fill[color=zzttqq,fill=zzttqq,fill opacity=0.1] (4,-2) -- (6,1) -- (8,-4) -- cycle;
		\fill[color=zzttqq,fill=zzttqq,fill opacity=0.1] (4,-2) -- (6,1) -- (4,-4) -- cycle;
		\fill[color=zzttqq,fill=zzttqq,fill opacity=0.1] (6,1) -- (4,-4) -- (8,-4) -- cycle;
		\fill[color=zzttqq,fill=zzttqq,fill opacity=0.1] (16,-4) -- (20,-4) -- (16,-2) -- cycle;
		\fill[color=zzttqq,fill=zzttqq,fill opacity=0.1] (16,-2) -- (18,1) -- (20,-4) -- cycle;
		\fill[color=zzttqq,fill=zzttqq,fill opacity=0.1] (18,1) -- (16,-4) -- (20,-4) -- cycle;
		\fill[color=zzttqq,fill=zzttqq,fill opacity=0.1] (16,-2) -- (12.56,-0.2) -- (16,-4) -- cycle;
		\fill[color=qqqqff,fill=qqqqff,fill opacity=0.1] (12.56,-0.2) -- (15.16,-1.08) -- (16,-4) -- cycle;
		\fill[color=qqqqff,fill=qqqqff,fill opacity=0.1] (15.16,-1.08) -- (16,-2) -- (12.56,-0.2) -- cycle;
		\fill[color=qqqqff,fill=qqqqff,fill opacity=0.1] (15.16,-1.08) -- (16,-4) -- (16,-2) -- cycle;
		\fill[color=qqqqff,fill=qqqqff,fill opacity=0.1] (12.56,-0.2) -- (13.78,-2.9) -- (16,-4) -- cycle;
		\fill[color=qqqqff,fill=qqqqff,fill opacity=0.1] (13.78,-2.9) -- (16,-2) -- (12.56,-0.2) -- cycle;
		\fill[color=qqqqff,fill=qqqqff,fill opacity=0.1] (13.78,-2.9) -- (16,-4) -- (16,-2) -- cycle;
		\fill[color=zzttqq,fill=zzttqq,fill opacity=0.15] (26,-6) -- (28,-3) -- (30,-8) -- cycle;
		\draw [color=zzttqq] (4,-4)-- (8,-4);
		\draw [color=zzttqq] (4,-2)-- (4,-4);
		\draw [color=zzttqq] (4,-2)-- (6,1);
		\draw [color=zzttqq] (6,1)-- (8,-4);
		\draw [dash pattern=on 1pt off 1pt,color=zzttqq] (8,-4)-- (4,-2);
		\draw [color=zzttqq] (4,-2)-- (6,1);
		\draw [color=zzttqq] (6,1)-- (4,-4);
		\draw [color=zzttqq] (4,-4)-- (4,-2);
		\draw [color=zzttqq] (6,1)-- (4,-4);
		\draw [color=zzttqq] (4,-4)-- (8,-4);
		\draw [color=zzttqq] (8,-4)-- (6,1);
		\draw [color=zzttqq] (16,-4)-- (20,-4);
		\draw [color=zzttqq] (16,-2)-- (16,-4);
		\draw [color=zzttqq] (16,-2)-- (18,1);
		\draw [color=zzttqq] (18,1)-- (20,-4);
		\draw [dash pattern=on 1pt off 1pt,color=zzttqq] (20,-4)-- (16,-2);
		\draw [color=zzttqq] (18,1)-- (16,-4);
		\draw [color=zzttqq] (16,-4)-- (20,-4);
		\draw [color=zzttqq] (20,-4)-- (18,1);
		\draw [color=zzttqq] (16,-2)-- (16.67,-2.33);
		\draw [shift={(16,-3)},dotted]  plot[domain=1.11:2.46,variable=\t]({1*4.47*cos(\t r)+0*4.47*sin(\t r)},{0*4.47*cos(\t r)+1*4.47*sin(\t r)});
		\draw [color=zzttqq] (12.56,-0.2)-- (16,-4);
		\draw [color=zzttqq] (16,-4)-- (16,-2);
		\draw [shift={(16,-3)},dotted]  plot[domain=0.66:1.98,variable=\t]({1*2.12*cos(\t r)+0*2.12*sin(\t r)},{0*2.12*cos(\t r)+1*2.12*sin(\t r)});
		\draw [color=qqqqff] (12.56,-0.2)-- (15.16,-1.08);
		\draw [color=qqqqff] (15.16,-1.08)-- (16,-4);
		\draw [color=qqqqff] (16,-4)-- (12.56,-0.2);
		\draw [color=qqqqff] (15.16,-1.08)-- (16,-2);
		\draw [color=qqqqff] (12.56,-0.2)-- (15.16,-1.08);
		\draw [color=qqqqff] (15.16,-1.08)-- (16,-4);
		\draw [color=qqqqff] (16,-4)-- (16,-2);
		\draw [color=qqqqff] (16,-2)-- (15.16,-1.08);
		\draw [color=qqqqff] (12.56,-0.2)-- (13.78,-2.9);
		\draw [color=qqqqff] (13.78,-2.9)-- (16,-4);
		\draw [color=qqqqff] (16,-4)-- (12.56,-0.2);
		\draw [dash pattern=on 1pt off 1pt,color=qqqqff] (16,-2)-- (12.56,-0.2);
		\draw [color=qqqqff] (12.56,-0.2)-- (13.78,-2.9);
		\draw [color=qqqqff] (13.78,-2.9)-- (16,-4);
		\draw [color=qqqqff] (16,-4)-- (16,-2);
		\draw [dash pattern=on 1pt off 1pt,color=qqqqff] (16,-2)-- (13.78,-2.9);
		\draw [color=zzttqq] (26,-8)-- (30,-8);
		\draw [color=zzttqq] (26,-6)-- (26,-8);
		\draw [color=zzttqq] (26,-6)-- (28,-3);
		\draw [color=zzttqq] (28,-3)-- (30,-8);
		\draw [color=zzttqq] (30,-8)-- (26,-6);
		\draw [color=zzttqq] (26,-6)-- (28,-3);
		\draw [color=zzttqq] (28,-3)-- (26,-8);
		\draw [color=zzttqq] (26,-8)-- (26,-6);
		\draw [color=zzttqq] (28,-3)-- (26,-8);
		\draw [color=zzttqq] (26,-8)-- (30,-8);
		\draw [color=zzttqq] (30,-8)-- (28,-3);
		\draw [dash pattern=on 1pt off 1pt,color=qqqqff] (30,-4)-- (26,-6);
		\draw [color=qqqqff] (30,-8)-- (30,-4);
		\draw [color=qqqqff] (28,-3)-- (30,-4);
		\draw [color=qqqqff] (30,-4)-- (30,-8);
		\draw [color=qqqqff] (30,-8)-- (28,-3);
		\draw [color=qqqqff] (26,-6)-- (28,-3);
		\draw [color=qqqqff] (28,-3)-- (30,-4);
		\draw [dash pattern=on 1pt off 1pt,color=ffqqqq] (28.46,-5.11)-- (27.23,-5.8);
		\draw [color=qqqqff] (28.67,-4.68)-- (30,-4);
		\draw [color=ffqqqq] (27.23,-5.8)-- (27.86,-5.45);
		\draw (5.56,-1.25) node[anchor=north west] {$\TT_i$};
		\draw (6.85,0.18) node[anchor=north west] {$\Gamma_{j_1}$};
		\draw (6.99,-2.49) node[anchor=north west] {$\Gamma_{j_2}$};
		\draw (4.02,-0.04) node[anchor=north west] {$\Gamma_{j_3}$};
		\draw (4.63,-4.18) node[anchor=north west] {$\Gamma_{j_4}$};
		\draw (14.4,0.1) node[anchor=north west] {$\QQ_{j_3}$};
		\draw (16.02,-0.08) node[anchor=north west] {$i$};
		\draw (17.82,-1.74) node[anchor=north west] {$\TT_i$};
		\draw (13.21,-3.03) node[anchor=north west] {$\mathbb{\wp}(i,j_3)$};
		\draw [->,dash pattern=on 1pt off 1pt] (28.02,-5.62) -- (28.66,-5.29);
		\draw (27.13,-5.78) node[anchor=north west] {$\ell(j)$};
		\draw (28.99,-4.46) node[anchor=north west] {$r(j)$};
		\draw (28.36,-5.41) node[anchor=north west] {$\vec{n_j}$};
		\draw (30.3,-7.62) node[anchor=north west] {$\Gamma_j$};
		\begin{scriptsize}
		\fill [color=uuuuuu] (4,-4) circle (1.0pt);
		\fill [color=uuuuuu] (8,-4) circle (1.0pt);
		\fill [color=uuuuuu] (4,-2) circle (1.0pt);
		\fill [color=uuuuuu] (6,1) circle (1.0pt);
		\fill [color=uuuuuu] (16,-4) circle (1.0pt);
		\fill [color=uuuuuu] (20,-4) circle (1.0pt);
		\fill [color=uuuuuu] (16,-2) circle (1.0pt);
		\fill [color=uuuuuu] (18,1) circle (1.0pt);
		\fill [color=uuuuuu] (12.56,-0.2) circle (1.0pt);
		\draw [color=ffqqqq] (17.68,-1.7) circle (1.5pt);
		\draw [color=ffqqqq] (15.16,-1.08) circle (1.5pt);
		\draw [color=ffqqqq] (13.78,-2.9) circle (1.5pt);
		\fill [color=uuuuuu] (26,-8) circle (1.0pt);
		\fill [color=uuuuuu] (30,-8) circle (1.0pt);
		\fill [color=uuuuuu] (26,-6) circle (1.0pt);
		\fill [color=uuuuuu] (28,-3) circle (1.0pt);
		\fill [color=qqqqff] (30,-4) circle (1.5pt);
		\draw[color=qqqqff] (0.19,5.94) node {$L$};
		\draw [color=ffqqqq] (28.46,-5.11) circle (1.5pt);
		\draw [color=ffqqqq] (27.23,-5.8) circle (1.5pt);
		\end{scriptsize}
		\end{tikzpicture}
		\caption{Tetrahedral element of the primary mesh with $S_i=\{j_1, j_2, j_3, j_4\}$ (left) and non-standard dual face-based hexahedral element associated to the face $j_3$ (right).}
		\label{fig.MESH_1}
	\end{center}
\end{figure}
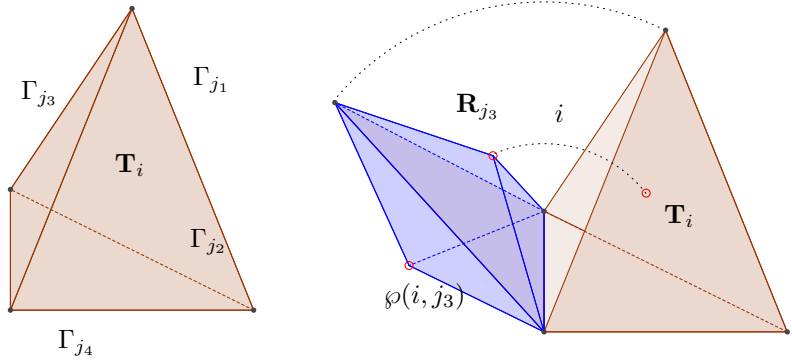
An example of the resulting main and dual grids in three space 
dimensions is reported in Figure \ref{fig.MESH_1}. The main grid 
consists of tetrahedral simplex elements, and the face-based dual
elements contain the three vertices of the common triangular face 
of two tetrahedra (a left and a right one), and the two barycenters 
of the two tetrahedra that share the same face. 
Therefore, in three space dimensions the dual grid consists of 
non-standard five-point hexahedral elements. The same face-based 
staggered dual mesh has also been used in 
\cite{THD09,DHCPT10,BFSV14,BFTVC17}.  

% % % % % % % % % % % % % % % % % % % % % % % % % % % % % %
% % % % % % % % % % % % % % % % % % % % % % % % % % % % % %
\subsection{Basis functions}\label{sec:basis_func}
The basis functions are chosen according to \cite{TD14,TD16}, thus, in the two dimensional case we first construct the
polynomial basis up to a generic polynomial of degree $p$ on some 
triangular and  quadrilateral reference elements with local 
coordinates $\xi$ and $\eta$. The reference triangle is considered
to be $T_{std}=\{(\xi,\eta) \in \R^{2} \,\, | \,\,  0 \leq \xi 
\leq 1, \,\,  0 \leq \eta \leq 1-\xi \}$ and the reference 
quadrilateral element is defined as $R_{std}=[0,1]^2$. Then, the standard 
nodal approach of conforming continuous finite elements yields $N_\phi=\frac{(p+1)(p+2)}{2}$ basis functions denoted with $\{\phi_k \}_{k \in [1,N_\phi]}$ on $T_{std}$, and 
$N_{\psi}=(p+1)^2$ basis functions referred to as $\{\psi_k \}_{k \in [1,N_\psi]}$ on $R_{std}$. The transformation between the reference coordinates 
$\boldsymbol{\xi}=(\xi,\eta)$ and the physical coordinates 
$\xx=(x,y)$ is performed by the maps $T_i:\TT_i \longrightarrow T_{std}$
for every $i =1 \ldots N_e$ and $T_j:\QQ_j \longrightarrow R_{std}$ 
for every $j =1 \ldots N_d$, with the associated inverse relations, namely 
$T_i^{-1}:\TT_i \longleftarrow T_{std}$ and $T_j^{-1}:\QQ_j \longleftarrow R_{std}$,
respectively.

For the three-dimensional tetrahedra, we use
again the standard nodal basis functions of conforming finite 
elements based on the reference element 
$T_{std}=\{(\xi,\eta,\zeta) \in \R^{3} \,\, | \,\,  0 \leq \xi
\leq 1, \,\,  0 \leq \eta \leq 1-\xi, \,\, 0 \leq \zeta \leq 1-\xi-\eta \}$ 
and then we define a map to connect the reference space, 
$\boldsymbol{\xi}=(\xi,\eta,\zeta)$, to the physical space, $\xx=(x,y,z)$, and vice-versa.  
Unfortunately, the non-standard five-point hexahedral
elements of the dual mesh entail the definition of a
polynomial basis directly in the physical space using
a simple modal basis based on rescaled Taylor monomials, 
such as the ones proposed in \cite{TD16}. We thus obtain 
$N_\phi=N_\psi=\frac{1}{6}(p+1)(p+2)(p+3)$ basis functions 
per element for both the main grid and the dual mesh. 

% % % % % % % % % % % % % % % % % % % % % % % % % % % % % %
% % % % % % % % % % % % % % % % % % % % % % % % % % % % % %
\subsection{Staggered semi-implicit DG scheme}\label{sec:semi_imp_dg} 
The discrete pressure and temperature are defined
on the main grid, that is $p_i(\xx,t)=p_h(\xx,t)|_{\TTst_i}$ 
and $\theta_i(\xx,t)=\theta_h(\xx,t)|_{\TTst_i}$, while
the discrete velocity is defined on the dual grid, namely $\vv_j(\xx,t)=\vv_h(\xx,t)|_{\QQst_j}$.

Therefore, the numerical solution of 
\eqref{eq:CS_2}-\eqref{eq:cons_eq_temperature}
is given within each spatial element 
by 
\begin{eqnarray}
p_i(\xx,t)      & = & \sum\limits_{l=1}^{\Nphist} \tphi_l^{(i)}(\xx)
\bpi_{l,i}(t)=:\tbphi^{(i)}(\xx)\bbpi_i(t), \label{eq:lc_pi} \\ 
\mathbf{v}_j(\xx,t)  & = & \sum\limits_{l=1}^{\Npsist} \tpsi_l^{(j)}(\xx)
\hbv_{l,j}(t)=:\tbpsi^{(j)}(\xx)\hbv_j(t),  \label{eq:lc_v}\\
\theta_i(\xx,t)      & = & \sum\limits_{l=1}^{\Nphist} \tphi_l^{(i)}(\xx)
\btheta_{l,i}(t)=:\tbphi^{(i)}(\xx)\bbtheta_i(t).  \label{eq:lc_theta} 
\end{eqnarray}
In the rest of the paper, we use the convention that variables indexed by $j$ are defined 
on the dual grid, while the index $i$ is used for the quantities which refer to the main grid. 
Furthermore, we will use the hat symbol to denote degrees of freedom on the dual 
grid, whereas bars are used to distinguish degrees of freedom on the primal mesh. 
The set of variables on the main grid will be denoted by $\{Q_i\}_{i \in [1,\Ni]}$, while $\{Q_j\}_{j \in [1,\Nj]}$ corresponds to the unknowns defined on the dual grid.
The vector of basis functions $\tbphi^{(i)}(\xx)$ is generated via the 
map $T_i^{-1}$ from $\tbphi(\boldsymbol{\xi})$ on $T_{std}$. 
The vector $\tbpsi^{(j)}(\xx)$ is generated from 
$\tbpsi(\boldsymbol{\xi})$ on $R_{std}$ through the mapping $T_j^{-1}$ in the two 
dimensional case, and it is directly defined in the physical space for each 
element in the three dimensional case, see \cite{TD16}. 

Multiplying equations \eqref{eq:CS_2} and \eqref{eq:cons_eq_temperature} 
by $\phi_{k}$, equation \eqref{eq:CS_2_2_0} by $\psi_{\kappa}$ and integrating on the related 
control volumes $\TTst_i$ and $\QQst_j$, respectively, we obtain the weak formulation of the incompressible model \eqref{eq:CS_2}-\eqref{eq:cons_eq_temperature} for every $i=1 \ldots \Ni$, $\kappa=1 \ldots \Nphist$, $j=1 \ldots \Nj$ and $k=1 \ldots \Npsist$:

\begin{gather}
\int_{\TTst_i} \tphi_{\kappa}^{(i)} \nabla \cdot \mathbf{v} \, \dxt=0,
\label{eq:weakmass_CS_4}\\
\int_{\QQst_j}{ \tpsi_k^{(j)}\left( \diff{\mathbf{v}}{t}+\nabla \cdot \TF^{\mathbf{v}} \right)  \, \dxt}+\int_{\QQst_j}\tpsi_k^{(j)} \nabla \np \,\dxt  
=\int_{\QQst_j}{\tpsi_k^{(j)}\left(  1- \beta \delta\theta \right) \mathbf{g}\, \dxt},
\label{eq:weakmomentum_CS_5}\\
\int_{\TTst_i} \tphi_{\kappa}^{(i)} \frac{\partial{\theta}}{\partial{t}}\, \dxt+ \int_{\TTst_i}\tphi_{\kappa}^{(i)}  \nabla \cdot \TFe\, \dxt = 0. \label{eq:weaktemperature}
\end{gather}

Integration by parts of \eqref{eq:weakmass_CS_4} leads to

\begin{gather}
\oint_{\partial \TTst_i}\tphi_{\kappa}^{(i)} \mathbf{v} \cdot \nv_{i} \, \dSt -\int_{\TTst_i}\nabla \tphi_{\kappa}^{(i)} \cdot \mathbf{v} \, \dxt =0,
\label{eq:CS_6}
\end{gather}
with $\nv_{i}$ indicating the outward pointing unit normal vector.
Next, taking into account the discontinuities of $p_h$ and $\theta_h$ on $\Gamma_{j}$ and $\mathbf{v}_h$ on the edges of $\QQst_j$, the weak form becomes

\begin{gather}
\sum_{j \in S_i}\left( \int_{\Gammast_{j}}{\tphi_{\kappa}^{(i)} \mathbf{v}_j \cdot \nextud \, \dSt}
-\int_{\TT_{i,j}}{\nabla \tphi_{\kappa}^{(i)} \cdot \mathbf{v}_j \, \dxt }  \right)=0,
\label{eq:mass_split_CS_8}\\
\int_{\QQst_j}{\tpsi_k^{(j)}\left( \diff{\mathbf{v}_j}{t}+\nabla \cdot \mathbf{\TF}^{\mathbf{v}}_{j} \right)  \, \dxt }
+ \int_{\TTlst} \tpsi_k^{(j)} \nabla \np_{\ell(j)} \, \dxt 
+ \int_{\TTrst}\tpsi_k^{(j)} \nabla \np_{r(j)} \, \dxt 
+\int_{\Gammast_{j}}{\tphi_k^{(j)} \left(\np_{r(j)}-\np_{\ell(j)}\right) \nstd \, \dSt}
\notag\\
=\int_{\QQst_j}{\tpsi_k^{(j)}  \mathbf{g}\, \dxt} 
- \int_{\TTlst} \tpsi_k^{(j)} \beta \delta\theta_{\ell(j)}\, \mathbf{g} \, \dxt
- \int_{\TTrst}{\tpsi_k^{(j)}  \beta \delta\theta_{r(j)} \, \mathbf{g} \, \dxt },\label{eq:momentum_split_CS_9}\\
\int_{\TTst_i} \tphi_{\kappa}^{(i)} \frac{\partial{\theta}_i}{\partial{t}}\, \dxt+ \int_{\TTst_i}\tphi_{\kappa}^{(i)}  \nabla \cdot \TFe_{i}\, \dxt = 0,\label{eq:temperature_split}
\end{gather}
where $\nextud=\nv_{i}|_{\Gammast_{j}}$. 
Note that the pressure has a discontinuity along $\Gammast_{j}$ 
inside the dual element $\QQst_j$ and hence the pressure gradient in 
\eqref{eq:weakmomentum_CS_5} needs to be interpreted in the sense of distributions, 
as in path-conservative finite volume schemes \cite{CGP06,Par06}. 
This leads to the jump terms present in \eqref{eq:momentum_split_CS_9}, see \cite{TD15}. 
Alternatively, the same jump term can be produced also via forward and backward 
integration by parts, see e.g. the well-known work of Bassi and Rebay \cite{BR97}.
Using definitions \eqref{eq:lc_pi}-\eqref{eq:lc_theta}, we rewrite the above equations as

\begin{gather}
\sum_{j \in S_i}\left( \int_{\Gammast_{j}}{\tphi_{\kappa}^{(i)} \tphi_l^{(j)} \nextud \, \dSt \cdot \hbv_{l,j}^{n+1}}
-\int_{\TT_{i,j}}{\nabla \tphi_{\kappa}^{(i)}  \tphi_l^{(j)}  \, \dxt \cdot \hbv_{l,j}^{n+1}}  \right)=0,
\label{eq:mass_lcs}\\
\int_{\QQst_j}\tpsi_k^{(j)}  \tpsi_l^{(j)}   \, \dxt \, \TL_{h}^{\mathbf{v}}(\hbv_{l,j})
+\int_{\TTlst} \tpsi_k^{(j)} \nabla \tphi_{l}^{(\ell(j))}  \, \dxt \,  \, \bpi_{l,\ell(j)}   
+\int_{\TTrst}  \tpsi_k^{(j)} \nabla \tphi_{l}^{(r(j))} \, \dxt \,  \, \bpi_{l,r(j)}
+\int_{\Gammast_{j}}\tpsi_k^{(j)} \tphi_{l}^{(r(j))}    \nstd \, \dSt \,  \, \bpi_{l,r(j)}\notag\\
-\int_{\Gammast_{j}} \tpsi_k^{(j)} \tphi_{l}^{(\ell(j))} \nstd \,\dSt \, \, \bpi_{l,\ell(j)}
=\int_{\QQst_j}{\tpsi_k^{(j)}  \mathbf{g}\, \dxt}
- \beta \int_{\TTrst} \tpsi_k^{(j)}  \tphi_{l}^{(r(j))}  \, \dxt\,\, \delta \btheta_{l,r(j)} \, \mathbf{g} 
-\beta \int_{\TTlst} \tpsi_k^{(j)} \tphi_{l}^{(\ell(j))}  \, \dxt \,\, \delta \btheta_{l,\ell(j)}\, \mathbf{g},\label{eq:momentum_lcs}\\
\int_{\TTst_{i}} \tphi_{\kappa}^{(i)} \tphi_{l}^{(i)}  \dxt \,  \TL_{h}^{\theta}(\btheta_{l,i},\bbv_{l,i}) = 0, \label{eq:temperature_lcs}
\end{gather}

\noindent where the Einstein summation convention over repeated indexes holds.
Moreover, $\TL_{h}^{\mathbf{v}}$ and $\TLe_{h}$ represent appropriate discretizations of the operators $\TL^{\mathbf{v}}$ and $\TLe$, which will be defined later.

For every $i$ and $j$ equations \eqref{eq:mass_lcs}-\eqref{eq:temperature_lcs} can be written in a compact matrix form as

\begin{gather}
\sum_{j \in S_i}\D_{i,j}\hbv_j=0 \label{eq:masseq_matrix_CS_12},\\
\Mpsi_{j} \TL_{h}^{\mathbf{v}}(\hbv_{j}) + \RM_j \bbpi_{r(j)}  - \LM_j \bbpi_{\ell(j)} = \Gg_j - \RM_j^{\theta}\, \delta\bbtheta_{r(j)} -\LM_j^{\theta}\, \delta\bbtheta_{\ell(j)}, \label{eq:momentumeq_matrix_CS_12_1}\\
\Mphi_i \, \TL_{h}^{\theta}(\bbtheta_{i},\bbv_{i}) =0,
\label{eq:temperatureeq_matrix}
\end{gather}
with the matrix definitions
\begin{eqnarray}
\D_{i,j}&=&\int_{\Gammast_{j}}\tphi_{\kappa}^{(i)}\tpsi_l^{(j)}\nextud dS dt-\int_{\TT_{i,j}^\st}\nabla \tphi_{\kappa}^{(i)}\tpsi_l^{(j)}\, d\xx  dt,
\label{eq:MD_3}\\
\Mpsi_j &=& \int_{\QQ_j}\tpsi_{k}^{(j)}\tpsi_l^{(j)}  \, d\xx, \label{eq:MD_2} \\
\RM_{j}&=&\int_{\Gammast_{j}}\tpsi_k^{(j)} \tphi_{l}^{(r(j))}\nstd \dSt+\int_{\TTrst}\tpsi_k^{(j)} \nabla \tphi_{l}^{(r(j))}  \, \dxt,
\label{eq:MD_4}\\
\LM_{j}&=&\int_{\Gammast_{j}}\tpsi_k^{(j)} \tphi_{l}^{(\ell(j))}\nstd \dSt-\int_{\TTlst}\tpsi_k^{(j)} \nabla \tphi_{l}^{(\ell(j))}  \, \dxt,
\label{eq:MD_5}\\
\Gg_j&=&\int_{\QQst_j}\tpsi_k^{(j)} \mathbf{g} \, \dxt,
\label{eq:MD_5_8}\\
\RM^{\theta}_{j}&=&\int_{\TTrst}\tpsi_k^{(j)}  \tphi_{l}^{(r(j))}  \, \dxt,
\label{eq:MD_6}\\
\LM^{\theta}_{j}&=&\int_{\TTlst}\tpsi_k^{(j)}  \tphi_{l}^{(\ell(j))}  \, \dxt,
\label{eq:MD_7}\\
\Mphi_i &=& \int_{\TT_{i}}\tphi_{\kappa}^{(i)} \tphi_l^{(i)} d \xx. \label{eq:vis11_1}  
\end{eqnarray}

The action of the matrices $\LM_j$ and $\RM_j$ can be generalized by introducing the new matrix $\Q_{i,j}$, defined as

\begin{equation}
\Q_{i,j}=\int_{\TTst_{i,j}}\tpsi_k^{(j)} \nabla \tphi_{l}^{(i)}  \, \dxt-\int_{\Gammast_j} \tpsi_k^{(j)} \tphi_{l}^{(i)}\sigma_{i,j} \nstd \dSt,
\label{eq:qforlr}
\end{equation}
with the sign function $\sigma_{i,j}$ given by

\begin{equation}
\sigma_{i,j}=\frac{r(j)-2i+\ell(j)}{r(j)-\ell(j)}.
\label{eq:SD_1}
\end{equation}
In this way, $\Q_{\ell(j),j}=-\LM_j$ and $\Q_{r(j),j}=\RM_j$, hence the momentum equation \eqref{eq:momentumeq_matrix_CS_12_1} writes

\begin{equation}
\Mpsi_{j} \TL_{h}^{\mathbf{v}}(\hbv_{j})  + \Q_{r(j),j} \bbpi_{r(j)} + \Q_{\ell(j),j} \bbpi_{\ell(j)} = \Gg_j - \RM_j^{\theta} \, \delta\bbtheta_{r(j)} -\LM_j^{\theta}\, \delta\bbtheta_{\ell(j)}.
\label{eq:CS_12_2}
\end{equation}

{\color{cr12} Time discretization using the theta ($\Theta$) method leads to}

\begin{gather}
\sum\limits_{j \in S_i}\D_{i,j}\hbv_j^{n+1}=0 \label{eq:masseq_matrixt},\\
\frac{1}{\Delta t}\Mpsi_{j} \left( \hbv_j^{n+1} - \mathbf{F}_{t}^{\mathbf{v}}\left( \mathbf{v}^{n}\right) \right) 
+\Q_{r(j),j} {\color{cr12} \bbpi_{r(j)}^{n+\Theta}} + \Q_{\ell(j),j} {\color{cr12} \bbpi_{\ell(j)}^{n+\Theta}}
=\Gg_j 
- \RM_j^{\theta}\, \delta\bbtheta_{r(j)}^{n+1} -\LM_j^{\theta}\, \delta\bbtheta_{\ell(j)}^{n+1}, \label{eq:momentumeq_matrixt}\\
\Mphi_i\, \bbtheta_i^{n+1} - \Mphi_i\, \Fte{\boldsymbol{\theta}^{n}}{\mathbf{v}^{n}}  =0,
\label{eq:temperatureeq_matrixt}
\end{gather}
where 
{\color{cr12} $\bbpi^{n+\Theta} = \Theta \bbpi^{n+1} + (1-\Theta) \bbpi^{n}$, with $\Theta$ an implicitness factor to be taken  in the range $\Theta \in \left[\frac{1}{2},1\right]$,} 
and $\mathbf{F}_{t}^{\mathbf{v}}\left( \mathbf{v}^{n}\right)$ and $\Fte{\boldsymbol{\theta}^{n}}{\mathbf{v}^{n}}$ are proper discretizations of the convective and diffusive 
terms of the momentum and energy equations, respectively, which read

\begin{eqnarray}
\mathbf{F}_{t}^{\mathbf{v}}\left( \mathbf{v}^{n}\right) =  \hbv_{j}^{n} - \Delta t \Mpsi_{j}^{-1} \hat{\boldsymbol{\Upsilon}}^{\mathbf{v}}_j, \quad
\hat{\boldsymbol{\Upsilon}}^{\mathbf{v}}_j = \int_{\QQst_j}\tpsi_k^{(j)} \nabla \cdot \TF^{\mathbf{v}} \, \dxt,\label{eq:ft}\\
\Fte{\boldsymbol{\theta}^{n}}{\mathbf{v}^{n}} = \bbtheta_{i}^{n}- \Delta t \Mphi_{i}^{-1} \overline{\boldsymbol{\Upsilon}}_i^{\theta}, \quad
\overline{\boldsymbol{\Upsilon}}_i^{\theta} = \int_{\TTst_i}\tphi_{\kappa}^{(i)} \nabla \cdot \TFe \, \dxt.\label{eq:fte}
\end{eqnarray}
Two different approaches are considered in this work to obtain a numerical approximation of the operators \eqref{eq:ft}-\eqref{eq:fte}: i) a fully Eulerian discretization and ii) {\color{cr1} an Eulerian-Lagrangian} method. 
The Eulerian scheme provides a fully conservative formulation, contrarily to the {\color{cr1} Eulerian-Lagrangian} approach \cite{TB19}. However, the latter scheme would be unconditionally stable, so that no CFL restrictions need to be considered for the time step, thus substantially reducing the computational cost of the algorithm.

% % % % % % % % % % % % % % % % % % % % % % % % % % % % % %
% % % % % % % % % % % % % % % % % % % % % % % % % % % % % %
\subsection{Pressure system}
Formal substitution of equation \eqref{eq:momentumeq_matrixt} 
into \eqref{eq:masseq_matrixt}, i.e. making use of the Schur complement, leads to a 
linear system in which the pressure is the only unknown, that is a \textit{scalar} quantity: 

\begin{equation}
 \sum\limits_{j \in S_i}\D_{i,j}\Mpsi_{j}^{-1}\left(  \Q_{r(j),j} {\color{cr12} \bbpi_{r(j)}^{n+\Theta}} + \Q_{\ell(j),j} {\color{cr12} \bbpi_{\ell(j)}^{n+\Theta} }  \right) 
= 
 \sum\limits_{j \in S_i}\D_{i,j}\Mpsi_{j}^{-1}\left(  \Gg_j 
- \RM_j^{\theta}\, \delta\bbtheta_{r(j)}^{n+1} -\LM_j^{\theta}\, \delta\bbtheta_{\ell(j)}^{n+1}\right) 
+\frac{1}{\Delta t} \sum\limits_{j \in S_i}\D_{i,j}\mathbf{F}^{\mathbf{v}}_t\left( \mathbf{v}^{n}\right) . \label{eq:pressure_final} 
\end{equation} 

The above system is symmetric and in general positive semi-definite, 
see \cite{TD15,TD17}, so that an efficient conjugate gradient method can be applied 
in order to obtain the new pressure $\bbpi^{n+1}$. The previous equation 
\eqref{eq:pressure_final} coupled with \eqref{eq:momentumeq_matrixt}-\eqref{eq:temperatureeq_matrixt} 
constitute the system of equations to be solved.

% % % % % % % % % % % % % % % % % % % % % % % % % % % % % %
% % % % % % % % % % % % % % % % % % % % % % % % % % % % % %
\subsection{Nonlinear advection and diffusion}\label{sec:advect_diff}
In the framework of semi-implicit schemes \cite{CG84,CC92,CC94,CW00,BFSV14,DC16}, the nonlinear convective terms are typically discretized 
\textit{explicitly}, while the pressure terms are treated \textit{implicitly}. For more details on how these numerical methods are related to flux-vector splitting schemes, see \cite{TV12}.

Following \cite{TD16}, we exploit the advantages of using staggered grids
to develop a suitable discretization of the nonlinear advection and diffusion terms. 
Therefore, the velocity field is first interpolated from the dual grid to the main grid:
\begin{eqnarray}
\bbv_i &=& \Mphi^{-1}_i \sum_{j \in S_i} \LL_{i,j} \hbv_j,
\label{eq:vinterpolation}
\end{eqnarray}
with
\begin{gather}
\LL_{i,j} = \int_{\TT_{i,j}} \tphi_{\kappa}^{(i)} \tpsi_l^{(j)} \, d\xx, \\
\hbv_j = {\Mpsi}_j^{\, -1} \left( \LL_{\ell(j),j}^\top \bbv_{\ell(j)} + \LL_{r(j),j}^\top \bbv_{r(j)}\right).
\label{eq:vis2}
\end{gather}
Next, the nonlinear convective terms can be easily discretized with 
a standard DG scheme on the main grid. Finally, the staggered mesh 
is used \textit{again} in order to define the gradient of the velocity 
on the dual elements, which yields a very simple and sparse system 
for the discretization of the viscous terms. 
A similar procedure is employed to discretize 
the convective and viscous terms of the energy equation. 
Let us remark that for this particular equation, the temperature is already 
defined on the primal elements, thus interpolation 
between the two meshes is avoided. Let us define two auxiliary variables for the diffusion terms, 
namely the stress tensor $\bsigma$ and the heat flux $\mathbf{q}$ as  
\begin{equation}
\bsigma=-\nu \nabla \mathbf{v}, \qquad \mathbf{q} =-\alpha\nabla\theta. \label{eqn.auxDiff}
\end{equation}
Then, the convective and viscous subsystems of the momentum and energy equations read 
\begin{eqnarray}
\frac{\partial{\mathbf{v}}}{\partial{t}}+\nabla \cdot \TF_c^{\mathbf{v}} + \nabla \cdot \bsigma & = & 0, \label{eq:convdifsubsyst_momentum1}\\
\bsigma &=&-\nu\nabla\mathbf{v}; \label{eq:convdifsubsyst_momentum2}\\[6pt]
\frac{\partial{\theta}}{\partial{t}}+\nabla \cdot \TFe_c + \nabla \cdot \mathbf{q} & = & 0, \label{eq:convdifsubsyst_temperature1}\\
\mathbf{q} &=&-\alpha\nabla\theta. \label{eq:convdifsubsyst_temperature2}
\end{eqnarray}
Defining the temperature and the velocity on the primal elements and the 
auxiliary variables \eqref{eqn.auxDiff} on the dual grid, we obtain a weak formulation for equations  
\eqref{eq:convdifsubsyst_momentum1}-\eqref{eq:convdifsubsyst_momentum2} and \eqref{eq:convdifsubsyst_temperature1}-\eqref{eq:convdifsubsyst_temperature2}, that is
\begin{gather}
\int_{\TTst_{i}} \tphi_{\kappa}^{(i)} \frac{\partial{\bov}}{\partial{t}} \dxt
+\int_{\partial \TTst_{i}} \tphi_{\kappa}^{(i)} \nabla \cdot \mathbf{G}_c^{\mathbf{v}} \left(\bbv^{-},\bbv^{+} \right) \dSt
-\int_{\TTst_{i}} \tphi_{\kappa}^{(i)}  \cdot \TF_c^{\mathbf{v}}\left(\bbv_{i} \right) \cdot \vec{n}_{i} \, \dSt
\nonumber\\
+\sum_{j \in S_i} \left( \int_{\Gammast_{j}} \tphi_{\kappa}^{(i)} \bsigma_{j} \cdot \vec{n}_{ij} \, \dSt
- \int_{\TTst_{i,j}}\nabla  \tphi_{\kappa}^{(i)} \cdot \bsigma_{j} \, \dxt\right) 
=  0, \label{eq:cdss_mom_weakproblem1}\\
\int_{\QQst_{j}} \tpsi_k^{(j)} \bsigma_{j} \dxt
= -\nu \left( \int_{\TTst_{\ell(j),j}} \tpsi_k^{(j)} \nabla\bbv_{\ell(j)} \dxt
+ \int_{\TTst_{r(j),j}} \tpsi_k^{(j)} \nabla \bbv_{r(j)} \dxt
+ \int_{\Gammast_{j}} \tpsi_k^{(j)} \left( \bbv_{r(j)} - \bbv_{\ell(j)} \right) \otimes \vec{n}_{j} \dSt \right);
\label{eq:cdss_mom_weakproblem2}
\\[6pt]
\int_{\TTst_{i}} \tphi_{\kappa}^{(i)} \frac{\partial{\theta}}{\partial{t}} \dxt
+\int_{\partial \TTst_{i}} \tphi_{\kappa}^{(i)} \nabla \cdot \mathbf{G}_c^{\theta}  \left(\btheta^{-},\btheta^{+},\bbv^{-},\bbv^{+} \right) \dSt
-\int_{\TTst_{i}} \tphi_{\kappa}^{(i)}  \cdot \TFe_c\left(\btheta_{i},\bbv_{i} \right) \cdot \vec{n}_{i}\dSt
\nonumber\\
+\sum_{j \in S_i} \left( \int_{\Gammast_{j}} \tphi_{\kappa}^{(i)} \mathbf{q}_{j} \cdot \vec{n}_{ij} \, \dSt
- \int_{\TTst_{i,j}}\nabla  \tphi_{\kappa}^{(i)} \cdot \mathbf{q}_{j} \, \dxt\right) 
=  0, \label{eq:cdss_temp_weakproblem1}\\
\int_{\QQst_{j}} \tpsi_k^{(j)} \mathbf{q}_{j} \dxt
= -\alpha \left( \int_{\TTst_{\ell(j),j}} \tpsi_k^{(j)} \nabla\btheta_{\ell(j)} \dxt
+ \int_{\TTst_{r(j),j}} \tpsi_k^{(j)} \nabla \btheta_{r(j)} \dxt
+ \int_{\Gammast_{j}} \tpsi_k^{(j)} \left( \btheta_{r(j)} - \btheta_{\ell(j)} \right)  \vec{n}_{j}\dSt \right) .
\label{eq:cdss_temp_weakproblem2}
\end{gather}
Accounting for the discretization in time, the former systems are expressed in matrix notation as
\begin{eqnarray}
\frac{1}{\Delta t}\Mphi_i \left( \bbv_{i}^{n+1} - \bbv_{i}^{n}\right)  + \boldsymbol{\Upsilon}_i^{\mathbf{v},c} + \sum_{j \in S_i}\D_{i,j} \bsigma^{n+1}_{j} & = & 0,\label{eq:matrixwp_momentum1}\\
\Mpsi_j \bsigma^{n+1}_{j} & = & -\nu \left( \Q_{r(j),j}  \bbv_{r(j)}^{n+1} + \Q_{\ell(j),j} \bbv_{\ell(j)}^{n+1} \right), \label{eq:matrixwp_momentum2}
\\[6pt]
\frac{1}{\Delta t}\Mphi_i \left( \bbtheta_{i}^{n+1} - \bbtheta_{i}^{n}\right)  + \boldsymbol{\Upsilon}_i^{\theta,c} + \sum_{j \in S_i}\D_{i,j} \mathbf{q}^{n+1}_{j} & = & 0,\label{eq:matrixwp_temperature1}\\
\Mpsi_j \mathbf{q}^{n+1}_{j} & = & -\alpha \left( \Q_{r(j),j}  \bbtheta_{r(j)}^{n+1} + \Q_{\ell(j),j} \bbtheta_{\ell(j)}^{n+1}\right), \label{eq:matrixwp_temperature2}
\end{eqnarray}
with
\begin{gather}
\boldsymbol{\Upsilon}_i^{\mathbf{v},c}= \int_{\partial \TTst_{i}} \tphi_{\kappa}^{(i)} \mathbf{G}_c^{\mathbf{v}} \left(\bbv^{-},\bbv^{+} \right) \cdot \vec{n}_{i} \, \dSt
-\int_{\TTst_{i}} \nabla \tphi_{\kappa}^{(i)}  \cdot \TF^{\mathbf{v}}_c\left(\bbv_{i} \right) \dxt,
\\
\boldsymbol{\Upsilon}_i^{\theta,c}= \int_{\partial \TTst_{i}} \tphi_{\kappa}^{(i)} \mathbf{G}_c^{\theta} \left(\btheta^{-},\btheta^{+},\bbv^{-},\bbv^{+} \right) \cdot \vec{n}_{i} \dSt
-\int_{\TTst_{i}} \nabla \tphi_{\kappa}^{(i)}  \cdot \TFe_c\left(\btheta_{i},\bbv_{i} \right)  \dxt.
\end{gather}
$\boldsymbol{\Upsilon}_i^{\mathbf{v},c}$ and $\boldsymbol{\Upsilon}_i^{\theta,c}$ are standard DG discretizations of the nonlinear convective terms, and 
$\mathbf{v}^-$, $\mathbf{v}^+$, $\boldsymbol{\theta}^-$, and $\boldsymbol{\theta}^+$ refer to the boundary extrapolated values from within 
the cell and from the neighbors, respectively. Furthermore, we use 
the Rusanov flux \cite{Rus62} as approximate Riemann solver:
\begin{gather}
\mathbf{G}_c^{\mathbf{v}} \left(\bbv^{-},\bbv^{+} \right) \cdot \vec{n}_i = 
\frac{1}{2}\left( \TF^{\mathbf{v}}_c(\bbv^{+}) +\TF^{\mathbf{v}}_c(\bbv^{-}) \right) \cdot \vec{n}_i \, - 
\frac{1}{2} s^{\mathbf{v}}_{\max} \left( \bbv^+ - \bbv^- \right),
\label{eq:rusanov_momentum} \\ 
\mathbf{G}_c^{\theta} \left(\btheta^{-},\btheta^{+},\bbv^{-},\bbv^{+} \right) \cdot \vec{n}_i = 
\frac{1}{2}\left( \TFe_c(\btheta^{+},\bbv^{+}) +\TFe_c(\btheta^{-},\bbv^{-}) \right) \cdot \vec{n}_i \, - 
\frac{1}{2} s_{\max}^{\theta} \left( \btheta^+ - \btheta^- \right),
\label{eq:rusanov_temperature}
\end{gather} 
where $s^{\mathbf{v}}_{\max} = \max\left( 2 |\bbv^+ \cdot \vec{n}_{i}|,  2|\bbv^- \cdot \vec{n}_{i}| \right)$  
and $s_{\max}^{\theta} = \max\left( |\bbv^+ \cdot \vec{n}_{i}|, |\bbv^- \cdot \vec{n}_{i}| \right)$ 
are the maximum eigenvalues of the convective operators 
$\TF^{\mathbf{v}}_c$ and $\TFe_c$, respectively. 
Finally, substituting \eqref{eq:matrixwp_momentum2} into 
\eqref{eq:matrixwp_momentum1}, and \eqref{eq:matrixwp_temperature2} into 
\eqref{eq:matrixwp_temperature1}, we obtain
\begin{gather}
\frac{1}{\Delta t}\Mphi_i \left( \bbv_{i}^{n+1} - \bbv_{i}^{n}\right) + \boldsymbol{\Upsilon}_i^{\mathbf{v},c}\left(\bbv^{n+1} \right)  - \nu \sum_{j \in S_i}\D_{i,j} \Mpsi_j^{-1} \left( \Q_{r(j),j} \bbv_{r(j)}^{n+1} +\Q_{\ell(j),j}  \bbv_{\ell(j)}^{n+1} \right) =0,\label{eq:matrixfinal_momentum}
\\[6pt]
\frac{1}{\Delta t}\Mphi_i \left( \bbtheta_{i}^{n+1} - \bbtheta_{i}^{n}\right) + \boldsymbol{\Upsilon}_i^{\theta,c}\left(\btheta^{n+1},\bbv^{n+1} \right)  - \alpha \sum_{j \in S_i}\D_{i,j} \Mpsi_j^{-1} \left( \Q_{r(j),j} \bbtheta_{r(j)}^{n+1} +\Q_{\ell(j),j}  \bbtheta_{\ell(j)}^{n+1} \right) =0.\label{eq:matrixfinal_temperature}
\end{gather}
In order to avoid the solution of a nonlinear system due to the 
presence of the nonlinear operator related to the convective terms, 
a fractional step scheme combined with an outer Picard iteration is used:
\begin{gather}
\frac{1}{\Delta t}\Mphi_i \bbv_{i}^{n+1,k+\frac{1}{2}} - \nu \sum_{j \in S_i}\D_{i,j} \Mpsi_j^{-1} \left( \Q_{r(j),j} \bbv_{r(j)}^{n+1,k+\frac{1}{2}} +\Q_{\ell(j),j}  \bbv_{\ell(j)}^{n+1,k+\frac{1}{2}} \right)=\frac{1}{\Delta t}\Mphi_i   \bbv_{i}^{n} \notag\\
-\boldsymbol{\Upsilon}_i^{\mathbf{v},c}\left(\bbv^{n+1,k} \right) - \sum_{j \in S_i} \LL_{i,j} \Mpsi_{j}^{-1}  \left(\Q_{r(j),j}  \hbpi_{r(j)}^{n+1,k} + \Q_{\ell(j),j} \hbpi_{\ell(j)}^{n+1,k} \right), \label{eq:matrixfinal_momentumpic}
\\[6pt]
\frac{1}{\Delta t}\Mphi_i \bbtheta_{i}^{n+1,k+\frac{1}{2}}  - \alpha \sum_{j \in S_i}\D_{i,j} \Mpsi_j^{-1} \left( \Q_{r(j),j} \bbtheta_{r(j)}^{n+1,k+\frac{1}{2}} +\Q_{\ell(j),j}  \bbtheta_{\ell(j)}^{n+1,k+\frac{1}{2}} \right) =\frac{1}{\Delta t}\Mphi_i  \bbtheta_{i}^{n} - \boldsymbol{\Upsilon}_i^{\theta,c}\left(\btheta^{n+1,k},\bbv^{n+1,k} \right).\label{eq:matrixfinal_temperaturepic}
\end{gather}
The previous procedure constitutes a discretization of the nonlinear 
convective and viscous terms on the main grid, both for the momentum 
and the energy equations, where $\boldsymbol{\sigma}$ 
and $\mathbf{q}$ are computed on the 
face-based dual mesh. To recover the contribution of the convective 
and viscous terms in the dual grid, as required in 
\eqref{eq:momentumeq_matrixt}, we perform the following projection:
\begin{gather}
\mathbf{F}^{\mathbf{v}}_t\left( \mathbf{v}^{n+1,k+\frac{1}{2}}\right)  = {\Mpsi}_j^{\, -1} \left( \LL_{\ell(j),j}^\top \bbv_{\ell(j)}^{n+1,k+\frac{1}{2}} + \LL_{r(j),j}^\top \bbv_{r(j)}^{n+1,k+\frac{1}{2}}\right). \label{eq:ftj_comp}
\end{gather}

% % % % % % % % % % % % % % % % % % % % % % % % % % % % % %
% % % % % % % % % % % % % % % % % % % % % % % % % % % % % %
\subsection{{\color{cr1} Eulerian-Lagrangian} approach}\label{sec:semilagrangian}
Instead of applying the Eulerian advection scheme illustrated in 
Section \ref{sec:advect_diff} for the approximation of 
the convection and diffusion terms, 
$\mathbf{F}^{\mathbf{v}}_t\left( \mathbf{v}^{n+1,k+\frac{1}{2}}\right) $ 
and $\Fte{\boldsymbol{\theta}^{n+1,k+\frac{1}{2}}}{\mathbf{v}^{n+1,k+\frac{1}{2}}}$, 
we may take into account the Lagrangian trajectory of the 
flow particles and use {\color{cr1} an Eulerian-Lagrangian} approach.  
{\color{cr1}Specifically, the departure point $x^*$ of the Lagrangian trajectory has to be determined in order to compute the corresponding value of the transferred quantities, namely velocity $\mathbf{v}^*$ and temperature $\boldsymbol{\theta}^*$. The Lagrangian trajectory is defined by the solution $\xx(\tau)$ of the \textit{trajectory equation}
	\begin{equation}
	\frac{d \xx}{d\tau} = -\vv(\xx(\tau)), \qquad \textnormal{with} \qquad \xx(0)=\xx_l^{n+1} \quad \textnormal{and} \quad \tau \in [0,\Delta t], 
	\label{eqn.LagTraj}
	\end{equation}
  	 where $\xx_l^{n+1}$ is the location of a generic quadrature point from which the backtracking of the trajectory
  	 is started. 
	 Furthermore, $\tau$ represents the rescaled time coordinate referred to the time step $\Delta t:=t^{n+1}-t^n$, and can be easily evaluated as $\tau = t - t^{n}$, while $\vv(\xx(\tau))$ is the local fluid velocity. The foot point of the characteristics, which is nothing but the sought departure point, is given by $\xx^*=\xx(\Delta t)$. In order to solve the system of ordinary differential equations (ODE) \eqref{eqn.LagTraj}, we rely on the approach presented in \cite{BDR13,BPR19}, hence using a high order Taylor method, which leads to the solution $\xx_{r+1}$ at the new time $\tau_{r+1}$:
	\begin{equation}
	\xx_{r+1} = \xx_{r} + \Delta \tau \frac{d\xx}{d\tau} + \frac{\Delta \tau^2}{2} \frac{d^2 \xx}{d\tau^2} 
	+ \frac{\Delta \tau^3}{6} \frac{d^3 \xx}{d\tau^3} + \mathcal{O}(4).  
	\label{eqn.taylor1} 
	\end{equation} 
	Expansion \eqref{eqn.taylor1} allows the scheme to be up to third order accurate in $\tau$. The index $r$ represents the iteration number if a sub-time stepping is chosen for the approximation of the time step interval $[0,\Delta t]$. High order time derivatives are then replaced by high order spatial derivatives using repeatedly the trajectory equation \eqref{eqn.LagTraj} via the \textit{Cauchy-Kovalevskaya} procedure, which is also typical for the ADER approach of Toro and 
	Titarev \cite{toro4,titarevtoro,Toro:2006a}. Thus, assuming that $\vv=\vv(\xx)=\vv(\xx,t^n)$ is frozen during one time step, one obtains 
	\begin{subequations}
		\begin{alignat}{1}
		\frac{d\xx}{d\tau} & = -\vv, \\
		\frac{d^2 \xx}{d\tau^2} & = \frac{d}{d \tau} \left( \frac{d \xx}{d\tau}\right) = -\frac{d \vv}{d \tau} = -\frac{\partial \vv}{\partial \xx} \frac{d \xx}{d\tau}  = \frac{\partial \vv}{\partial \xx} \vv , \\     
		\frac{d^3 \xx}{d\tau^3} & = \frac{d}{d \tau} \left( \frac{\partial \vv}{\partial \xx} \vv \right) = 
		\frac{d}{d \tau} \left( \frac{\partial \vv}{\partial \xx} \right) \vv + \frac{\partial \vv}{\partial \xx} \frac{d \vv}{d \tau}  = \frac{\partial}{\partial \xx} \left( \frac{\partial \vv}{\partial \xx} \right) \frac{d \xx}{d\tau}  \vv - \left( \frac{\partial \vv}{\partial \xx} \right)^2  \vv \nonumber \\
		& = -  \left( \frac{\partial^2 \vv}{\partial \xx^2} \right) \vv \vv - \left( \frac{\partial \vv}{\partial \xx} \right)^2  \vv.
		\end{alignat}	
\end{subequations} 
In this work we follow the methodology detailed in \cite{TB19}, in which a staggered high order DG scheme has been used together with the Eulerian-Lagrangian technique previously illustrated. In the DG context, the Eulerian-Lagrangian scheme is used to compute the nonlinear convective terms inside a \textit{weak formulation} of the governing PDE, therefore the starting points for the trajectory equation \eqref{eqn.LagTraj} are given by the Gaussian quadrature points in space within each control volume $T_i$. In order to guarantee sufficient accuracy, we integrate backward in time \textit{twice} the minimum number of Gaussian points that ensures the formal order of accuracy of the scheme, hence performing over-integration. According to \cite{TB19}, the integration of the ODE \eqref{eqn.LagTraj} is carried out in a reference system with local coordinates $(\xi,\eta,\zeta)$, so that element and physical boundaries can be easily identified and treated. Finally, the high order spatial discretization of the DG scheme is employed to compute the values of the velocity and the 
temperature at the foot point $\xx^*=\xx(\Delta t)$ of each trajectory. Further details on the aforementioned methodology can be found in \cite{TB19}, whereas in \cite{Mcg93} a study for determining the departure points of trajectories is proposed. If stiff problems are considered, in which the solution rapidly changes in time, a complete space-time method would be necessary to capture properly the flow trajectory and which will be subject of future work. However, for the applications considered in this work, that are concerned with natural convection problems, this is typically not the case, so that our Eulerian-Lagrangian algorithm can be adopted for the discretization of the convection and diffusion terms, 
$\mathbf{F}^{\mathbf{v}}_t\left( \mathbf{v}^{n+1,k+\frac{1}{2}}\right) $ 
and $\Fte{\boldsymbol{\theta}^{n+1,k+\frac{1}{2}}}{\mathbf{v}^{n+1,k+\frac{1}{2}}}$.  
}
The semi-implicit 
scheme for the Navier-Stokes equations together with the {\color{cr1} Eulerian-Lagrangian}
approach becomes unconditionally stable for arbitrary 
high order of accuracy and thus allows large time steps. For a detailed analysis 
in the case of semi-implicit finite difference schemes, see {\color{cr1} \cite{CC94}}.

% % % % % % % % % % % % % % % % % % % % % % % % % % % % % %
% % % % % % % % % % % % % % % % % % % % % % % % % % % % % %
\subsection{Overall method}\label{sec:overall}
Given $\bbpi_{i}^{n}$, $\hbv_{j}^{n}$ and $\bbtheta_{i}^{n}$, the final algorithm reads as follows.
\begin{enumerate}
	\item The temperature at the new iteration, $\bbtheta^{n+1,k+\frac{1}{2}}_{i}$, is obtained solving \eqref{eq:matrixfinal_temperaturepic} when the fully Eulerian method is selected. Otherwise, the {\color{cr1} Eulerian-Lagrangian} approach determines the term $\bar{\mathbf{F}}^{\theta}_{t}$ in equation \eqref{eq:temperatureeq_matrixt} from which the temperature is updated.
	
	\item The nonlinear convective and viscous terms for the momentum equation, $\mathbf{F}_{t}^{\mathbf{v}}\left(\mathbf{v}^{n+1,k+\frac{1}{2}} \right)$, are computed. Let us remark that within this term we are accounting for the contribution of the pressure at the previous Picard iteration, see Eqn. \eqref{eq:matrixfinal_momentumpic}.
	
	\item The pressure, $\bbpi_{i}^{n+1,k+1}$, results from solving system \eqref{eq:pressure_final} after substitution of $\mathbf{F}_{t}^{\mathbf{v}}\left(\mathbf{v}^{n+1,k+\frac{1}{2}} \right)$ and  $\bbtheta^{n+1,k+\frac{1}{2}}_{i}$:
	\begin{gather}
	{\color{cr12} \Theta\Delta t \sum\limits_{j \in S_i}\D_{i,j}\Mpsi_{j}^{-1}\left[ \Q_{r(j),j}\left(  \bbpi_{r(j)}^{n+1,k+1}-\bbpi_{r(j)}^{n+1,k} \right) + \Q_{\ell(j),j} \left( \bbpi_{\ell(j)}^{n+1,k+1}-\bbpi_{\ell(j)}^{n+1,k}\right)  \right]= \sum\limits_{j \in S_i}\D_{i,j}\mathbf{F}^{\mathbf{v}}_t\left( \mathbf{v}^{n+1,k+\frac{1}{2}}\right)}\notag \\
	{\color{cr12} +\Delta t \sum\limits_{j \in S_i}\D_{i,j}\Mpsi_{j}^{-1}\left( \Gg_j 
	-\RM_j^{\theta}\, \delta\bbtheta_{r(j)}^{n+1,k+\frac{1}{2}} -\LM_j^{\theta}\, \delta\bbtheta_{\ell(j)}^{n+1,k+\frac{1}{2}}\right)
	-\left(1-\Theta \right) \Delta t \sum\limits_{j \in S_i}\D_{i,j}\Mpsi_{j}^{-1}\left[ \Q_{r(j),j}  \bbpi_{r(j)}^{n} + \Q_{\ell(j),j}  \bbpi_{\ell(j)}^{n}  \right]}
	. \label{eq:pressure_finalpic} 
	\end{gather} 
	
	\item The velocity, $\hbv_{j}^{n+1,k+1}$, is then updated from \eqref{eq:momentumeq_matrixt}:
	\begin{equation}
	\hbv_j^{n+1,k+1} =
	\mathbf{F}_{t}^{\mathbf{v}}\left(\mathbf{v}^{n+1,k+\frac{1}{2}} \right)  + 
	\Delta t \Mpsi_{j}^{-1} \left[ -{\color{cr12}\Q_{r(j),j} \left( \bbpi_{r(j)}^{n+\Theta,k+1}  - \bbpi_{r(j)}^{n+\Theta,k}\right) - \Q_{\ell(j),j} \left(  \bbpi_{\ell(j)}^{n+1,k+1} - \bbpi_{\ell(j)}^{n+1,k}\right) }\right.\notag\\ \left.
	+\Gg_j 
	- \RM_j^{\theta}\, \delta\bbtheta_{r(j)}^{n+1,k+\frac{1}{2}} -\LM_j^{\theta}\, \delta\bbtheta_{\ell(j)}^{n+1,k+\frac{1}{2}} \right] .
	\end{equation}
\end{enumerate}
{\color{cr12}
\begin{remark}
	For the sake of simplicity, the former algorithm has been presented assuming the theta method is used for time discretization. As a consequence, the accuracy of the resulting scheme is of arbitrary order in space and only up to second order in time.
	Besides, the space-time extension for the Eulerian approach has been developed following \cite{TD15,TD16}, hence obtaining arbitrary high order semi-implicit DG schemes at the aid of test and basis functions that depend on both space and time.
\end{remark}
}

% % % % % % % % % % % % % % % % % % % % % % % % % % % % % %
% % % % % % % % % % % % % % % % % % % % % % % % % % % % % %
\section{Numerical method for the compressible model}\label{sec:numerical_method_cns}
The staggered semi-implicit discontinuous Galerkin scheme described 
in the previous section for the incompressible model has been extended 
in \cite{TD17} to solve the compressible Navier-Stokes equations at all 
Mach numbers. Concerning semi-implicit finite volume schemes for all 
and low Mach number flows, we refer the reader to \cite{KleinMach,Klein2001,Munz2003,RussoAllMach,CordierDegond}.  

To simulate natural convection problems, some modifications are needed in order to incorporate the gravitational terms. In what follows, we will provide the details associated to their inclusion.
For well-balanced schemes for the compressible Euler equations with gravity source terms, see \cite{BottaKlein,Kapelli2014,Klingenberg2015,GCD18,KlingenbergPuppo}.

% % % % % % % % % % % % % % % % % % % % % % % % % % % % % %
% % % % % % % % % % % % % % % % % % % % % % % % % % % % % %
\subsection{Staggered semi-implicit DG scheme}
The computational domain is discretized as already explained in Section \ref{sec:mesh} and the basis functions are defined according to Section \ref{sec:basis_func}. Then, the discrete pressure $p_h$, the fluid density $\rho_h$ and the discrete total energy density $(\rhoEE)_h$ are computed on the main grid, while the discrete 
velocity vector field $\vv_h$, the discrete momentum density $(\rhov)_h$, and the discrete specific enthalpy $H_h$ are defined on the dual grid.

The numerical solution of \eqref{eq:CNS_mass}-\eqref{eq:CNS_energy} at a given time $t$ is represented inside the control volumes of the primal and the dual grids by piecewise spatial polynomials. The discrete pressure is approximated using \eqref{eq:lc_pi}, while the total energy as well as the density  on the main mesh and the momentum on the dual mesh read
\begin{eqnarray}
\rho E_i(\xx,t)      & = & \sum\limits_{l=1}^{\Nphist} \tphi_l^{(i)}(\xx)
\brE_{l,i}(t)=:\tbphi^{(i)}(\xx)\bbrE_i(t), \label{eq:lc_rhoE} \\
\rho_i(\xx,t)      & = & \sum\limits_{l=1}^{\Nphist} \tphi_l^{(i)}(\xx)
\brho_{l,i}(t)=:\tbphi^{(i)}(\xx)\bbrho_i(t),  \label{eq:lc_rho}\\ 
\mathbf{v}_j(\xx,t)  & = & \sum\limits_{l=1}^{\Npsist} \tpsi_l^{(j)}(\xx)
\hrv_{l,j}(t)=:\tbpsi^{(j)}(\xx,t)\hbrv_j(t).  \label{eq:lc_rhov} 
\end{eqnarray}

Similarly to what has been done in Section \ref{sec:semi_imp_dg} for the incompressible system, a weak formulation of \eqref{eq:CNS_mass}-\eqref{eq:CNS_energy} is obtained by multiplication of the governing equations by appropriate test functions and integration over the associated control volumes:
\begin{gather}
\int_{\TT_{i}} \phi_{\kappa}^{(i)} \frac{\partial \rho_{i}}{\partial t} \dx
= - \sum_{j \in S_i}\left( \int_{\Gammast_{j}}{\tphi_{\kappa}^{(i)} \left( \rho\mathbf{v}\right)_{j} \cdot \nextud \, \dSt}
-\int_{\TTst_{i,j}}{\nabla \tphi_{\kappa}^{(i)} \cdot \left( \rho\mathbf{v}\right)_{j} \, \dxt }  \right)
\label{eq:mass_cns_weak} \\
\int_{\QQst_j}{ \tpsi_k^{(j)}\left( \frac{\partial \left( \rho\mathbf{v}\right)_{j}}{\partial t}+\nabla \cdot \TF^{\rho\mathbf{v}}_{j} \right)  \, \dxt}
+ \int_{\TTlst} \tpsi_k^{(j)} \nabla \np_{\ell(j)} \, \dxt 
+ \int_{\TTrst}\tpsi_k^{(j)} \nabla \np_{r(j)} \, \dxt 
+\int_{\Gammast_{j}}{\tphi_k^{(j)} \left(\np_{r(j)}-\np_{\ell(j)}\right) \nstd \, \dSt}  
=\int_{\QQst_j}{\tpsi_k^{(j)} \rho_{j} \mathbf{g}\, \dxt},
\label{eq:momentum_cns_weak}\\
\int_{\TT_i}{\phi_{\kappa}^{(i)}\frac{\partial \left( \rhoEE\right)_{i}}{\partial t} \dxt} 
+	\int_{\TT_i}{\phi_{\kappa}^{(i)}\nabla \cdot \left[  k_{i} (\rho \vv)_{i}\right]  \dxt} 
+ 	\int_{\TT_i}{\phi_{\kappa}^{(i)} \nabla \cdot \left[ H_{i} (\rho \vv)_{i}\right] \, \dxt}
= \int_{\TT_i}{\phi_{\kappa}^{(i)}\nabla \cdot \left[ \boldsymbol{\sigma}_{i}  \vv_{i} + \kappa \nabla \theta_{i} \right] \dxt} 
+ \int_{\TT_i} \phi_{\kappa}^{(i)} \left( \rho\mathbf{v}\right)_{i} \cdot \mathbf{g} \, \dxt
\label{eq:energy_cns_weak}.
\end{gather}
The above weak formulation of the governing PDE accounts for the discontinuities of pressure and momentum  along the boundaries of the primal and dual cells, respectively. Let $\mathbf{w}_{j}:= \boldsymbol{\sigma}_{j} \mathbf{v}_{j} $ be the work of the stress tensor in the energy equation and let  $\mathbf{q}_{j} := \kappa \nabla \theta_{j}$ define the heat flux vector. Using the polynomial approximations \eqref{eq:lc_rhoE}-\eqref{eq:lc_rhov} in the semi-discrete system \eqref{eq:mass_cns_weak}-\eqref{eq:energy_cns_weak} leads to
\begin{gather}
\int_{\TT_{i}} \phi_{\kappa}^{(i)}  \phi_{l}^{(i)} \dx \frac{\partial \rho_{l,i}}{\partial t}
= - \sum_{j \in S_i}\left( \int_{\Gammast_{j}}{\tphi_{\kappa}^{(i)}  \tpsi_{l}^{(j)}  \nextud \, \dSt \cdot \hrv_{l,j}}
-\int_{\TTst_{i,j}}{\nabla \tphi_{\kappa}^{(i)}  \tpsi_{l}^{(j)} \, \dxt \cdot \hrv_{l,j} }  \right)
\label{eq:mass_cns_weak2} \\
\int_{\QQst_j}\tpsi_k^{(j)}  \tpsi_l^{(j)}   \, \dxt \, \TL^{\rho\mathbf{v}}_{h}(\hrv_{l,j})
+\int_{\TTlst} \tpsi_k^{(j)} \nabla \tphi_{l}^{(\ell(j))}  \, \dxt \,  \, \bpi_{l,\ell(j)}    
+\int_{\TTrst}  \tpsi_k^{(j)} \nabla \tphi_{l}^{(r(j))} \, \dxt \,  \, \bpi_{l,r(j)}\notag\\
+\int_{\Gammast_{j}}\tpsi_k^{(j)} \tphi_{l}^{(r(j))}    \nstd \, \dSt \,  \, \bpi_{l,r(j)}
-\int_{\Gammast_{j}} \tpsi_k^{(j)} \tphi_{l}^{(\ell(j))} \nstd \,\dSt \, \, \bpi_{l,\ell(j)}
=\int_{\QQst_j}{\tpsi_k^{(j)} \tpsi_l^{(j)}  \, \dxt}\;  \hrho_{l,j} \mathbf{g} ,\label{eq:momentum_cns_weak2}\\
\int_{\TT_i}{\phi_{\kappa}^{(i)} \phi_{l}^{(i)}  \dxt} \frac{\partial \left( \rhoEE\right)_{l,i}}{\partial t}
+ \sum_{j \in S_i}\left( \int_{\Gamma_j} \phi_{\kappa}^{(i)}\psi_{l}^{(j)}\psi_{r}^{(j)}   \vec{n}_{i,j} \, \dSt \cdot H_{l,j} (\rho \vv)_{r,j} 
-\int_{\TT_{i,j}} \nabla \phi_{\kappa}^{(i)}\psi_{l}^{(j)}\psi_{r}^{(j)}  \dxt \cdot H_{l,j} (\rho \vv)_{r,j} \right) \notag \\
+	\int_{\TT_i}{\phi_{\kappa}^{(i)}\nabla \cdot \TF^{k}_c\left(  k, (\rho \mathbf{v})\right)  \dxt} =\sum_{j \in S_i}\left( \int_{\Gamma_j} \phi_k^{(i)} \psi_{l}^{(j)}   \vec{n}_{i,j} \, \dSt \cdot\mathbf{w}_{l,j}
-\int_{\TT_{i,j}} \nabla \phi_k^{(i)} \psi_{l}^{(j)}  \, \dxt\cdot \mathbf{w}_{l,j}  \right) \notag \\
+\sum_{j \in S_i}\left( \int_{\Gamma_j}   \phi_k^{(i)} \psi_{l}^{(j)} \vec{n}_{i,j} \, \dSt \cdot \mathbf{q}_{l,j} 
-\int_{\TT_{i,j}}{\nabla \phi_k^{(i)} \psi_{l}^{(j)} \, \dxt \cdot \mathbf{q}_{l,j}}  \right)
+ \int_{\TT_i} \phi_{\kappa}^{(i)}\phi_{l}^{(i)}   \, \dxt \,\left(\rho\mathbf{v}\right)_{l,i} \cdot \mathbf{g}.
\label{eq:energy_cns_weak2}
\end{gather}

Finally, discretization in time of the above system yields

\begin{gather}
\int_{\TT_{i}} \phi_{\kappa}^{(i)} \phi_{l}^{(i)} \dx \,\brho_{l,i}^{n+1}
= \int_{\TT_{i}} \phi_{\kappa}^{(i)}  \phi_{l}^{(i)} \dx \,\brho_{l,i}^{n} - \Delta t \, \sum_{j \in S_i}\left( \int_{\Gammast_{j}}{\tphi_{\kappa}^{(i)} \tpsi_{l}^{(j)}  \nextud \, \dSt \cdot \hrv_{l,j}^{n}}
-\int_{\TTst_{i,j}} \nabla \tphi_{\kappa}^{(i)} \tpsi_{l}^{(j)} \, \dxt \cdot \hrv_{l,j}^{n}   \right)
\label{eq:mass_cns_weak_time2} \\
\frac{1}{\Delta t}\int_{\QQst_j}\tpsi_k^{(j)}  \tpsi_l^{(j)}   \, \dxt \, \hrv^{n+1}_{l,j}
-\frac{1}{\Delta t} \int_{\QQst_j}\tpsi_k^{(j)}  \tpsi_l^{(j)}  \, \dxt \, \hat{\TF}^{\rho\mathbf{v}}_{t}(\hbrv_{l,j}^{n})
+\int_{\TTlst} \tpsi_k^{(j)} \nabla \tphi_{l}^{(\ell(j))}  \, \dxt \,  \, \bpi_{l,\ell(j)}^{n+1}\notag\\    
+\int_{\TTrst}  \tpsi_k^{(j)} \nabla \tphi_{l}^{(r(j))} \, \dxt \,  \, \bpi_{l,r(j)}^{n+1}
+\int_{\Gammast_{j}}\tpsi_k^{(j)} \tphi_{l}^{(r(j))}    \nstd \, \dSt \,  \, \bpi_{l,r(j)}^{n+1}
-\int_{\Gammast_{j}} \tpsi_k^{(j)} \tphi_{l}^{(\ell(j))} \nstd \,\dSt \, \, \bpi_{l,\ell(j)}^{n+1}
=\int_{\QQst_j}{\tpsi_k^{(j)} \tpsi_l^{(j)}  \, \dxt}\;  \hrho_{l,j}^{n} \mathbf{g} ,\label{eq:momentum_cns_weak_time2}\\
\frac{1}{\Delta t}\int_{\TT_i}{\phi_{\kappa}^{(i)} \phi_{l}^{(i)}  \dxt}  \left( \overline{\rhoEE}\right)_{l,i}^{n+1}
-\frac{1}{\Delta t}\int_{\TT_i}{\phi_{\kappa}^{(i)} \phi_{l}^{(i)}  \dxt}  \left( \overline{\rhoEE}\right)_{l,i}^{n}
+ \sum_{j \in S_i}\left( \int_{\Gamma_j} \phi_{\kappa}^{(i)}\psi_{l}^{(j)}\psi_{r}^{(j)}   \vec{n}_{i,j} \, \dSt \cdot \hat{H}_{l,j}^{n+1} (\hrv)_{r,j}^{n+1} 
-\int_{\TT_{i,j}} \nabla \phi_{\kappa}^{(i)}\psi_{l}^{(j)}\psi_{r}^{(j)}  \dxt \cdot \hat{H}_{l,j}^{n+1} (\hrv)_{r,j}^{n+1} \right) \notag \\
+	\int_{\TT_i} \phi_{\kappa}^{(i)}\nabla \cdot \TF^{k}_c\left(  k^{n+1}, (\rho \mathbf{v})^{n+1}\right) \dxt 
=\sum_{j \in S_i}\left( \int_{\Gamma_j} \phi_k^{(i)} \psi_{l}^{(j)}   \vec{n}_{i,j} \, \dSt \cdot\hat{\mathbf{w}}_{l,j}^{n+1}
-\int_{\TT_{i,j}} \nabla \phi_k^{(i)} \psi_{l}^{(j)}  \, \dxt\cdot \hat{\mathbf{w}}_{l,j}^{n+1}  \right) \notag \\
+\sum_{j \in S_i}\left( \int_{\Gamma_j}   \phi_k^{(i)} \psi_{l}^{(j)} \vec{n}_{i,j} \, \dSt \cdot \hat{\mathbf{q}}_{l,j}^{n+1} 
-\int_{\TT_{i,j}} \nabla \phi_k^{(i)} \psi_{l}^{(j)} \, \dxt \cdot \hat{\mathbf{q}}_{l,j}^{n+1}  \right)
+ \int_{\TT_i} \phi_{\kappa}^{(i)}\phi_{l}^{(i)}   \, \dxt \,\left(\brv\right)_{l,i}^{n+1} \cdot \mathbf{g},
\label{eq:energy_cns_weak_time2}
\end{gather}
where the convective and diffusive terms
\begin{gather}
\hat{\TF}^{\rho\mathbf{v}}_{t}(\hbrv_{j}^{n}) = \hbrv_{j}^{n} - \Delta t \MM^{-1}_j \boldsymbol{\Upsilon}_{j}^{\rho\mathbf{v}}, \quad \boldsymbol{\Upsilon}_{j}^{\rho\mathbf{v}} = \int_{\QQst_{j}} \psi_{k}^{(i)} \nabla\cdot \mathbf{F}^{\rho\mathbf{v}},\\
\int_{\TT_i} \phi_{\kappa}^{(i)}\nabla \cdot \TF^{k}_c\left(  k^{n+1}, (\rho \mathbf{v})^{n+1}\right) \dxt = \boldsymbol{\Upsilon}_{i}^{k,c},
\end{gather}
are evaluated following the procedure introduced in Section \ref{sec:advect_diff}.
The density on the dual mesh, needed in \eqref{eq:momentum_cns_weak_time2}, is recovered from its value on the primal mesh as 
\begin{gather}
\hat{\boldsymbol{\rho}}_j^{n} = \Mpsi_j^{-1}\left( \LL^\top_{\ell(j),j} \bar{\boldsymbol{\rho}}_{\ell(j)}^{n} + \LL^\top_{r(j),j}\bar{\boldsymbol{\rho}}_{r(j)}^{n} \right). \label{eq:recrho_pic}
\end{gather}
By introducing
\begin{gather}
\MEPS_{ij}= \int\limits_{\Gamma_{j}}{\phi_{k}^{(i)}\psi_l^{(j)}\psi_r^{(j)}\vec{n}_{i,j}\dSt}-\int\limits_{\TT_{ij}}{\nabla \phi_k^{(i)}\psi_l^{(j)}\psi_r^{(j)}\dxt},
\label{eq:Dtilde}
\end{gather}
and using the matrix definitions \eqref{eq:MD_3}-\eqref{eq:vis11_1}, the above system is written more compactly as
\begin{gather}
\Mphi_i   \bbrho_i^{n+1}  = \Mphi_i  \bbrho_i^{n} - \Delta t \, \sum_{j \in S_i} \D_{i,j}\hbrv_j^{n},
\label{eq:mass_mat}\\
\Mpsi_j (\rhovh)_j^{n+1} 
= \Mpsi_j(\rhovh)_j^{n} + \Delta t\,\left(- \hat{\boldsymbol{\Upsilon}}_{j}^{\rho\mathbf{v},n} - \Q_{r(j),j}  \etah_{r(j)}^{n+1} - \Q_{\ell(j),j} \etah_{\ell(j)}^{n+1} + \Mpsi_j \hbrho_{j}^{n} \mathbf{g}\right) ,
\label{eq:momentum_mat} \\
\Mphi_i(\rhoEEh)_i^{n+1} 
=   \Mphi_i(\rhoEEh)_i^n + \Delta t\,\left[ -\bar{\boldsymbol{\Upsilon}}_{i}^{k,c,n+1}  - \sum_{j \in S_i}{\MEPS_{ij} \hh_j^{n+1} (\rhovh)_j^{n+1}} + \sum_{j \in S_i}{\D_{i,j} \left( \hat{\mathbf{w}}_j^{n+1}  + \hat{\mathbf{q}}_j^{n+1} \right) }  +  \Mphi_i \bbrv_{i}^{n+1} \cdot\mathbf{g}       \right].
\label{eq:energy_mat} 
\end{gather}

% % % % % % % % % % % % % % % % % % % % % % % % % % % % % %
\subsection{Pressure system and Picard iteration}\label{sec:picard_cns}
The pressure appearing in the momentum equation \eqref{eq:momentum_mat} is discretized implicitly as well as the momentum in the energy equation. Then, a pressure equation can be derived by 
formal substitution of the discrete momentum equation \eqref{eq:momentum_mat} into the discrete energy equation \eqref{eq:energy_mat}:
\begin{gather}
\Mphi_i\left[ (\rhoeh)_i^{n+1}+(\rhokh)_i^{n+1} \right] -  \Delta t\sum_{j \in S_i}\MEPS_{ij} \hh_j^{n+1} \left[ \Mpsi_j^{-1}\,\left( \Q_{r(j),j}  \etah_{r(j)}^{n+1} + \Q_{\ell(j),j} \etah_{\ell(j)}^{n+1} \right) \right] 
=   \Mphi_i(\rhoEEh)_i^n + \Delta t\,\left[ -\bar{\boldsymbol{\Upsilon}}_{i}^{k,c,n+1}
\right. \notag\\\left. - \sum_{j \in S_i}\MEPS_{ij} \hh_j^{n+1} \Mpsi_j^{-1}\left(  \Mpsi_j(\rhovh)_j^{n} - \Delta t\,\left( \hat{\boldsymbol{\Upsilon}}_{j}^{\rho\mathbf{v},n}  + \Mpsi_j \hbrho_{j}^{n} \mathbf{g}\right) \right) + \sum_{j \in S_i}{\D_{i,j} \left( \hat{\mathbf{w}}_j^{n+1}  + \hat{\mathbf{q}}_j^{n+1} \right) }  +  \Mphi_i \bbrv_{i}^{n+1} \cdot\mathbf{g}       \right].
\label{eq:pressure1_cns}
\end{gather} 
Here, $(\rhoeh)_i^{n+1} = \boldsymbol{\rho} \mathbf{e}\left( \bar{\mathbf{p}}_i^{n+1}, \bar{\boldsymbol{\rho}}_i^{n+1} \right)$ is to be understood 
as a componentwise evaluation of the internal energy density $\rho e$ for each component of the input vectors $\bar{\mathbf{p}}_i^{n+1}$ and $\bar{\boldsymbol{\rho}}_i^{n+1}$. 
Due to the dependency of the enthalpy on the pressure, and due to the presence of the kinetic energy in the total energy, 
system \eqref{eq:pressure1_cns} is highly nonlinear. In order to obtain a \textit{linear} system for the pressure, a simple but very efficient 
Picard iteration is used, as suggested in \cite{DBTM08,DZ09,CZ09,DC16,TD14,TD16}. Consequently,  some nonlinear terms are discretized at the previous Picard iteration, and thus 
become essentially explicit. Specifically, only pressure remains fully implicit. Therefore, at each Picard iteration, $m=1,\ldots, N_{Pic}$, we need to solve 
\begin{gather}
\Mphi_i \, \left[ {\boldsymbol{\rho}} \mathbf{e}\left( \bar{\mathbf{p}}_i^{n+1,m+1}, \bar{\boldsymbol{\rho}}_i^{n+1} \right)  \right] - \Delta t\, \sum_{j \in S_i}{\MEPS_{ij} \hh_j^{n+1,m} \left[ \Delta t\, \Mpsi_j^{-1}\left( \Q_{r(j)j} \pph_{r(j)}^{n+1,m+1}  + \Q_{\ell(j)j} \pph_{\ell(j)}^{n+1,m+1} \right) \right]} \nonumber \\ 
= \Mphi_i^-(\rhoEEh)_i^n +\Delta t\, \left\lbrace - \Mphi_i  (\rhokh)_i^{n+1,m} - \overline{\boldsymbol{\Upsilon}}^{k,c,n+1,m}_i  + \sum_{j \in S_i}{\D_{i,j} \left( \hat{\mathbf{w}}_j^{n+1,m}  + \hat{\mathbf{q}}_j^{n+1,m} \right) } 
\right. \notag\\\left.
- \sum_{j \in S_i}{\MEPS_{ij} \hh_j^{n+1,m} \left[ \Mpsi_j^{-1}\left( \Mpsi_j^-(\rhovh)_j^{n} - \Delta t\, \hat{\boldsymbol{\Upsilon}}_{j}^{\rho\mathbf{v},n}\right) + \Delta t\,\hbrho_{j}^{n}  \mathbf{g}\right]}+  \Mphi_i \bbrv_{i}^{n+1,m} \cdot\mathbf{g}  \right\rbrace,
\label{eq:pressure_pic}\\
\Mpsi_j (\rhovh)_j^{n+1,m+1} +\Delta t\,\left(  \Q_{r(j),j}  \etah_{r(j)}^{n+1,m+1} + \Q_{\ell(j),j} \etah_{\ell(j)}^{n+1,m+1} \right)
= \Mpsi_j(\rhovh)_j^{n} + \Delta t\,\left( -\hat{\boldsymbol{\Upsilon}}_{j}^{\rho\mathbf{v},n} + \Mpsi_j \hbrho_{j}^{n} \mathbf{g}\right) ,
\label{eq:momentum_pic} \\
\Mphi_i(\rhoEEh)_i^{n+1} 
=   \Mphi_i(\rhoEEh)_i^n + \Delta t\,\left[ -\overline{\mathbf{F}^{k}}^{n+1,m+1}_i  - \sum_{j \in S_i}{\MEPS_{ij} \hh_j^{n+1,m} (\rhovh)_j^{n+1,m+1}} + \sum_{j \in S_i}{\D_{i,j} \left( \hat{\mathbf{w}}_j^{n+1,m}  + \hat{\mathbf{q}}_j^{n+1,m} \right) }  +  \Mphi_i \bbrv_{i}^{n+1,m+1} \cdot\mathbf{g}       \right].
\label{eqn:energy_pic} 
\end{gather}
We typically use a total number of $N_{Pic}=2$ iterations. {\color{cr12} The above equations introduce an implicit discretization for pressure and momentum in the momentum equation and in the energy equation, respectively. Second order of accuracy in time can be easily achieved by taking  $\bar{\mathbf{p}}_i^{n+\Theta,m+1} $ and $ (\rhovh)_j^{n+\Theta,m+1}$ in the momentum and energy equations with $\Theta=0.5$, which yields a Crank-Nicolson discretization.}

% % % % % % % % % % % % % % % % % % % % % % % % % % % % % %
%        Overall method
% % % % % % % % % % % % % % % % % % % % % % % % % % % % % %
\subsection{Overall method}\label{sec:overall_comp}
Given $\bbrho_{i}^{n}$,  $\bbpi_{i}^{n}$, $\hbrv_{j}^{n}$, and $\bbrE_{i}^{n}$, the staggered semi-implicit DG algorithm for solving the compressible Navier-Stokes equations \eqref{eq:CNS_mass}-\eqref{eq:CNS_energy} 
can therefore be summarized as follows.
\begin{enumerate}
	\item The density at the new time time step, $\bbrho_{i}^{n+1}$, is obtained using \eqref{eq:mass_mat}. This value is then projected onto the dual mesh, $\hbrho_{j}^{n+1}$, by means of \eqref{eq:recrho_pic}.
	\item System \eqref{eq:pressure_pic}-\eqref{eq:momentum_pic} is solved relying on a Picard iteration procedure in order to evaluate the pressure and the momentum at the new time step, $\bbpi_{i}^{n+1,m+1}$ and $\hbrv_{j}^{n+1,m+1}$, respectively.
	\item The updated total energy density, $\bbrE_{i}^{n+1}$, is computed from \eqref{eqn:energy_pic}  once the Picard loop has finished.
\end{enumerate}

{\color{cr1}
\section{Time step restriction}
The maximum time step is restricted by a CFL-type condition based on the local flow velocity:
\begin{equation}
\Delta t_{\mathrm{max}} = \frac{\mathrm{CFL}}{2 p+1} \cdot \frac{ h_{\mathrm{min}}}{2 \left| \mathbf{v}_{\mathrm{max}}\right|}\label{eq:timestep_restriction}
\end{equation}
with $\mathrm{CFL}<1/d$, $d$ the space dimension, $ h_{\mathrm{min}}$ the smallest insphere diameter (in 3D) or incircle radius (in 2D) and
$\mathbf{v}_{\mathrm{max}}$ is the maximum convective speed. If viscous terms are present, the eigenvalues of the viscous operator have to be considered as well (see \cite{Dum10,TD17}).

As commented in Section \ref{sec:semilagrangian}, if we employ the Eulerian-Lagrangian approach, the scheme becomes unconditionally stable for inviscid fluids, so that the time restriction is no more determined by the above CFL condition and $CFL\geq 1$ may be chosen. However, the parallel version of the code requires a safety factor to be specified, in order to ensure that the Lagrangian trajectories never exit the MPI neighborhood of each region (see \cite{TB19} for further details).
}

% % % % % % % % % % % % % % % % % % % % % % % % % % % % % %
% % % % % % % % % % % % % % % % % % % % % % % % % % % % % %
%               Numerical results                         %
% % % % % % % % % % % % % % % % % % % % % % % % % % % % % %
% % % % % % % % % % % % % % % % % % % % % % % % % % % % % %
\section{Numerical test problems}\label{sec:num_res}
In this section, classical benchmarks for natural convection problems 
are used in order to verify the validity and the efficiency of the novel algorithms presented in this work. Moreover, these tests allow us to analyze the strengths and drawbacks of our numerical schemes.

% % % % % % % % % % % % % % % % % % % % % % % % % % % % % %
% % % % % % % % % % % % % % % % % % % % % % % % % % % % % %
{\color{cr12}
	\subsection{Taylor-Green Vortex with gravity}\label{sec:TGV2D}
	To analyse the accuracy of the proposed schemes, we consider a modification of the Taylor-Green vortex benchmark by including the gravity term in the momentum equation. The exact solution of this test case reads
	\begin{gather}
	u(x,y,t) = \sin (x) \cos (y) e^{-2\nu t}, \label{eq:uvel_TGV}\\ 
	v(x,y,t) = - \cos (x) \sin (y) e^{-2\nu t} + gt, \label{eq:vvel_TGV}\\
	p(x,y,t) = \frac{1}{4}\left(  \cos (2x) + \cos (2y)\right)  e^{-4\nu t}, \label{eq:press_TGV}
	\end{gather}
	with $g=-9.81$. In order to satisfy the governing equations \eqref{eq:CS_2}-\eqref{eq:CS_2_2_0} with the definitions \eqref{eq:uvel_TGV}-\eqref{eq:press_TGV}, the following source terms need to be added to the right hand side of the momentum equations:
	\begin{gather}
	\mathbf{s}_{u} = -g t \sin(x)\sin(y)  e^{-2\nu t},\\
	\mathbf{s}_{v} = g t \cos(x)\cos(y)  e^{-2\nu t}.
	\end{gather}
	The simulations are run on the computational domain $\Omega=[0,2\pi]^{2}$ with periodic boundaries on a sequence of successively refined unstructured grids. Two settings are considered with different viscosity coefficient, namely $\nu=0$ and $\nu=0.1$. The convergence results at $t_{\mathrm{end}}=0.1$ are shown in Table \ref{tab:TGV}. 
	We observe that the optimal convergence rates are achieved for this non-trivial test with gravity and viscosity terms in a transient regime. The space-time DG discretization of \cite{TD15,TD16} has been employed with $p$ and $p_\gamma$ denoting the polynomial degree in space and time, respectively. From the obtained results we can conclude that the scheme converges with the expected convergence rate of at least $p+\frac{1}{2}$. 
\begin{table}
	{\color{cr12}\begin{center}
			\begin{tabular}{|c||c|c|c|c||c|c|c|c|}
				\hhline{-||----||----}
				\multirow{3}{*}{$N_{e}$}& \multicolumn{4}{c||}{$\nu=0$}  & \multicolumn{4}{c|}{$\nu=0.1$}  \\\hhline{|~||-|-|-|-||-|-|-|-|}
				& \multicolumn{2}{c|}{$(p,p_{\gamma})=(1,1)$}  & \multicolumn{2}{c||}{$(p,p_{\gamma})=(2,2)$} & \multicolumn{2}{c|}{$(p,p_{\gamma})=(1,1)$} &  \multicolumn{2}{c|}{$(p,p_{\gamma})=(2,2)$} \\\hhline{|~||-|-|-|-||-|-|-|-|} 		 
				& $\epsilon\left(\mathbf{v}\right)$ & $\mathcal{O}$ &  $\epsilon\left(\mathbf{v}\right)$ & $\mathcal{O}$  &  $\epsilon\left(\mathbf{v}\right)$ & $\mathcal{O}$  &  $\epsilon\left(\mathbf{v}\right)$ & $\mathcal{O}$  \\ \hhline{=::====::====} 	
				$44  $ & $ 2.81E-01$ &  & $6.72E-02 $ &  & $2.62E-01 $ & & $4.45E-02 $ &  \\ \hhline{-||----||----}  
				$176 $ & $ 9.56E-02$ & $ 1.6$ & $1.33E-02 $ & $2.3 $ & $7.43E-02 $ & $1.8 $ & $5.02E-03 $ & $3.1 $ \\ \hhline{-||----||----} 
				$396 $ & $ 5.32E-02$ & $ 1.4$ & $5.25E-03 $ & $2.3 $ & $3.34E-02 $ & $2.0 $ & $1.43E-03 $ & $3.1 $ \\ \hhline{-||----||----} 
				$704 $ & $ 3.54E-02$ & $ 1.4$ & $2.65E-03 $ & $2.4 $ & $1.86E-02 $ & $2.0 $ & $5.91E-04 $ & $3.1 $ \\ \hhline{-||----||----} 
				$1100$ & $ 2.59E-02$ & $ 1.4$ & $1.54E-03 $ & $2.4 $ & $1.18E-02 $ & $2.1 $ & $2.97E-04 $ & $3.1 $ \\ \hhline{-||----||----} 
				$1584$ & $ 1.99E-02$ & $ 1.4$ & $9.80E-04 $ & $2.5 $ & $8.15E-03 $ & $2.0 $ & $1.70E-04 $ & $3.1 $ \\ \hhline{-||----||----} 
				$2156$ & $ 1.59E-02$ & $ 1.5$ & $6.66E-04 $ & $2.5 $ & $5.96E-03 $ & $2.0 $ & $1.06E-04 $ & $3.0 $ \\ \hhline{-||----||----} 
			\end{tabular}
			\caption{Taylor-Green vortex. Numerical convergence results for the velocity vector field.} \label{tab:TGV}
	\end{center}	}
\end{table}
}

% % % % % % % % % % % % % % % % % % % % % % % % % % % % % %
% % % % % % % % % % % % % % % % % % % % % % % % % % % % % %
\subsection{Differentially heated cavity}\label{sec:DHC}

The differentially heated cavity test (DHC) has been proposed in \cite{VD83} to assess the performance of numerical methods used to solve the incompressible Navier-Stokes equations with heat transport. The problem is defined on a squared shaped domain $\Omega = [0,L]^2$ with two opposite differentially heated walls
and characteristic length $L=1$. 
Due to the small temperature gap between the walls $\delta \theta$ the Boussinesq assumption can be applied, which means that changes in the density can be neglected everywhere in the incompressible Navier-Stokes equations, apart from the buoyancy forces (gravity source term) in the momentum equation. 

\begin{figure}[h]
	\centering
	\includegraphics[width=0.34\linewidth]{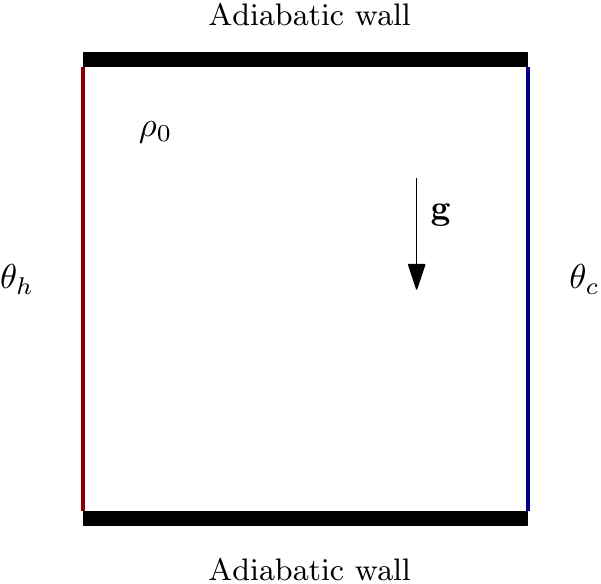}
	\caption{Differentially heated cavity test. Computational domain and boundary conditions.}
	\label{fig:domain}
\end{figure}

We consider the computational domain defined in Figure \ref{fig:domain}.   
Five different sets of parameters have been set according to values of the Rayleigh number $Ra = \frac{\beta \delta \theta g L^3}{\nu \alpha}$ between $10^{3}$ and $10^{7}$ with Prandtl number $Pr=0.71$ (see Table \ref{tab:dhc_opcond} for further details  on the operating conditions). Initially, we assume a constant temperature $\theta(\mathbf{x},0)=\theta_{0}=0$ and a fluid at rest, i.e. $\mathbf{v}(\mathbf{x},0)=0$. Adiabatic boundary conditions are prescribed at the bottom and top walls, whereas the exact temperature is imposed in the two heated walls, which is $\theta_{h}=0.5$ on the left and $\theta_{c}=-0.5$ on the right, hence $\delta \theta = \theta_h - \theta_c = 1$.

\begin{table}[H]
	\begin{center}
		\begin{tabular}{|c||c|c|c|c|c|}\hline
			Rayleigh &  $\mathbf{g}$ & $\rho_{0}$ & $\alpha$ & $\beta$ & $\nu$ \\\hline\hline
			$10^{3}$ & \multirow{5}{*}{$\left(0,-9.81\right)^{T}$} & \multirow{5}{*}{$1$} & $6.836 \cdot 10^{-3}$ & \multirow{5}{*}{$3.4112 \cdot 10^{-3}$} & $4.874 \cdot 10^{-3}$ \\\hhline{-||~|~|-|~|-|}
			$10^{4}$ & &  & $2.171 \cdot 10^{-3}$ & & $1.541 \cdot 10^{-3}$ \\\hhline{-||~|~|-|~|-|}
			$10^{5}$ & &  & $6.836 \cdot 10^{-4}$ & & $4.874 \cdot 10^{-4}$ \\\hhline{-||~|~|-|~|-|}
			$10^{6}$ & &  & $2.171 \cdot 10^{-4}$ & & $1.541 \cdot 10^{-4}$ \\\hhline{-||~|~|-|~|-|}
			$10^{7}$ & &  & $6.836 \cdot 10^{-5}$ & & $4.874 \cdot 10^{-5}$ \\\hline
		\end{tabular}
		\caption{Differentially heated cavity test. Operating conditions.}\label{tab:dhc_opcond}
	\end{center}
\end{table}

The staggered mesh employed has $5172$ primal elements. The simulations were run considering the numerical parameters $p=2$, $p_{\gamma}=0$, i.e. we use piecewise quadratic polynomials in space and a first order scheme in time. The steady state stopping criterion is defined as
\begin{equation}
\frac{1}{\Delta t}\left\|\mathbf{v}^{n+1}-\mathbf{v}^{n}\right\|_{L^{2}\left(\Omega\right)}\leq 10^{-4}.
\end{equation}
In order to analyze and compare the numerical results with available data in the literature, let us introduce the Nusselt number at the heated walls:
\begin{equation}
Nu_{\Gamma}=\frac{1}{\left|\Gamma\right|} \int_{ {\Gamma}} Nu_{\mathrm{loc}} \, dA, \quad Nu_{\mathrm{loc}} = \frac{\kappa\, L}{\kappa_{0} \left( \theta_{h}-\theta_{c}\right) }\frac{\partial \theta}{\partial\vec{n}},
\end{equation}
where $\Gamma$ refers to one of the heated walls, $L$ denotes the characteristic length
and $\kappa$ is the thermal conductivity with $\kappa_{0}=\kappa\left(\theta_{0}\right)$.

The Nusselt numbers obtained at both heated walls using the fully Eulerian and
the {\color{cr1} Eulerian-Lagrangian} schemes are shown in \mbox{Table \ref{tab:nusselt}}.
The average values presented in \cite{VD83}, \cite{MNZ98}, {\color{cr1} \cite{MUC00}}, \cite{WPW01}
and \cite{BB11} have been included for comparison purposes.
{\color{cr1} To avoid overestimation of the Nusselt number by the {\color{cr1} Eulerian-Lagrangian} scheme
we need to bound the time step. More precisely, the time step given by \eqref{eq:timestep_restriction} is halved, thus allowing the small structures embedded in the flow to be properly tracked.}

\begin{table}[H]
	\begin{center}
		\begin{tabular}{|c||c|c|c|c|c|c|c|c|c|}
			\hline
			\multirow{2}{*}{Ra}    & \multicolumn{2}{c|}{STIN2D Eu.} & \multicolumn{2}{c|}{STIN2D EL} &  Ref. \cite{VD83} & Ref. \cite{MNZ98} &  Ref. \cite{MUC00} & Ref. \cite{WPW01} &  Ref. \cite{BB11} 
			\\\hhline{|~||-|-|-|-|-|-|-|-|-|}
			& $\Gamma_h$ & $\Gamma_c$  & $\Gamma_h$ & $\Gamma_c$   &  Average &  Average  &  Average   & Average & Average \\\hline\hline
			$10^{3}$   &  $1.1199$     & $1.1199 $ & {\color{cr1} $1.1212 $} & {\color{cr1} $1.1212 $} & $1.118$ & $1.117$ & $1.1149 $ & $1.117 $ & $1.112$ \\\hline
			$10^{4}$   &  $2.2471$     & $2.2471 $ & {\color{cr1} $2.2486 $} & {\color{cr1} $2.2486 $} & $2.243$ & $2.243$ & $2.2593 $ & $2.254 $ & $2.198$ \\\hline
			$10^{5}$   &  $4.5283$     & $4.5302 $ & {\color{cr1} $4.6605 $} & {\color{cr1} $4.5302 $} & $4.519$ & $4.521$ & $4.4832 $ & $4.598 $ & $4.465$ \\\hline
			$10^{6}$   &  $8.8651$     & $8.8655 $ & {\color{cr1} $8.8592 $} & {\color{cr1} $8.8595 $} & $8.800$ & $8.806$ & $8.8811 $ & $8.976 $ & $8.783$ \\\hline
			$10^{7}$   &  $16.8394$    & $16.8428$ & {\color{cr1} $16.7684$} & {\color{cr1} $16.7714$} & $  --  $ & $16.40$ & $16.3869 $ & $16.656$ & $16.46$ \\\hline
		\end{tabular}
	\end{center}	
	\caption{Differentially heated cavity (DHC) test. Comparison of the Nusselt number obtained with the Eulerian (Eu) and {\color{cr1} Eulerian-Lagrangian} (EL) advection scheme with available reference solutions from the literature.}\label{tab:nusselt}
\end{table}	

\begin{table}[H]{\color{cr1}
	\begin{center}
		\begin{tabular}{|c||c|c|c|c|c|c|}
			\hline
			Ra  & STIN2D Eu. & STIN2D EL &  Ref. \cite{VD83} & Ref. \cite{MNZ98} & Ref. \cite{MUC00} & Ref. \cite{WPW01}   \\\hline\hline
			$10^{3}$   &  $3.7227  $    & {\color{cr1} $3.7324  $} & $3.679  $ & $3.692 $ & $3.6962   $ & $3.686  $ \\\hline
			$10^{4}$   &  $19.6342 $    & {\color{cr1} $19.6372 $} & $19.51  $ & $19.63 $ & $19.6177  $ & $19.79  $ \\\hline
			$10^{5}$   &  $68.9334 $    & {\color{cr1} $68.9526 $} & $68.22  $ & $68.85 $ & $68.6920  $ & $70.63  $ \\\hline
			$10^{6}$   &  $220.603 $    & {\color{cr1} $220.212 $} & $216.75 $ & $221.6 $ & $220.8331 $ & $227.11 $ \\\hline
			$10^{7}$   &  $701.051 $    & {\color{cr1} $700.679 $} & $--$      & $702.3 $ & $703.2526 $ & $714.48 $ \\\hline
		\end{tabular}
	\end{center}	
	\caption{Differentially heated cavity (DHC) test. Comparison of the normalized maximum vertical velocity at $y=0.5$ obtained with the Eulerian (Eu) and {\color{cr1} Eulerian-Lagrangian} (EL) advection schemes with available reference solutions from the literature.}\label{tab:vvelocity}}
\end{table}

\begin{table}[H]{\color{cr1}
	\begin{center}
		\begin{tabular}{|c||c|c|c|c|c|}
			\hline
			Ra  & STIN2D Eu. & STIN2D EL &  Ref. \cite{VD83} & Ref. \cite{MUC00} & Ref. \cite{WPW01}    \\\hline\hline
			$10^{3}$   &  $3.6736  $    & {\color{cr1} $3.6828  $} & $3.634 $ & $3.6493   $ & $3.489  $ \\\hline
			$10^{4}$   &  $16.1856 $    & {\color{cr1} $16.1877 $} & $16.2  $ & $16.1798  $ & $16.12  $ \\\hline
			$10^{5}$   &  $34.8869 $    & {\color{cr1} $34.896 $} & $34.81 $ & $34.7741  $ & $33.39  $ \\\hline
			$10^{6}$   &  $64.8511 $    & {\color{cr1} $64.9075 $} & $65.33 $ & $64.6912  $ & $65.40  $  \\\hline
			$10^{7}$   &  $148.681 $    & {\color{cr1} $149.561 $} & $--$     & $145.2666 $ & $143.56 $ \\\hline
		\end{tabular}
	\end{center}	
	\caption{ Differentially heated cavity (DHC) test. Comparison of the normalized maximum horizontal velocity at $x=0.5$ obtained with the Eulerian (Eu) and {\color{cr1} Eulerian-Lagrangian} (EL) advection schemes with available reference solutions from the literature.}\label{tab:hvelocity}}
\end{table}

\begin{figure}
	\begin{center}
		\includegraphics[width=0.4\linewidth]{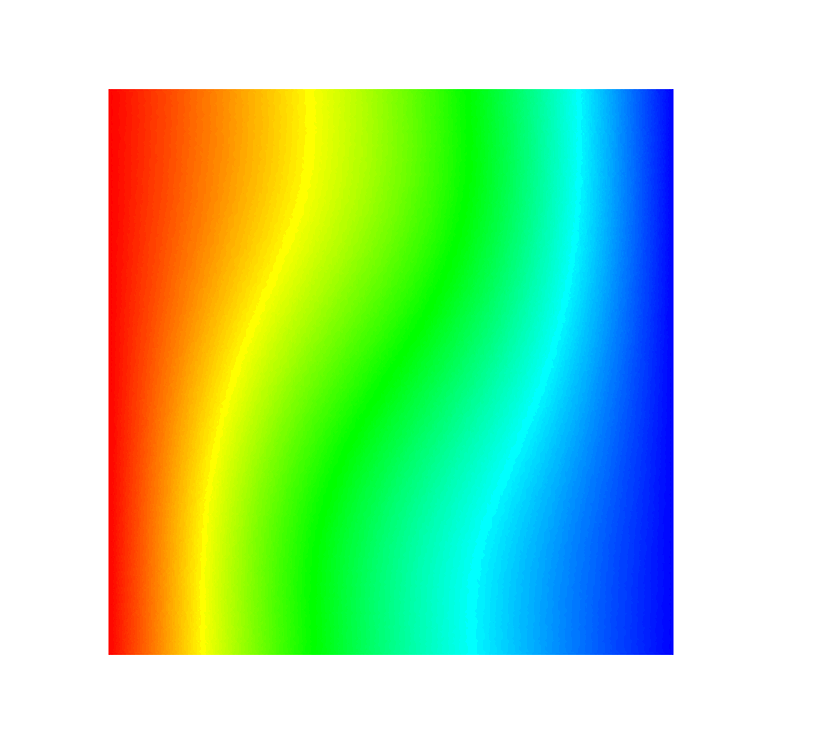}\hspace*{-0.07\linewidth}
		\includegraphics[width=0.4\linewidth]{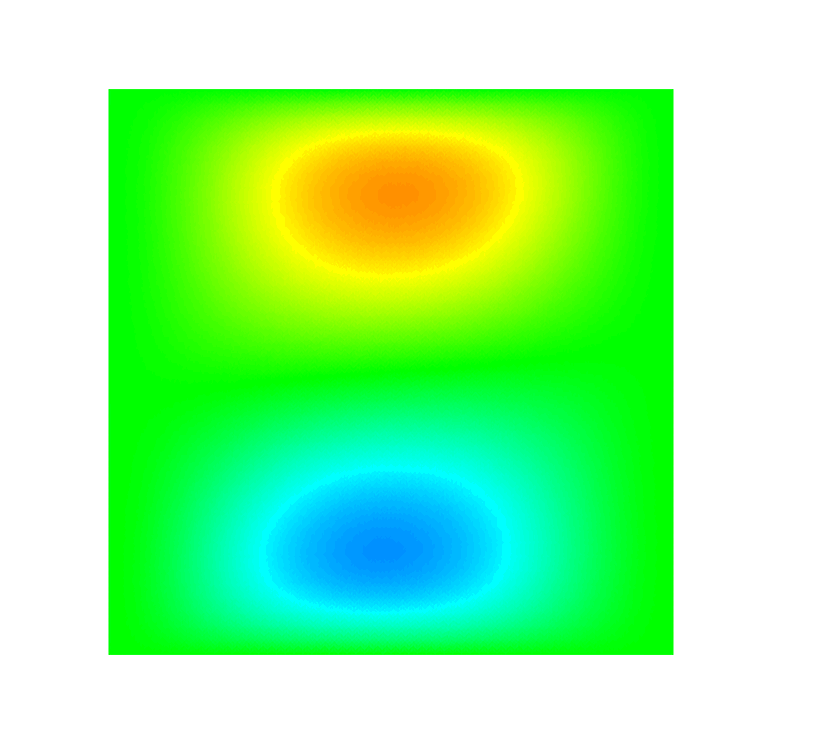}\hspace*{-0.07\linewidth}
		\includegraphics[width=0.4\linewidth]{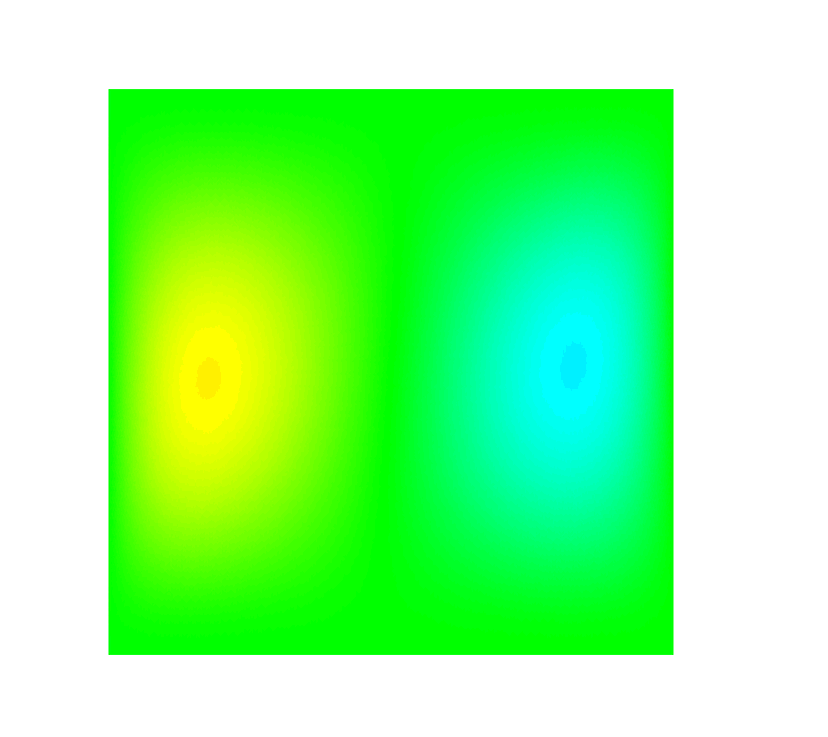}\\
		\includegraphics[width=0.4\linewidth]{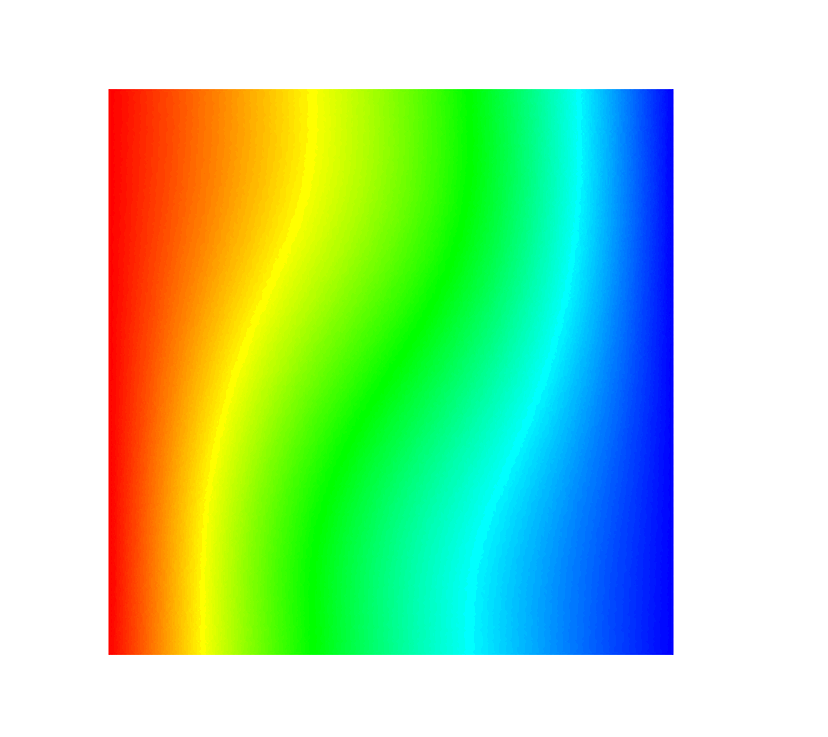}\hspace*{-0.07\linewidth}
		\includegraphics[width=0.4\linewidth]{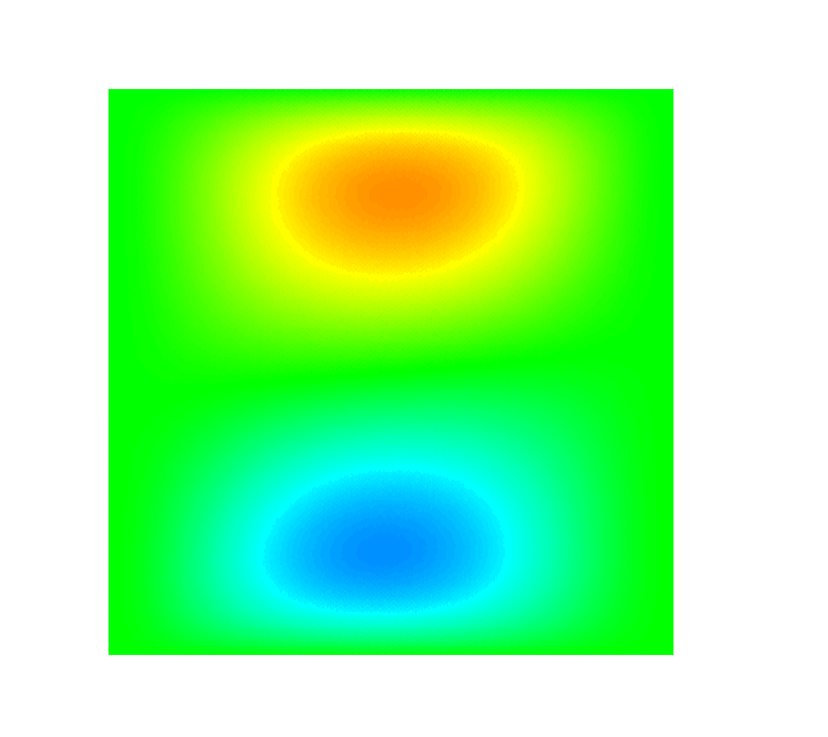}\hspace*{-0.07\linewidth}
		\includegraphics[width=0.4\linewidth]{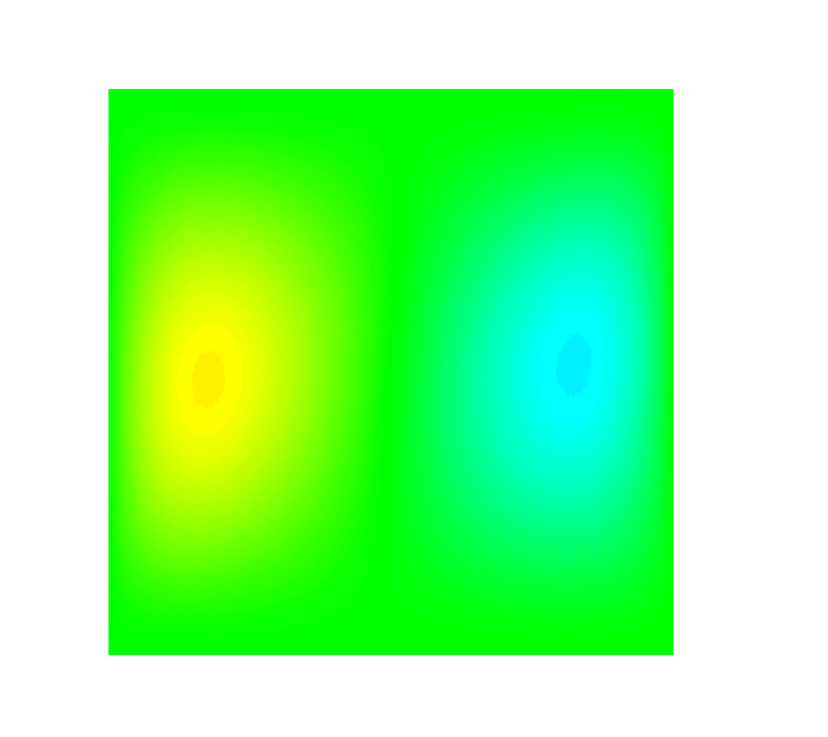}\\
		\includegraphics[width=0.36\linewidth]{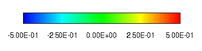}\hspace*{-0.02\linewidth}
		\includegraphics[width=0.36\linewidth]{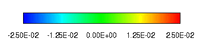}\hspace*{-0.02\linewidth}
		\includegraphics[width=0.36\linewidth]{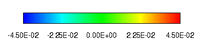}
	\end{center}
	\caption{DHC. From left to right: temperature, horizontal and vertical velocity contours at $Ra=10^{3}$. Top: Eulerian advection scheme. Bottom: {\color{cr1} Eulerian-Lagrangian} advection scheme.}\label{fig:dhc_1e3}
\end{figure}

\begin{figure}
	\begin{center}
		\includegraphics[width=0.4\linewidth]{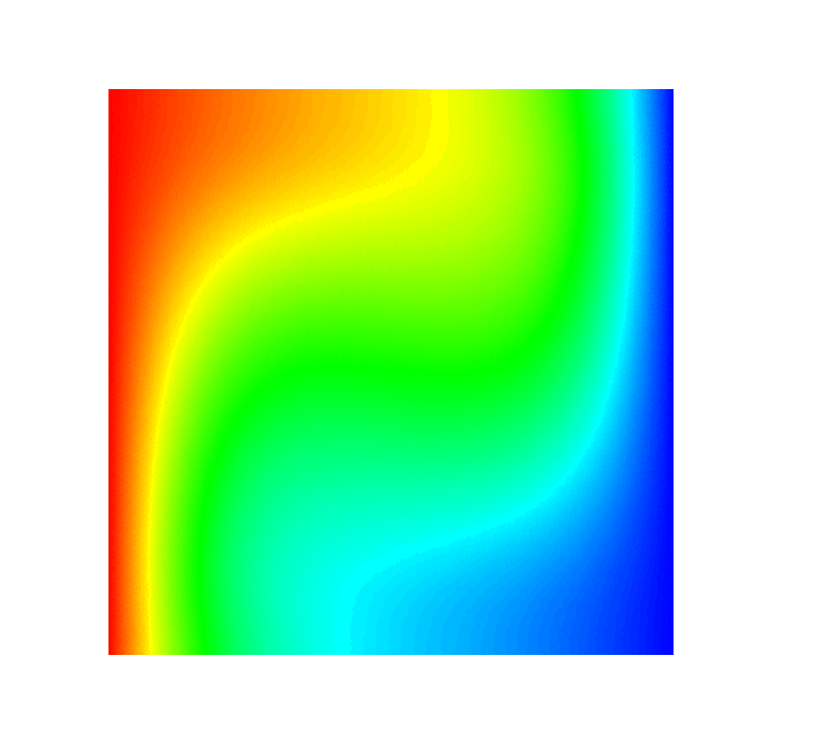}\hspace*{-0.07\linewidth}
		\includegraphics[width=0.4\linewidth]{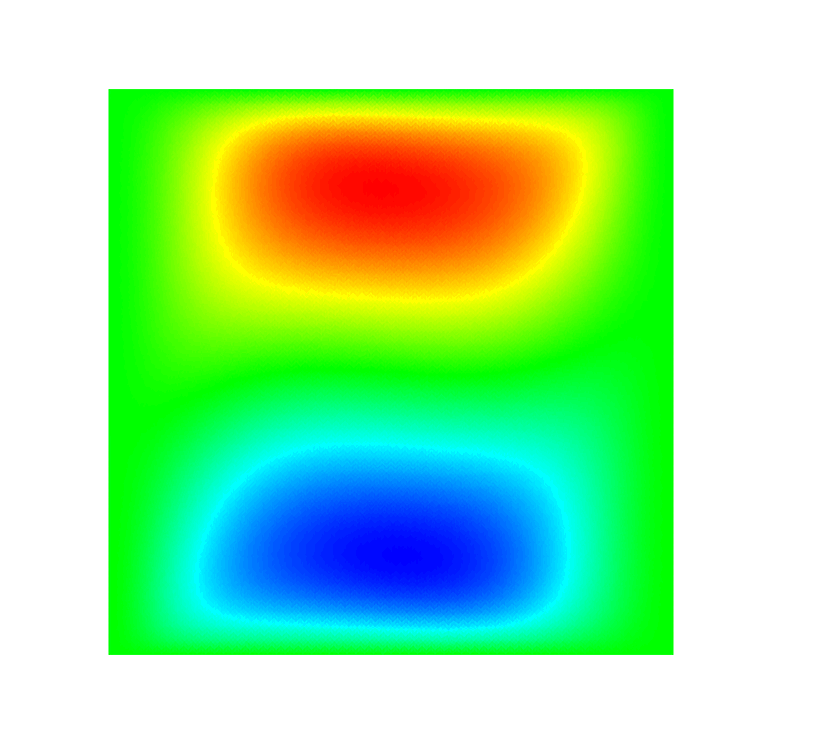}\hspace*{-0.07\linewidth}
		\includegraphics[width=0.4\linewidth]{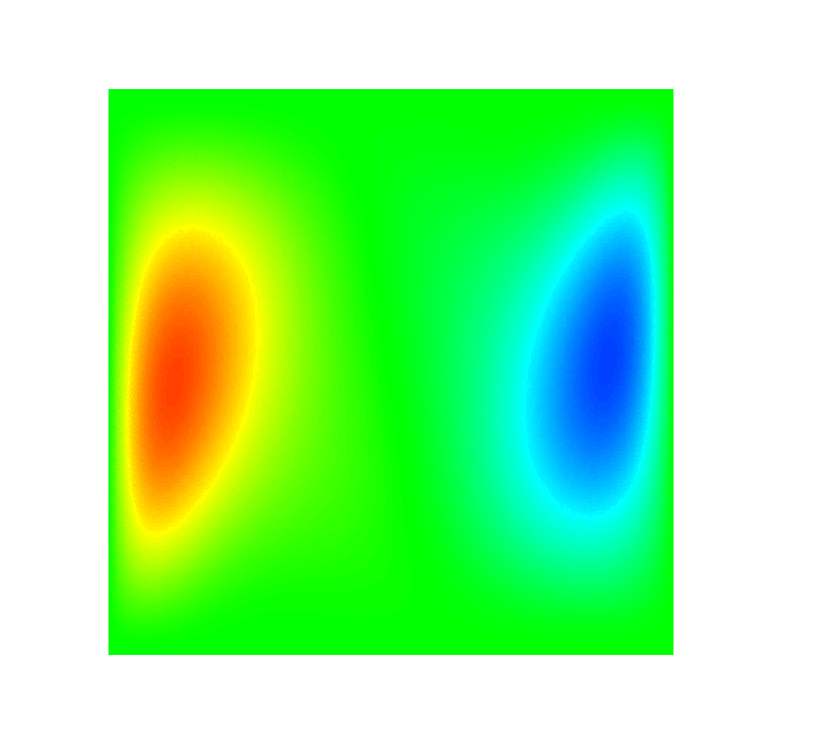}\\
		\includegraphics[width=0.4\linewidth]{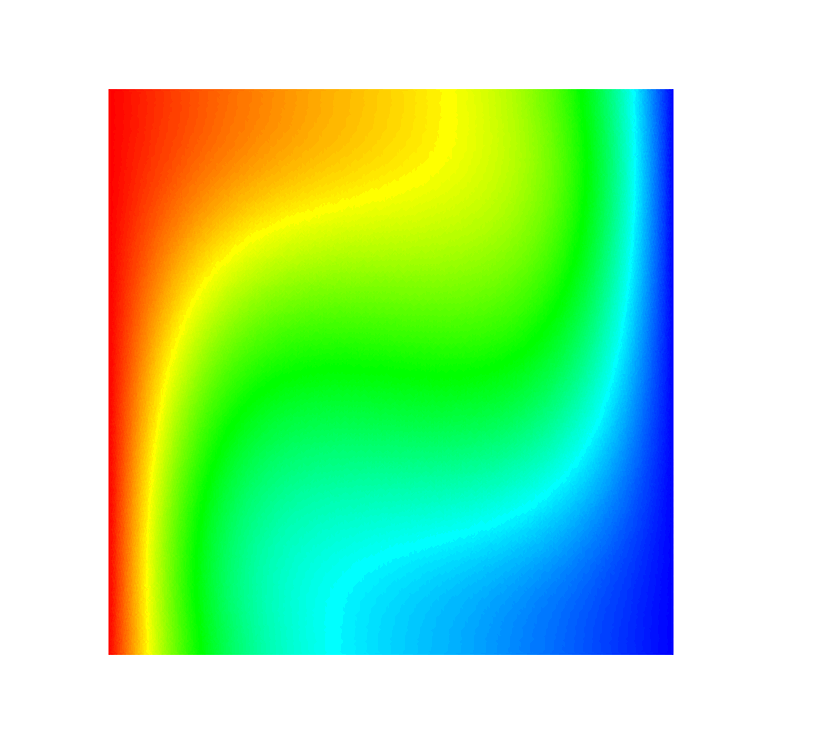}\hspace*{-0.07\linewidth}
		\includegraphics[width=0.4\linewidth]{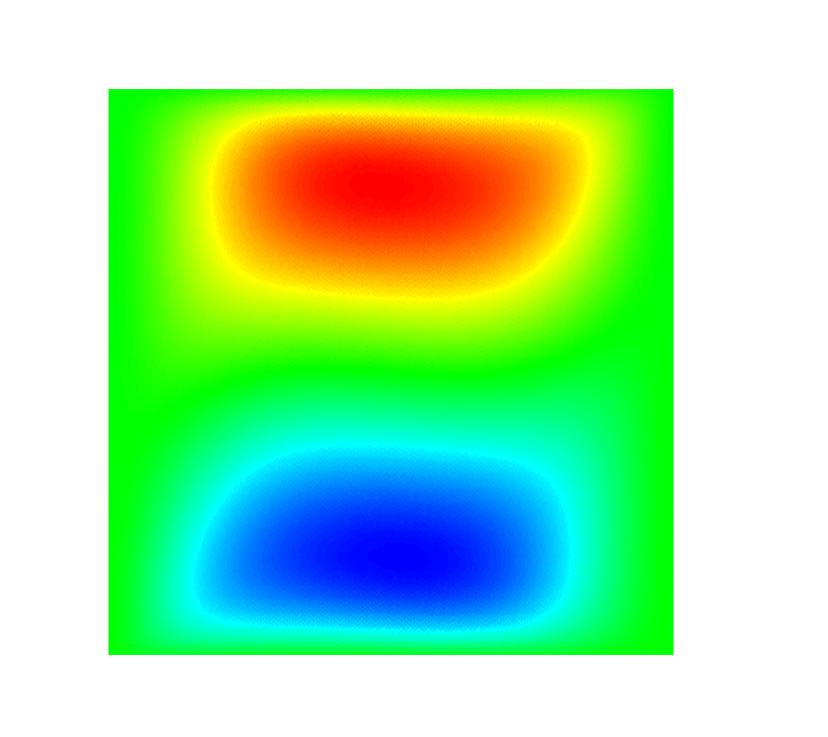}\hspace*{-0.07\linewidth}
		\includegraphics[width=0.4\linewidth]{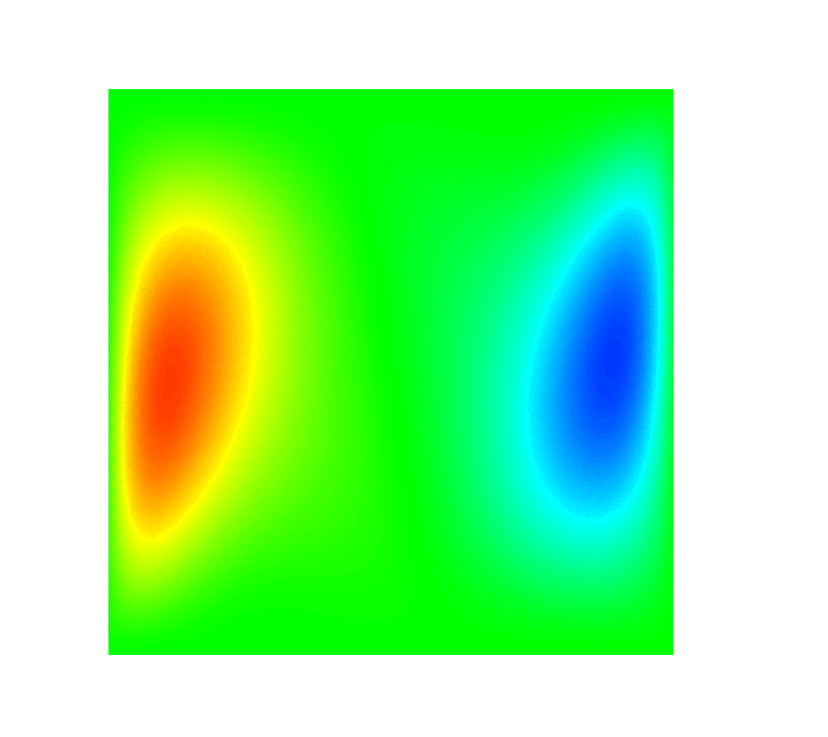}\\
		\includegraphics[width=0.36\linewidth]{./legend_temp.png}\hspace*{-0.02\linewidth}
		\includegraphics[width=0.36\linewidth]{./legend_velx.png}\hspace*{-0.02\linewidth}
		\includegraphics[width=0.36\linewidth]{./legend_vely.png}
	\end{center}
	\caption{DHC. From left to right: temperature, horizontal and vertical velocity contours at $Ra=10^{4}$. Top: Eulerian advection scheme. Bottom: {\color{cr1} Eulerian-Lagrangian} advection scheme.}\label{fig:dhc_1e4}
\end{figure}

\begin{figure}
	\begin{center}
		\includegraphics[width=0.4\linewidth]{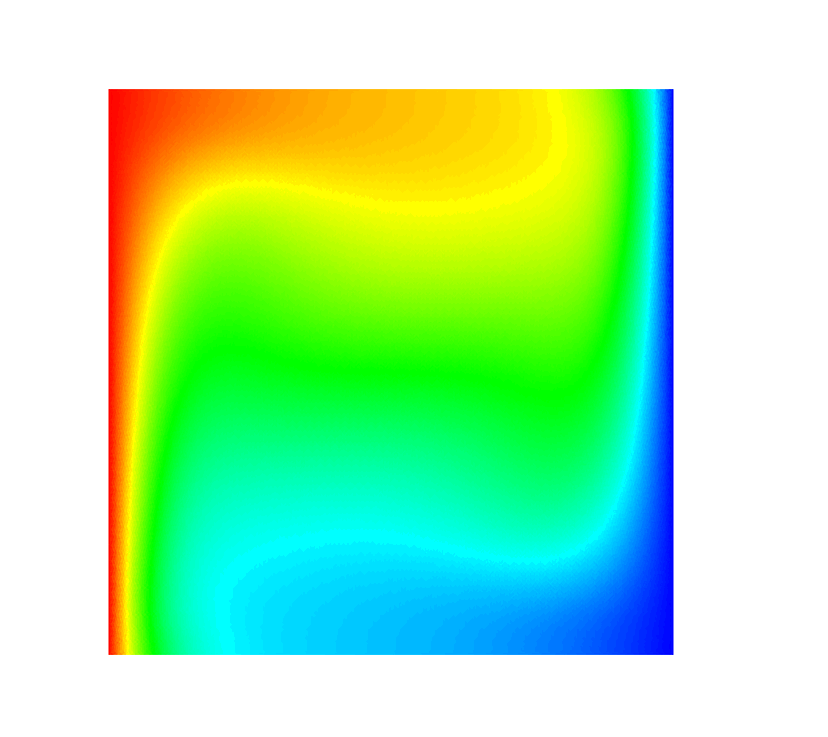}\hspace*{-0.07\linewidth}
		\includegraphics[width=0.4\linewidth]{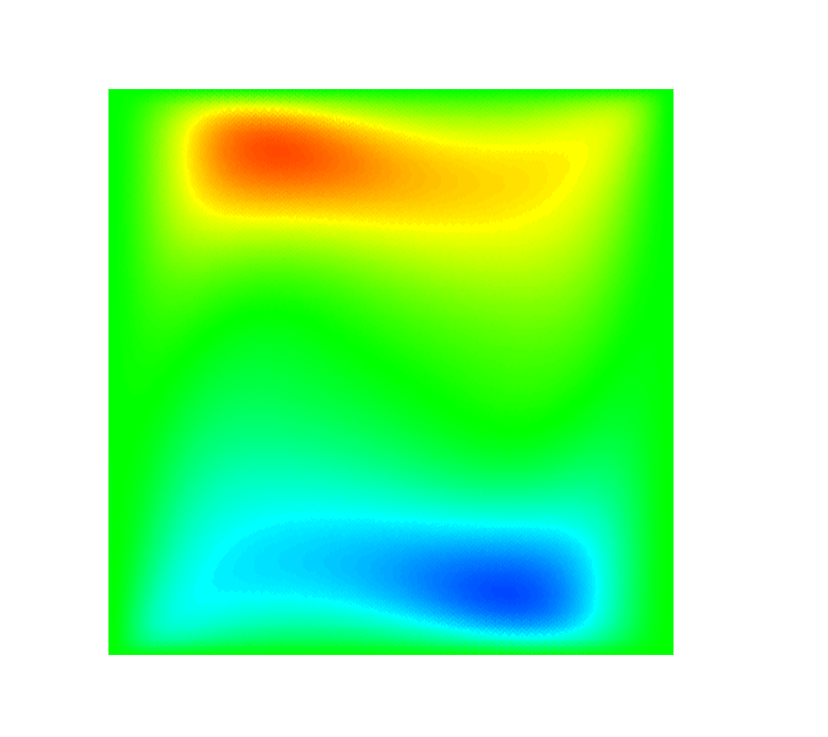}\hspace*{-0.07\linewidth}
		\includegraphics[width=0.4\linewidth]{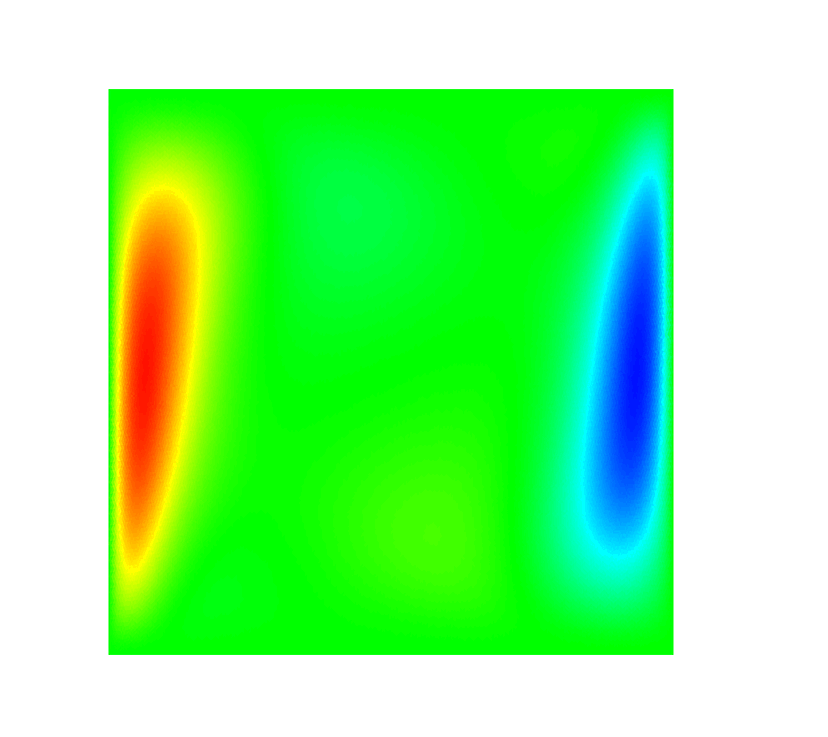}\\
		\includegraphics[width=0.4\linewidth]{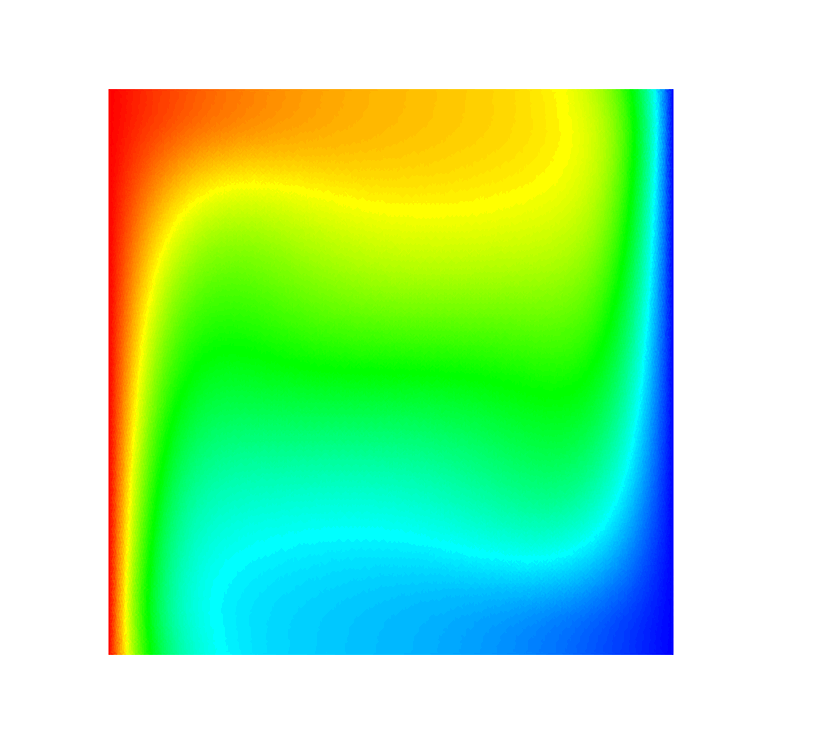}\hspace*{-0.07\linewidth}
		\includegraphics[width=0.4\linewidth]{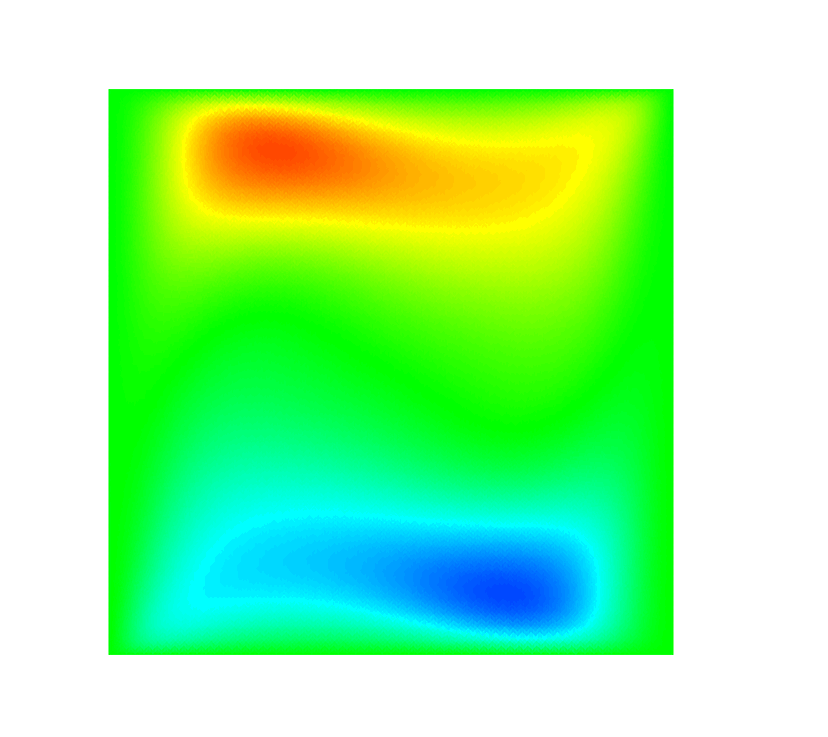}\hspace*{-0.07\linewidth}
		\includegraphics[width=0.4\linewidth]{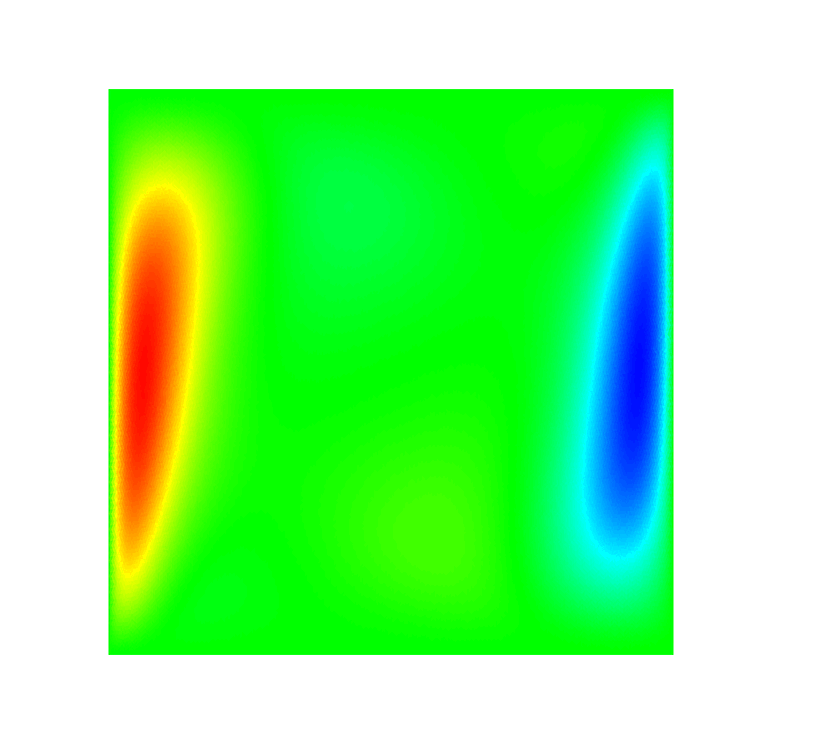}\\
		\includegraphics[width=0.36\linewidth]{./legend_temp.png}\hspace*{-0.02\linewidth}
		\includegraphics[width=0.36\linewidth]{./legend_velx.png}\hspace*{-0.02\linewidth}
		\includegraphics[width=0.36\linewidth]{./legend_vely.png}
	\end{center}
	\caption{DHC. From left to right: temperature, horizontal and vertical velocity contours, $Ra=10^{5}$. Top: Eulerian advection scheme. Bottom: {\color{cr1} Eulerian-Lagrangian} advection scheme.}\label{fig:dhc_1e5}
\end{figure}

\begin{figure}
	\begin{center}
		\includegraphics[width=0.4\linewidth]{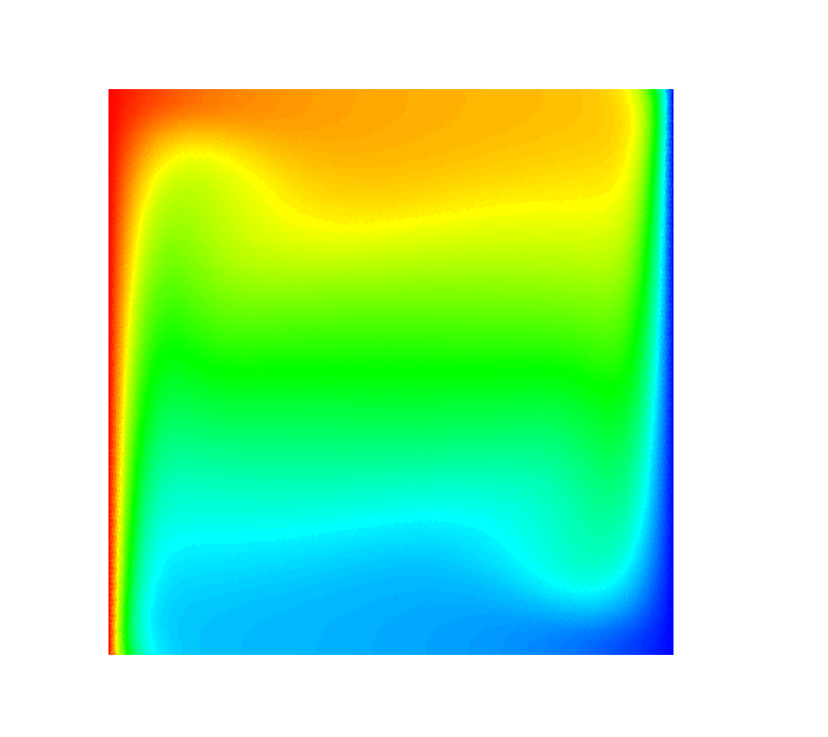}\hspace*{-0.07\linewidth}
		\includegraphics[width=0.4\linewidth]{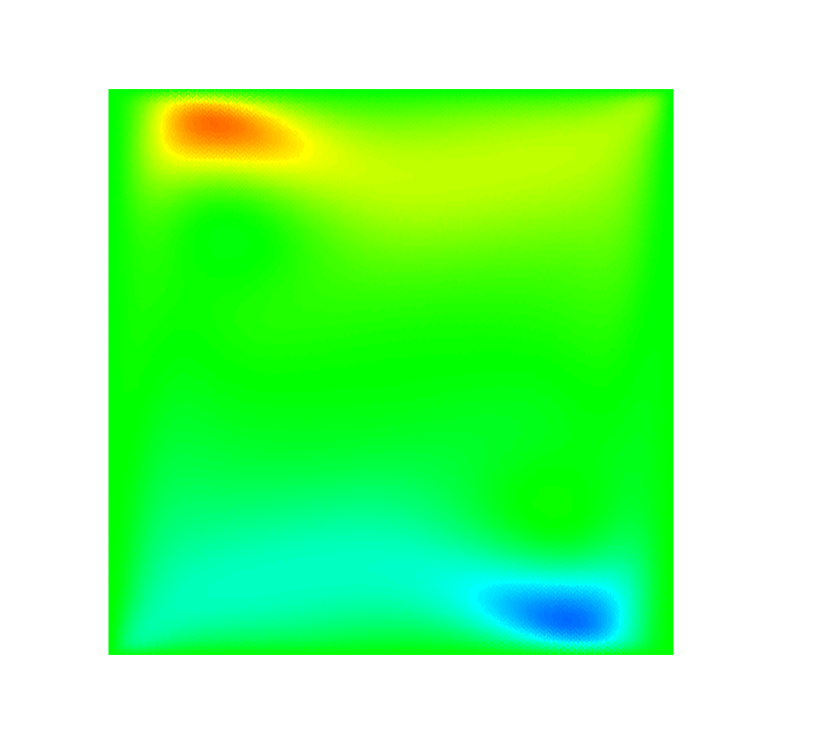}\hspace*{-0.07\linewidth}
		\includegraphics[width=0.4\linewidth]{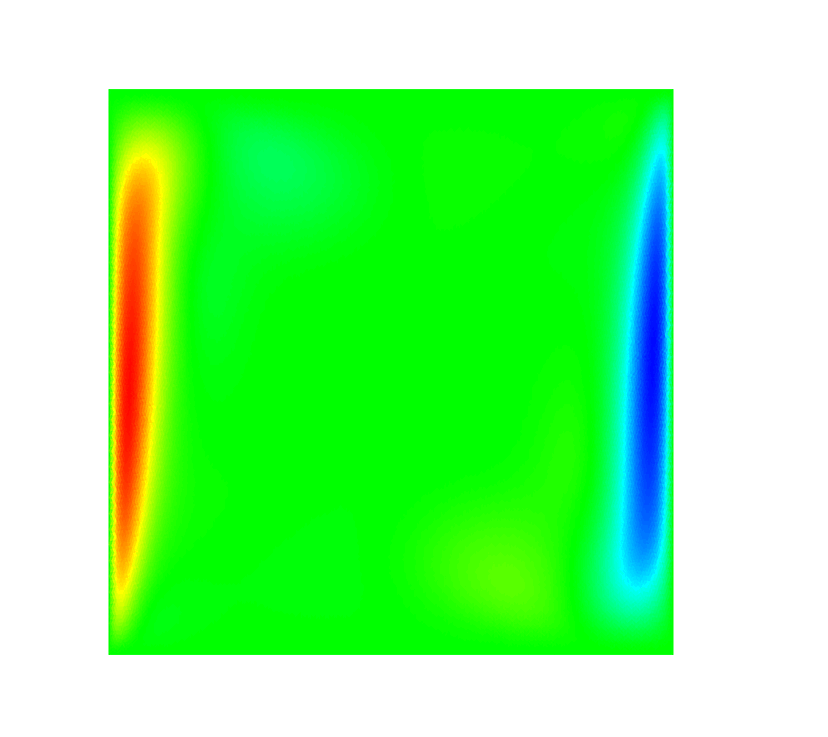}\\
		\includegraphics[width=0.4\linewidth]{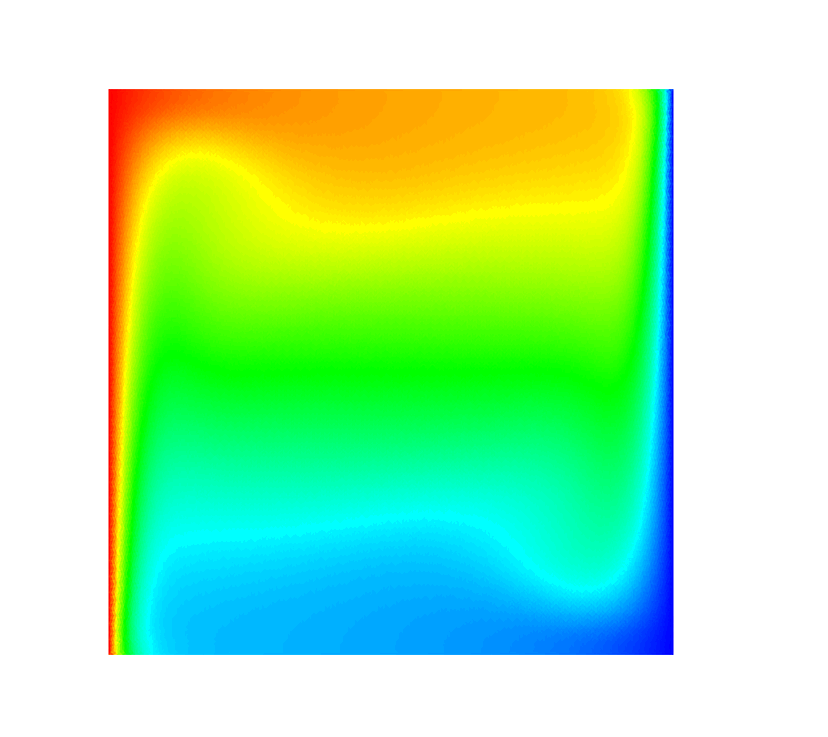}\hspace*{-0.07\linewidth}
		\includegraphics[width=0.4\linewidth]{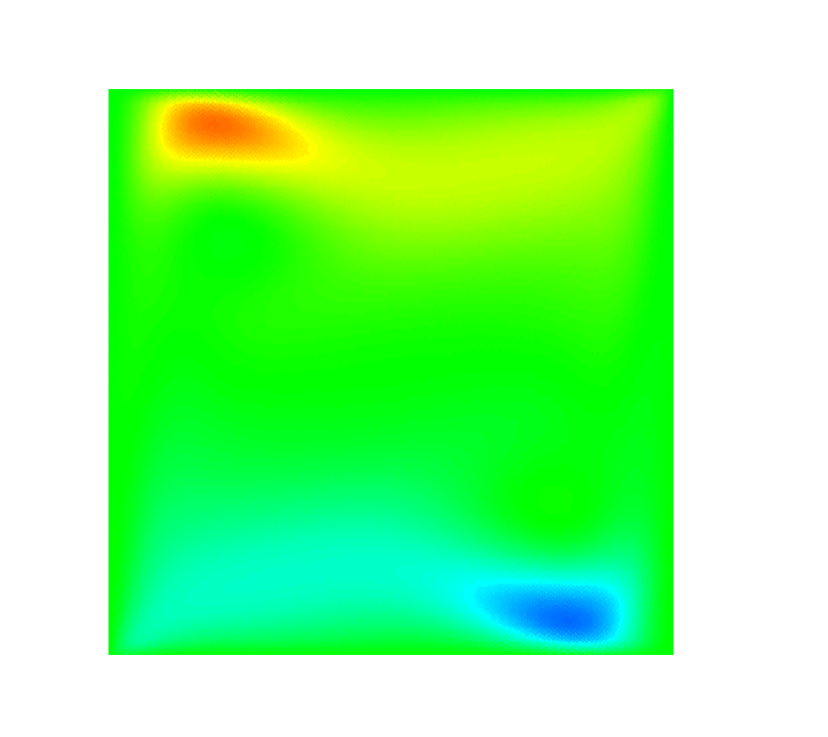}\hspace*{-0.07\linewidth}
		\includegraphics[width=0.4\linewidth]{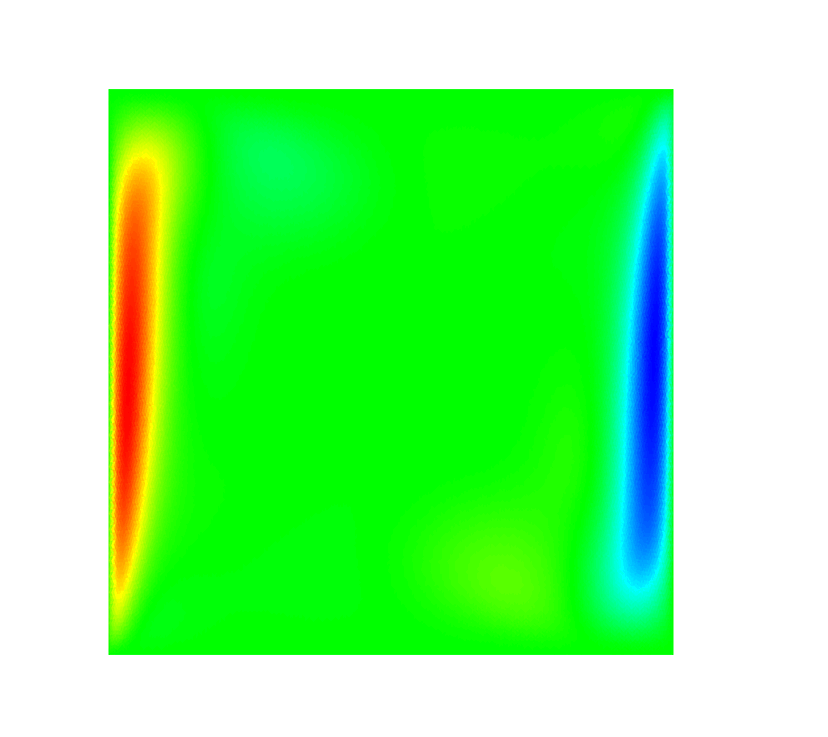}\\
		\includegraphics[width=0.36\linewidth]{./legend_temp.png}\hspace*{-0.02\linewidth}
		\includegraphics[width=0.36\linewidth]{./legend_velx.png}\hspace*{-0.02\linewidth}
		\includegraphics[width=0.36\linewidth]{./legend_vely.png}
	\end{center}
	\caption{DHC. From left to right: temperature, horizontal and vertical velocity contours, $Ra=10^{6}$. Top: Eulerian advection scheme. Bottom: {\color{cr1} Eulerian-Lagrangian} advection scheme.}\label{fig:dhc_1e6}
\end{figure}

\begin{figure}
	\begin{center}
		\includegraphics[width=0.4\linewidth]{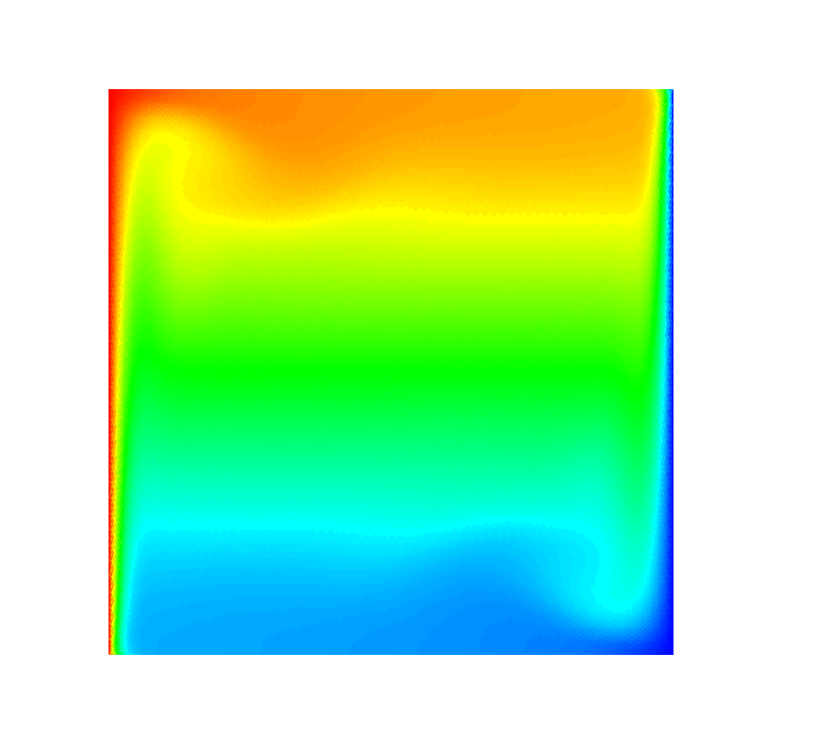}\hspace*{-0.07\linewidth}
		\includegraphics[width=0.4\linewidth]{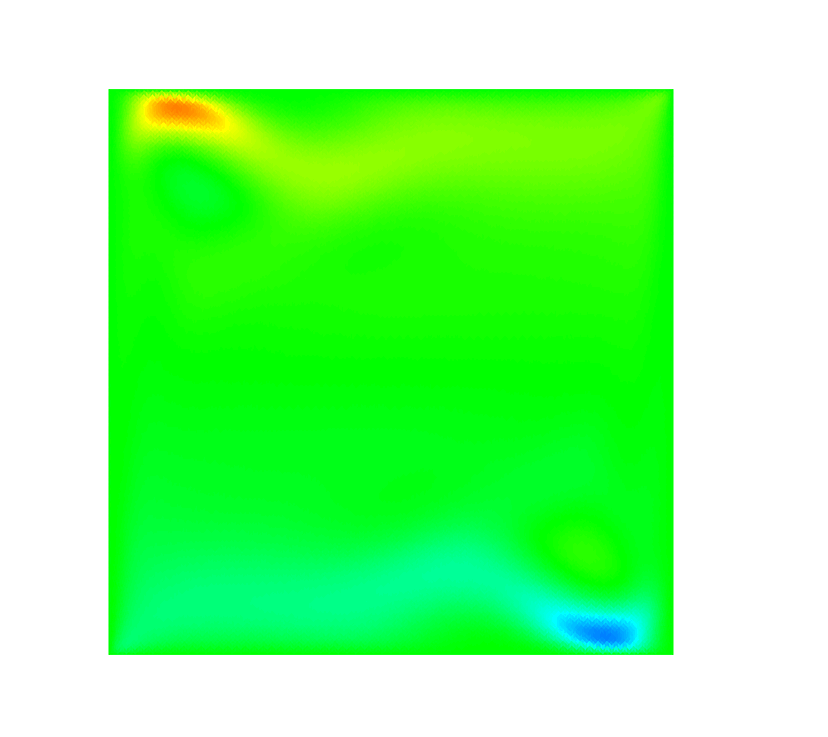}\hspace*{-0.07\linewidth}
		\includegraphics[width=0.4\linewidth]{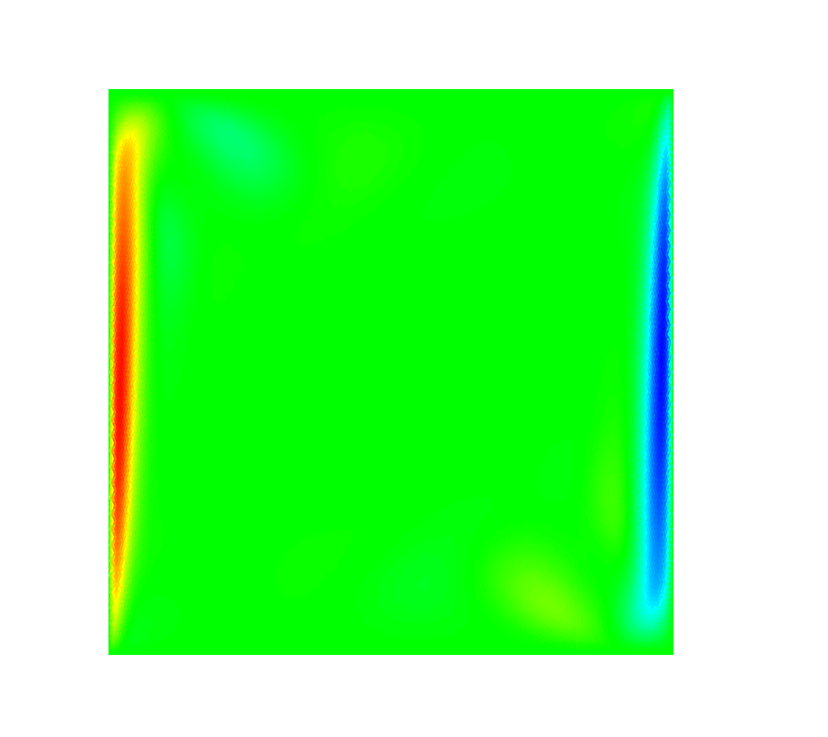}\\
		\includegraphics[width=0.4\linewidth]{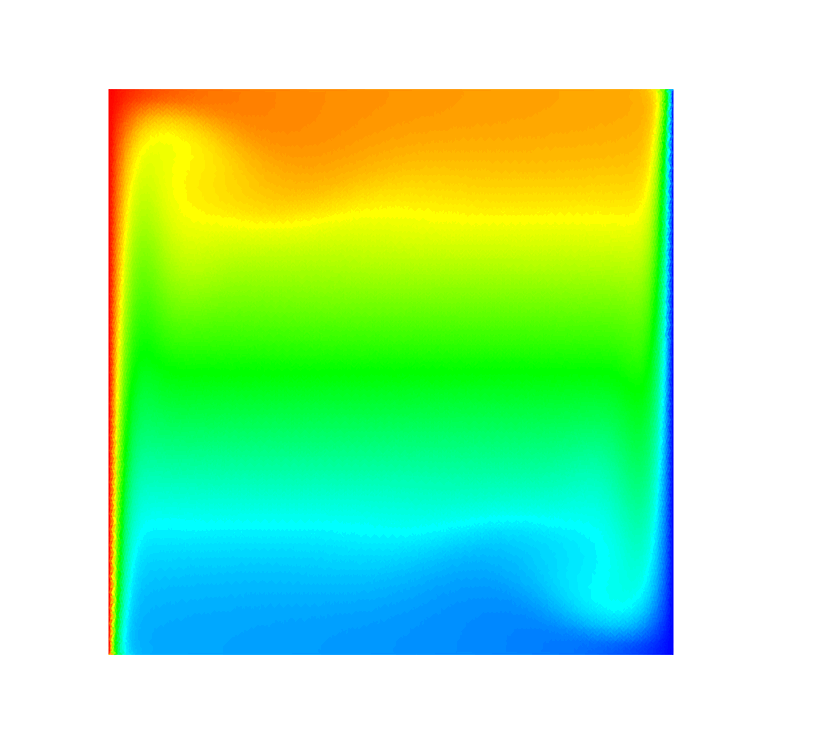}\hspace*{-0.07\linewidth}
		\includegraphics[width=0.4\linewidth]{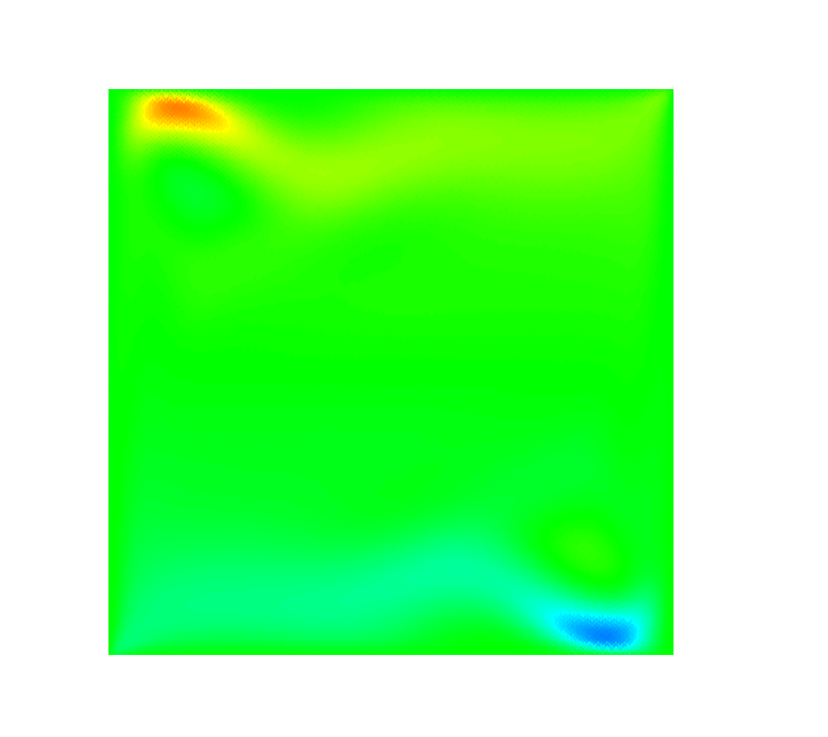}\hspace*{-0.07\linewidth}
		\includegraphics[width=0.4\linewidth]{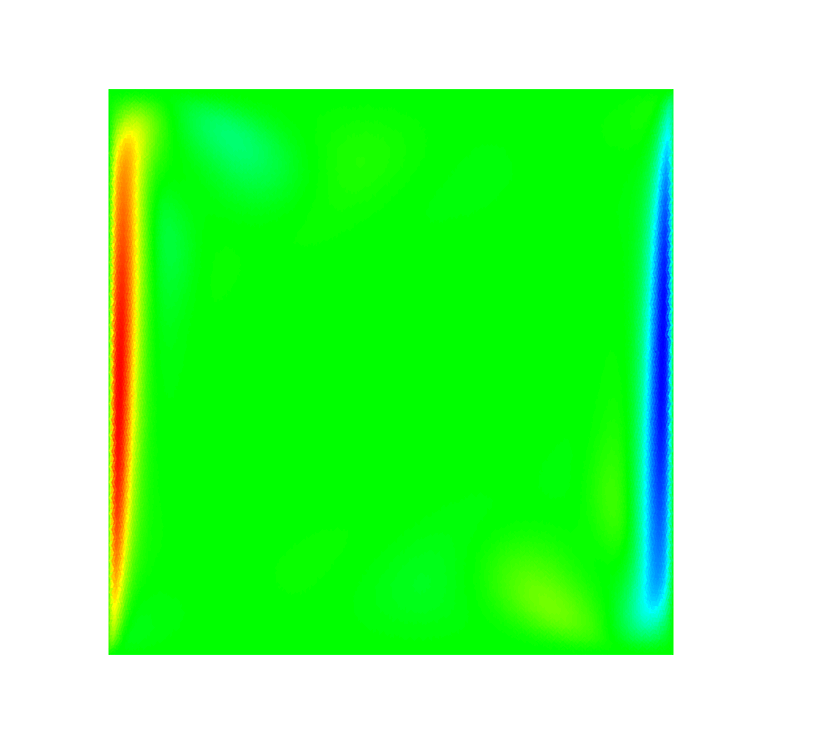}\\
		\includegraphics[width=0.36\linewidth]{./legend_temp.png}\hspace*{-0.02\linewidth}
		\includegraphics[width=0.36\linewidth]{./legend_velx.png}\hspace*{-0.02\linewidth}
		\includegraphics[width=0.36\linewidth]{./legend_vely.png}
	\end{center}
	\caption{DHC. From left to right: temperature, horizontal and vertical velocity contours, $Ra=10^{7}$. Top: Eulerian advection scheme. Bottom: {\color{cr1} Eulerian-Lagrangian} advection scheme.}\label{fig:dhc_1e7}
\end{figure}

\begin{figure}
	\begin{center}
		{\small $Ra=10^{3}$ \hspace{0.2\linewidth} $Ra=10^{4}$} \\
		\includegraphics[width=0.4\linewidth]{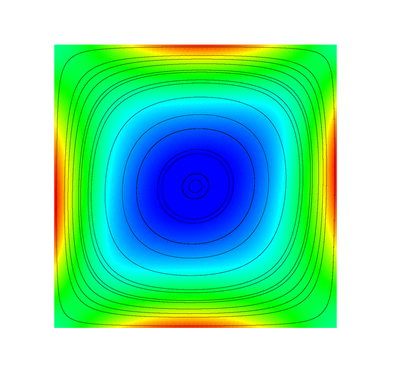}\hspace*{-0.07\linewidth}
		\includegraphics[width=0.4\linewidth]{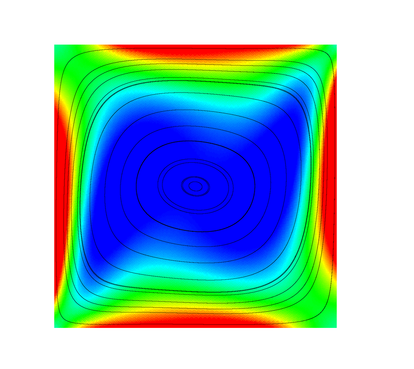}\\
		{\small$Ra=10^{5}$  \hspace{0.2\linewidth} $Ra=10^{6}$ \hspace{0.2\linewidth} $Ra=10^{7}$}\\
		\hspace*{-0.03\linewidth}\includegraphics[width=0.4\linewidth]{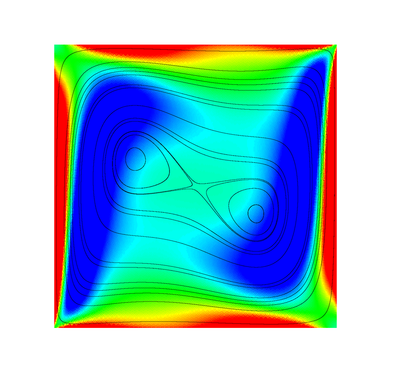}\hspace*{-0.07\linewidth}
		\includegraphics[width=0.4\linewidth]{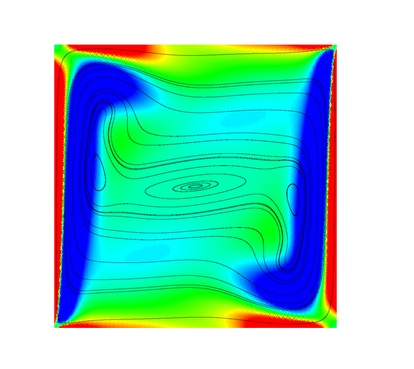}\hspace*{-0.07\linewidth}
		\includegraphics[width=0.4\linewidth]{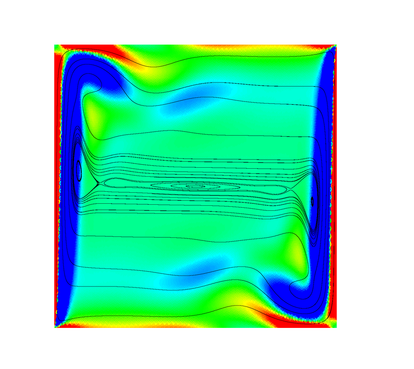}
	\end{center}
	\caption{DHC. Streamlines and vorticity contours obtained using the {\color{cr1} Eulerian-Lagrangian} approach for advection.}\label{fig:dhc_vorticity}
\end{figure}

Figures \ref{fig:dhc_1e3}-\ref{fig:dhc_1e7} depict the numerical solution of temperature and
velocity. Moreover, streamlines and vorticity contours are
shown in Figure \ref{fig:dhc_vorticity}. A good agreement is observed 
between the results computed with our two different advection schemes (Eulerian upwind scheme and {\color{cr1} Eulerian-Lagrangian} scheme)
and with the data reported in 
literature (see \cite{MD84,MNZ98,SX98,Man99,MUC00,WPW01,DT02,BB11}). 
Comparison with available data is also carried out considering the
vertical and horizontal velocities in the mid planes, which have been plotted
in Figures \ref{fig:vvel}-\ref{fig:hvel}. {\color{cr1} Furthermore, the normalized maximum velocities in the mid plane are reported in Tables \ref{tab:vvelocity}-\ref{tab:hvelocity}.}


%% Without normalizing
\begin{minipage}{\textwidth}
	\begin{multicols}{2}
		\begin{figure}[H]
			\begin{center}
				\includegraphics[width=\linewidth]{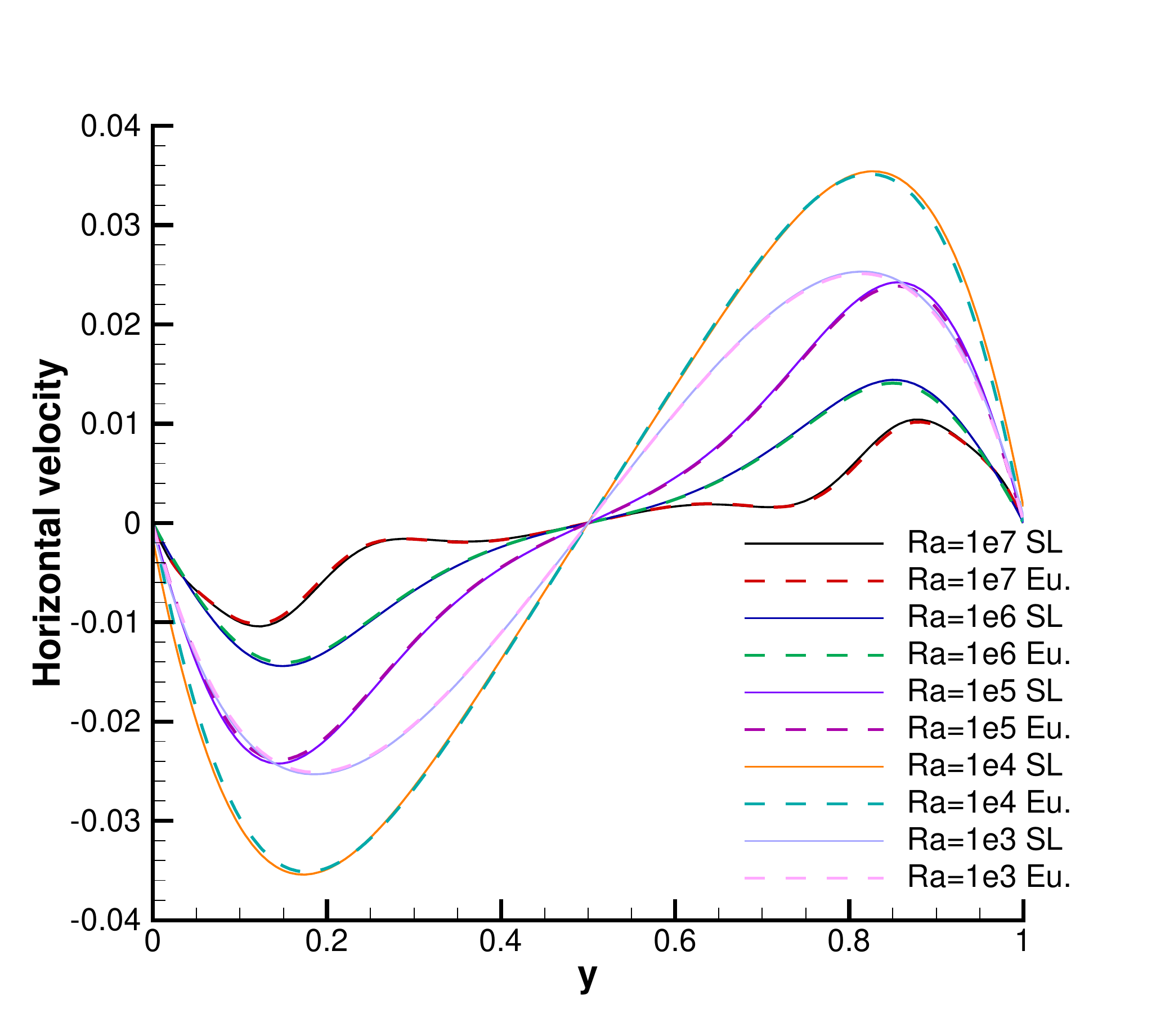}
			\end{center}
			\caption{DHC. Horizontal velocity, $x=0.5$.}\label{fig:hvel}
		\end{figure}
		\begin{figure}[H]
			\begin{center}
				\includegraphics[width=\linewidth]{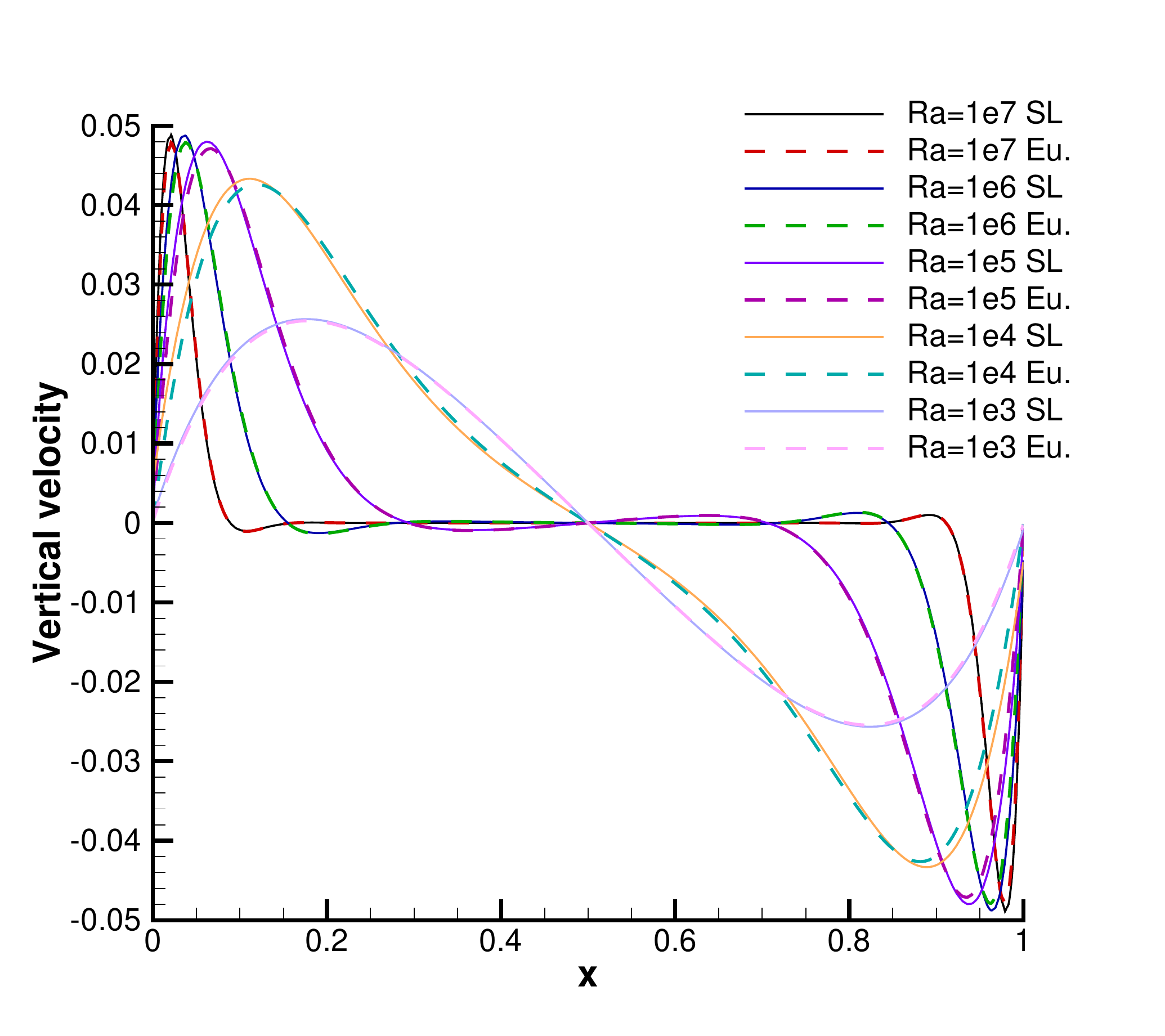}
			\end{center}
			\caption{DHC. Vertical velocity, $y=0.5$.} \label{fig:vvel}
		\end{figure}
	\end{multicols}
\end{minipage}

% % % % % % % % % % % % % % % % % % % % % % % % % % % % % %
% % % % % % % % % % % % % % % % % % % % % % % % % % % % % %
{\color{cr1}
\subsection{Cavity with differentially heated cylinders}\label{sec:DHCC}
We propose a modification to the previous test case on a more complex geometry where six cylinders of radius $r=0.1$ are subtracted from the square cavity, see Figure \ref{fig:dhcc_domain}. 
\begin{figure}[h]
	\centering
	\includegraphics[width=0.44\linewidth]{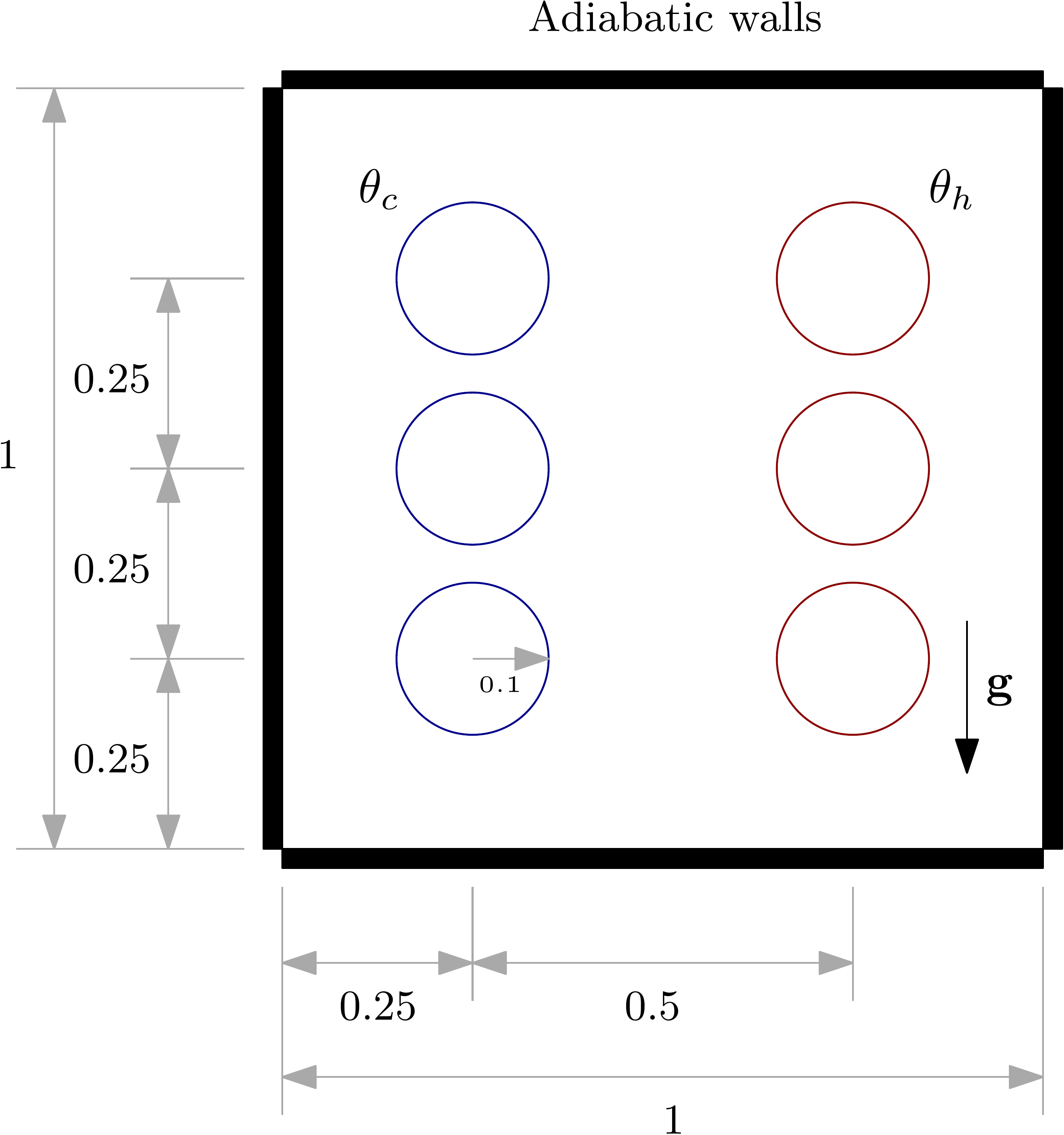}
	{\color{cr1} \caption{Cavity with differentially heated cylinders. Computational domain and boundary conditions.}}
	\label{fig:dhcc_domain}
\end{figure}
We assume an initial fluid at rest with constant temperature $\theta_{0}=293.15K$, density $\rho_{0}=1$ and $\mathbf{g}=\left(0,-9.81\right)^{T}$. 
Dirichlet boundary conditions are imposed on the cylinder walls, where the exact temperature is prescribed. Specifically, the left cylinders are cooled with $\theta_{c}=292.65K$, whereas the right ones are heated with $\theta_{h}=293.65K$. The remaining walls are assumed to be adiabatic. The computational domain is paved with a primal mesh of $18396$ elements and two different settings are considered, according to the diffusion coefficients: $\left(\mu_{T_{1}},\alpha_{T_{1}}\right)=\left(4.874\cdot 10^{-4},6.86\cdot 10^{-4}\right) $ and $\left(\mu_{T_{2}},\alpha_{T_{2}}\right)=\left(4.874\cdot 10^{-6},6.865\cdot 10^{-6}\right)$. 

Figure \ref{fig:DHCC_ra5} shows the contour plots of temperature, horizontal and vertical velocity obtained for the first setup using the fully Eulerian and the Eulerian-Lagrangian schemes with $p=2$, $\Theta = 0.51$. A good agreement is observed between the two approaches. To illustrate the mesh convergence, the results corresponding to a mesh refinement factor of 2 are included as well.
\begin{figure}
	\begin{center}
		\includegraphics[width=0.30\linewidth]{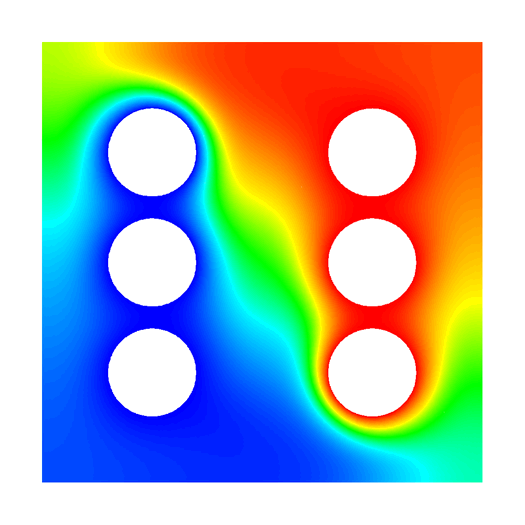} 
		\includegraphics[width=0.30\linewidth]{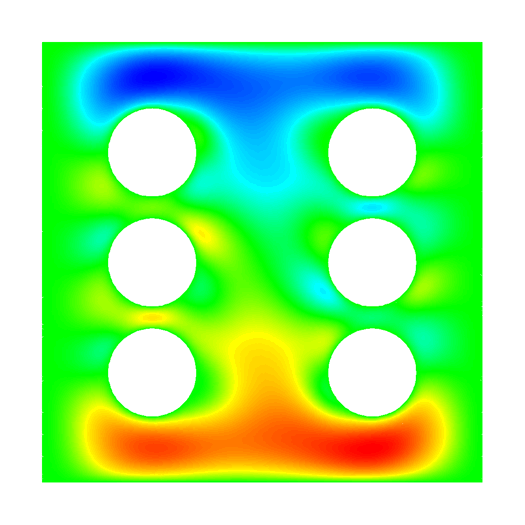}
		\includegraphics[width=0.30\linewidth]{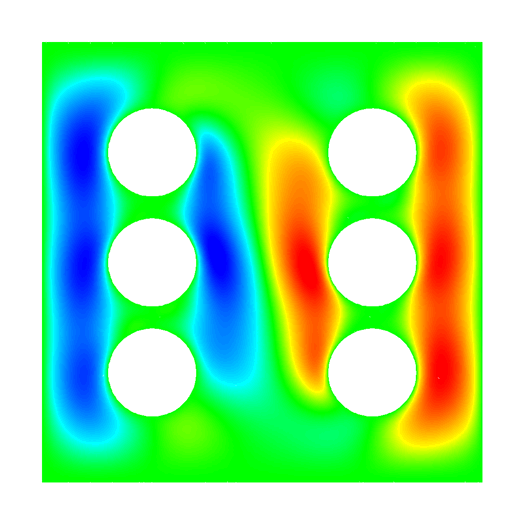}
		
		\includegraphics[width=0.30\linewidth]{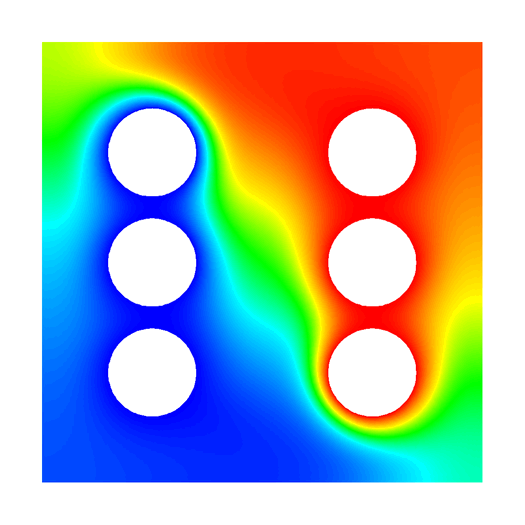} 
		\includegraphics[width=0.30\linewidth]{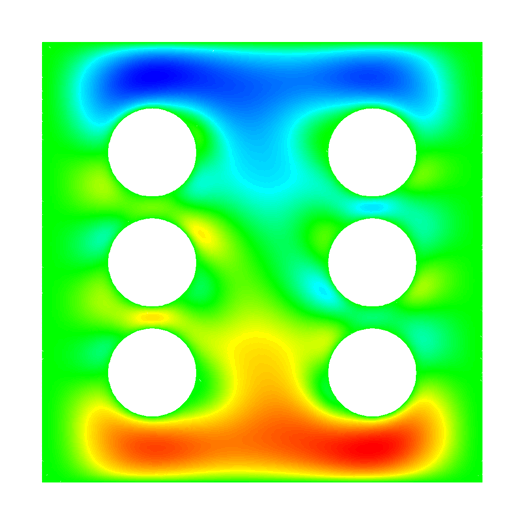}
		\includegraphics[width=0.30\linewidth]{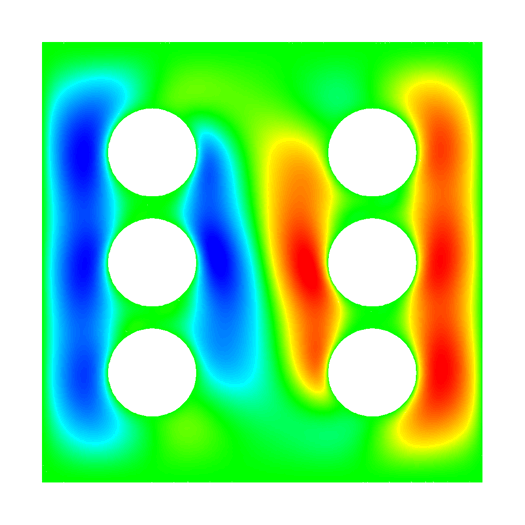}
		
		\includegraphics[width=0.30\linewidth]{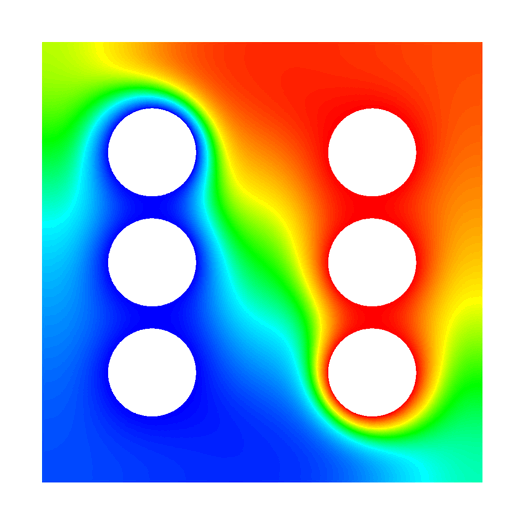} 
		\includegraphics[width=0.30\linewidth]{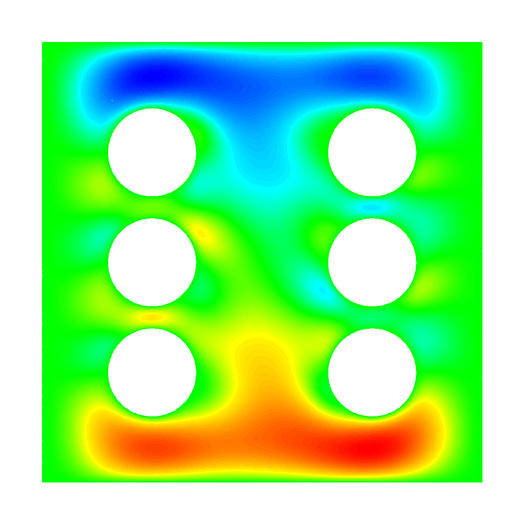}
		\includegraphics[width=0.30\linewidth]{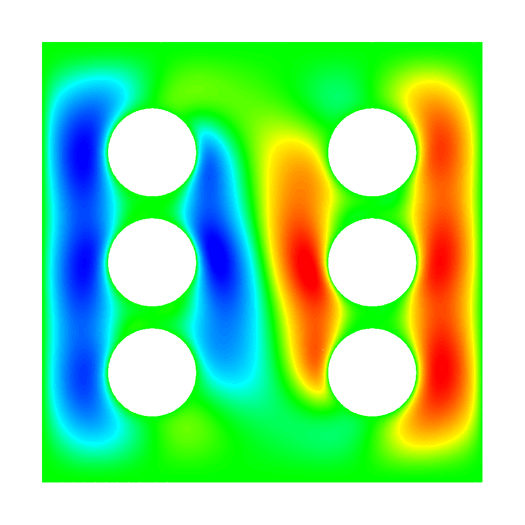}
		
		\includegraphics[width=0.25\linewidth]{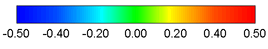}\hspace{0.05\textwidth}
		\includegraphics[width=0.25\linewidth]{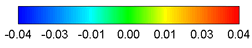}\hspace{0.05\textwidth}
		\includegraphics[width=0.25\linewidth]{./HE5_eu_legend_vel.png}

{\color{cr1}	\caption{Cavity with differentially heated cylinders. From left to right: temperature, horizontal and vertical velocity contours for $\left(\mu,\alpha\right)=\left(4.874\cdot 10^{-4},6.86\cdot 10^{-4}\right)$. Top: Eulerian advection scheme. Center: Eulerian-Lagrangian advection scheme. Bottom: Eulerian advection scheme with mesh refinement factor 2.}}\label{fig:DHCC_ra5}
	\end{center}
\end{figure}

The reduced viscosity coefficient defined in the second setup leads to an unsteady flow field, developing many secondary vortices and flow instabilities. The simulation was run relying on the Eulerian-Lagrangian scheme which reduces the computational cost. Nevertheless, to capture the sub-scale structures of the flow, we have fixed the time step to be $5/2$ times greater than the one computed from \eqref{eq:timestep_restriction}. In Figure \ref{fig:DHCC_ra9} we depict the results obtained at different output times.
\begin{figure}
	\begin{center}
		\includegraphics[width=0.3\linewidth]{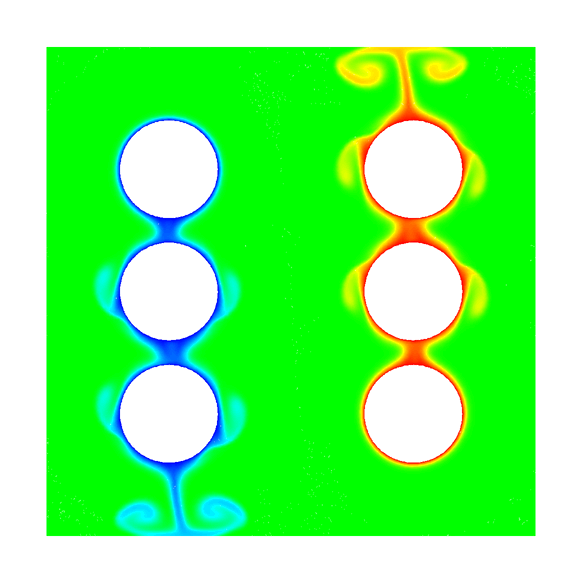} 
		\includegraphics[width=0.3\linewidth]{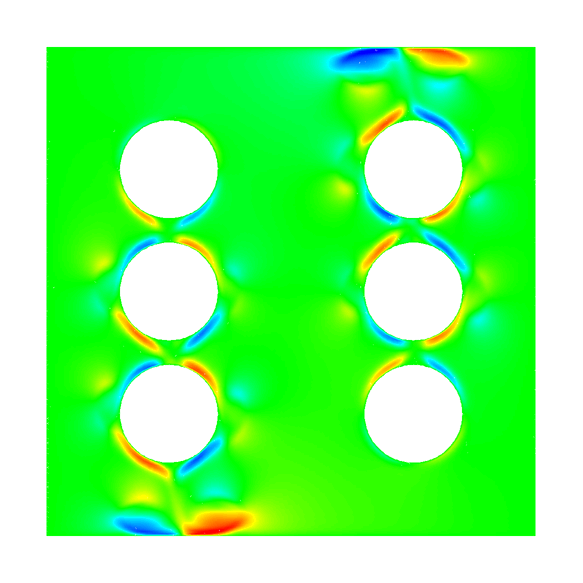}
		\includegraphics[width=0.3\linewidth]{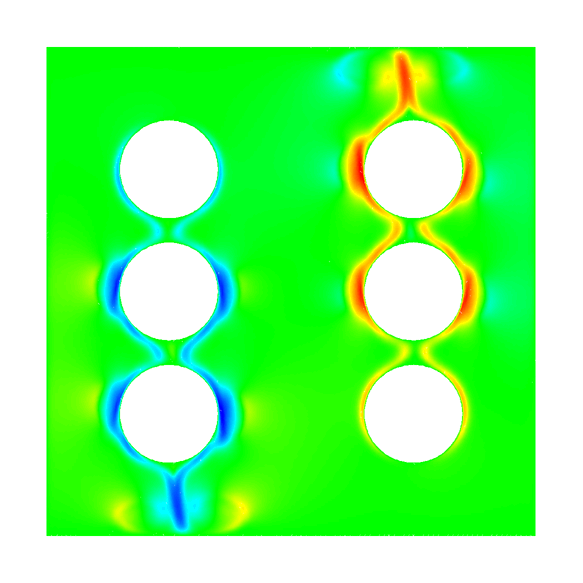}
		
		\includegraphics[width=0.3\linewidth]{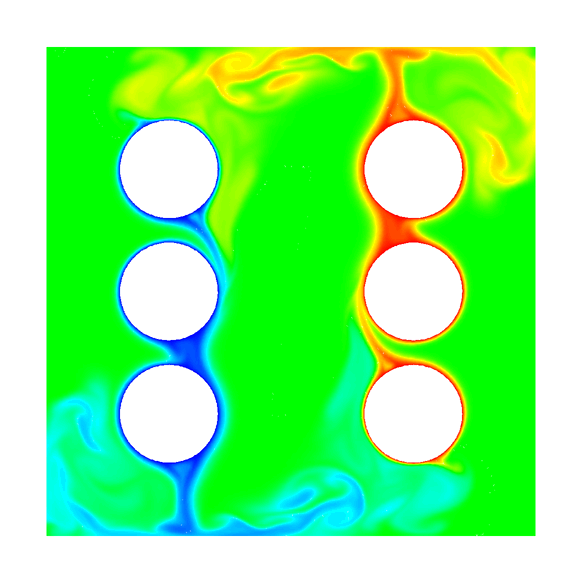}
		\includegraphics[width=0.3\linewidth]{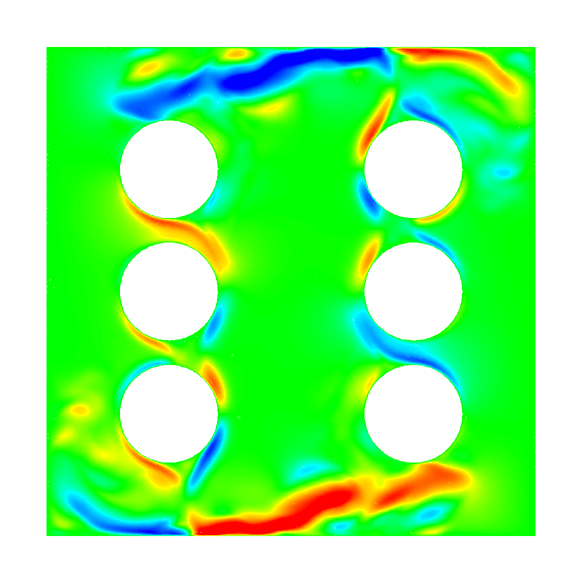}
		\includegraphics[width=0.3\linewidth]{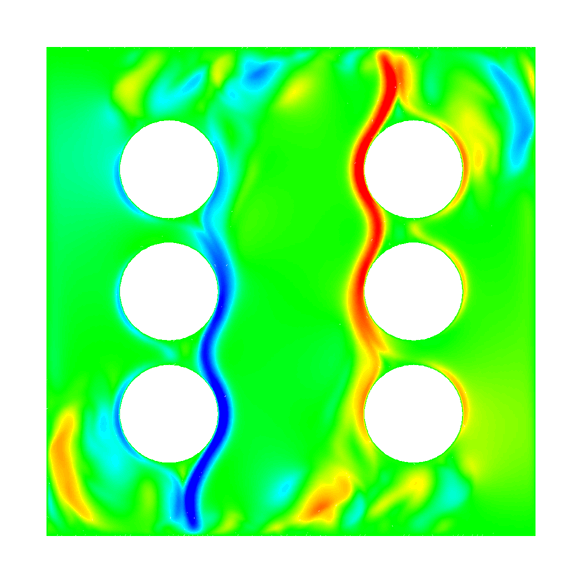} 
		
		\includegraphics[width=0.3\linewidth]{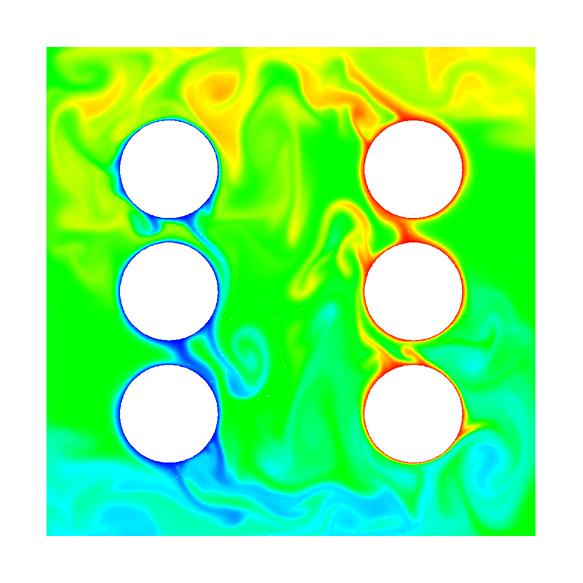}		
		\includegraphics[width=0.3\linewidth]{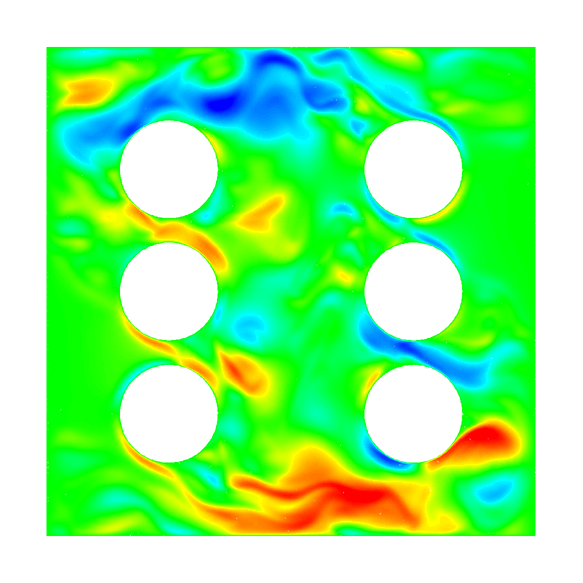}		
		\includegraphics[width=0.3\linewidth]{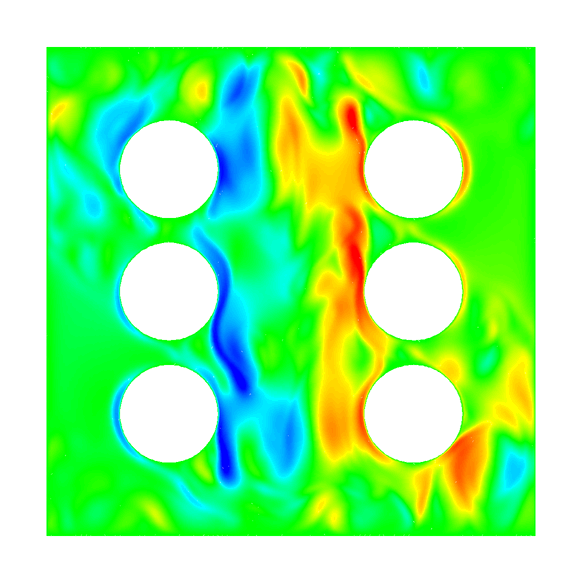}		
				
		\includegraphics[width=0.3\linewidth]{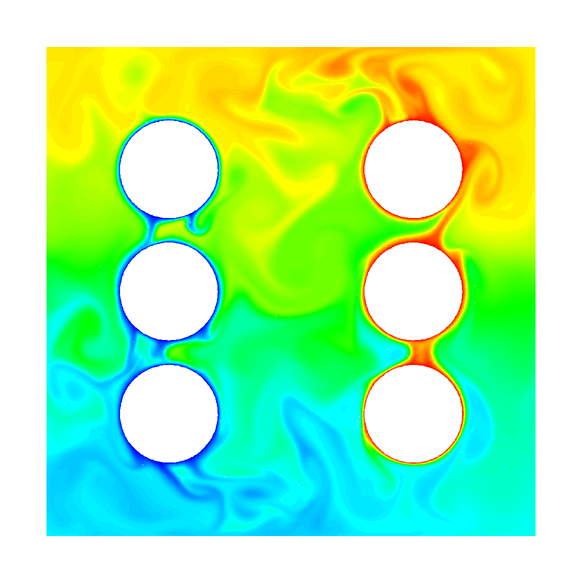}
        \includegraphics[width=0.3\linewidth]{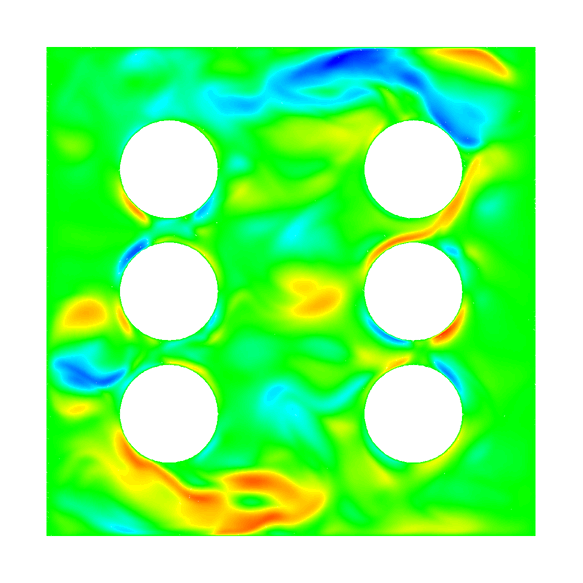}
		\includegraphics[width=0.3\linewidth]{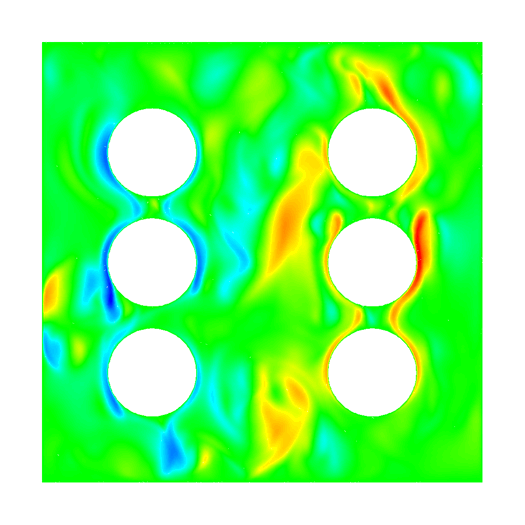}
		
		\includegraphics[width=0.25\linewidth]{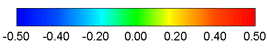}\hspace{0.05\textwidth}
		\includegraphics[width=0.25\linewidth]{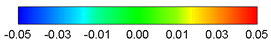}\hspace{0.05\textwidth}
		\includegraphics[width=0.25\linewidth]{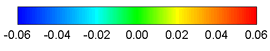}
		{\color{cr1}\caption{Cavity with differentially heated cylinders. From left to right: temperature, horizontal and vertical velocity contours for $\left(\mu,\alpha\right)=\left(4.874\cdot 10^{-4},6.86\cdot 10^{-4}\right)$. From top to bottom: output time $20s$, $50s$, $80s$, $200s$. }}\label{fig:DHCC_ra9}
	\end{center}
\end{figure}

}
% % % % % % % % % % % % % % % % % % % % % % % % % % % % % %
% % % % % % % % % % % % % % % % % % % % % % % % % % % % % %
\subsection{Warm bubble in two space dimensions}\label{sec:WBT2D}

As {\color{cr12} fourth} test case we propose to solve the smooth rising bubble benchmark problem 
introduced in \cite{GR08}, which has been widely used to test numerical solvers of thermal convection problems
(see \cite{MBGW13,YMLGW14,BLY17,Yi18}). % hay muchos mas con distintas variaciones del test
This problem assumes an initial fluid at rest and a warm bubble embedded in it.
During the simulation, the bubble will rise and deform gradually acquiring a mushroom-type shape and
later also developing secondary Kelvin-Helmholtz-type flow instabilities.
Although no exact solution is known, the analysis of the results can be qualitatively
performed taking into account the symmetry of the bubble as well as other numerical
results available in the literature.

The computational domain is a square of $1$km side length.
The bubble is initially placed at $\mathbf{x}_b=(0.5,0.35)$ km and is assigned the following 
initial temperature fluctuation: 
\begin{gather}
\theta(\mathbf{x},0)  - \theta_0 = \left\lbrace
\begin{array}{lr}
0, & \mathrm{if}\; r >r_{b},\\
\dfrac{\theta_{b}}{2} \left[1+ \cos\left( \dfrac{\pi r}{r_{b}} \right) \right],  & \mathrm{if}\; r \leq r_{b},
\end{array}
\right.\label{eq:WBT:_temperaturefun}
\end{gather}
with $r = \left\| \mathbf{x} - \mathbf{x}_b \right\|$ representing the distance from the center of the bubble, $r_{b}=0.25$km being its radius and
$\theta_{b}=0.5$K denoting the maximal initial amplitude of the perturbation. The temperature perturbation \eqref{eq:WBT:_temperaturefun} is imposed over a background temperature which is assumed to be $\theta_{0}=303.15$K. 

\begin{figure}
	\vspace{-1.5cm}	
	\begin{center}
		\includegraphics[width=0.38\linewidth]{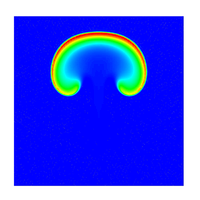} 
		\includegraphics[width=0.38\linewidth]{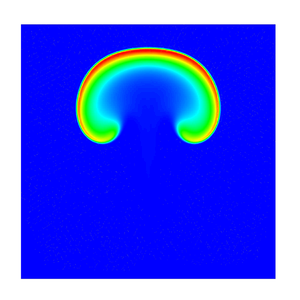}
		
		\vspace{-0.8cm}
		\includegraphics[width=0.38\linewidth]{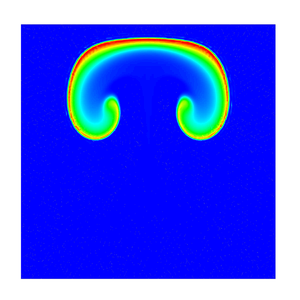} 
		\includegraphics[width=0.38\linewidth]{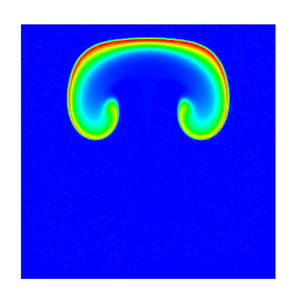}
		
		\vspace{-0.8cm}	
		\includegraphics[width=0.38\linewidth]{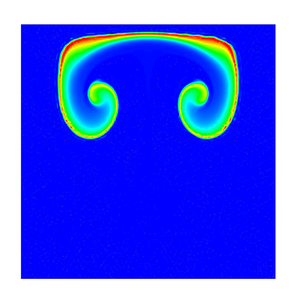}
		\includegraphics[width=0.38\linewidth]{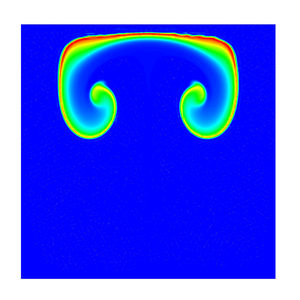}
		
		\vspace{-0.8cm}
		\includegraphics[width=0.38\linewidth]{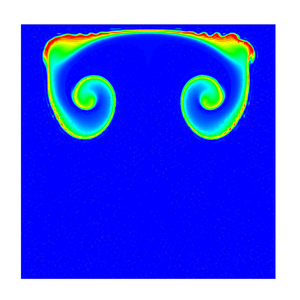}
		\includegraphics[width=0.38\linewidth]{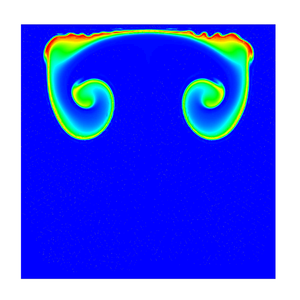} 
	\end{center}
	\vspace{-1cm}	
	\caption{Warm bubble test in 2D: temperature contours. $\mu=10^{-6}$, {\color{cr12} $p=4$, $\Theta = 0.51$}. Left: Eulerian advection scheme. Right: {\color{cr1} Eulerian-Lagrangian} advection scheme. From top to bottom: output time $600$s, $700$s, $800$s, $900$s.}\label{fig:WBT_sl_mu1e6}
\end{figure}

The use of the Eulerian scheme, without any limiter, to solve Euler equations 
may produce high velocity gradients in the boundary of the warm bubble.
To avoid the growth of instabilities in the numerical solution, we may add a small artificial viscosity.
However, as already discussed in \cite{MBGW13,YMLGW14,TDg16},
this leads to a smoothness of the boundary of the bubble, that is affected by numerical dissipation.
Figure \ref{fig:WBT_sl_mu1e6} shows the results obtained at time instants
$t\in\left\lbrace 600,700,800,900\right\rbrace$s using a constant artificial viscosity with $\mu=10^{-6}$ and a very coarse primal mesh composed of only $5172$ triangles. We run the test problem using piecewise  polynomials of degree $p=4$ in order to represent the discrete solution. 
We observe that the evolution of the main shape of the bubble agrees with the 
results available in literature. Moreover, for large times,
long-wave oscillations arise on top of the main structure.

An alternative strategy to get a stable solution with very little numerical dissipation is the use of a {\color{cr1} Eulerian-Lagrangian} advection scheme. In this case we do not need to add any artificial viscosity, even for high order DG discretization. Moreover, the use of this methodology yields to a reduction on the computational cost of the algorithm. For the test case shown in this section, the implementation based on the {\color{cr1} Eulerian-Lagrangian} advection scheme was overall {\color{cr1} 4.3 times faster} than the implementation based on the classical Eulerian advection scheme. 
This improvement is related to the increase of the time step size allowed by the {\color{cr1} Eulerian-Lagrangian} approach.  

\begin{figure}
	\vspace{-1.5cm}	
	\begin{center}
		\includegraphics[width=0.38\linewidth]{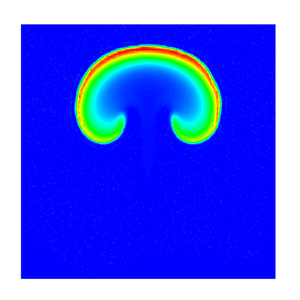}
		\includegraphics[width=0.38\linewidth]{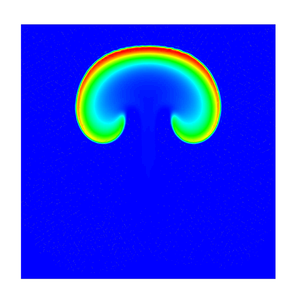}
		
		\vspace{-0.8cm}
		\includegraphics[width=0.38\linewidth]{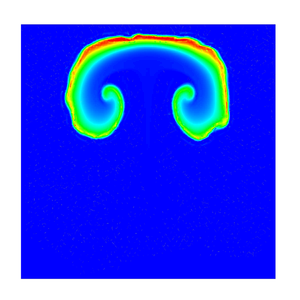} 
		\includegraphics[width=0.38\linewidth]{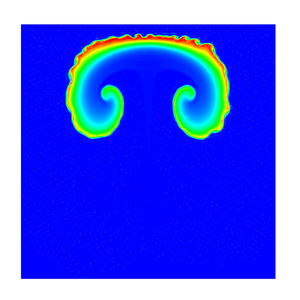}
		
		\vspace{-0.8cm}	
		\includegraphics[width=0.38\linewidth]{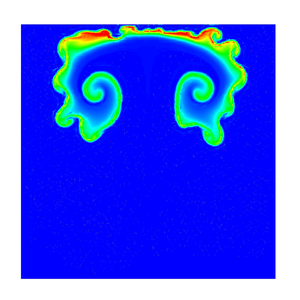}
		\includegraphics[width=0.38\linewidth]{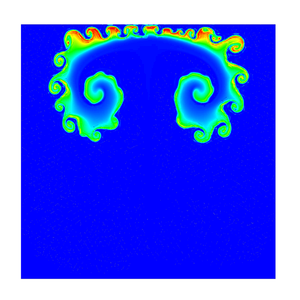}
		
		\vspace{-0.8cm}
		\includegraphics[width=0.38\linewidth]{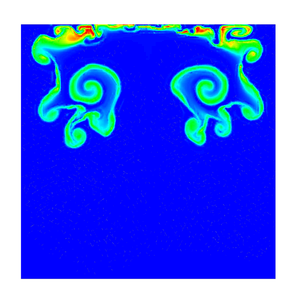} 
		\includegraphics[width=0.38\linewidth]{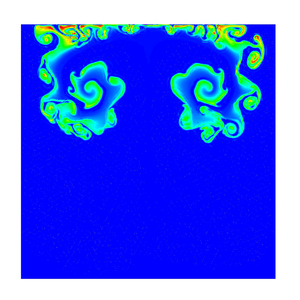} 
	\end{center}
	\vspace{-1cm}	
	\caption{Warm bubble test in 2D: temperature contours. {\color{cr1} Eulerian-Lagrangian} advection scheme, $\mu=0$. Left:  {\color{cr12} $p=2$, $\Theta = 0.51$}. Right: {\color{cr12} $p=4$, $\Theta = 0.51$}. From top to bottom: output time $600$s, $700$s, $800$s, $900$s.} \label{fig:WBT_sl_mu0}
\end{figure}

The results obtained using the {\color{cr1} Eulerian-Lagrangian} approach for $\mu=0$ are depicted in Figure \ref{fig:WBT_sl_mu0}  for the intermediate
time instants $t\in \left\lbrace 600, 700, 800, 900 \right\rbrace$s.
We observe that the solution at the beginning of the simulation is almost symmetric,
where the small discrepancies which arise are due to the unstructured and non-symmetric mesh that is used.
After $800$s mushroom-shaped structures start to grow from the cap of the main shape. 
Moreover, for long times small vortex structures develop on the top of the
long waves resulting in secondary Kelvin-Helmholtz instabilities.
At this step the flow becomes completely turbulent and therefore the initial symmetry is lost.
The comparison against the results depicted in Figure \ref{fig:WBT_sl_mu1e6}, where a non-zero artificial viscosity was employed, show that the viscosity required by the Eulerian scheme
to provide a stable solution is already too high to properly capture the small vortexes generated within the fluid flow. 
Therefore, the use of a {\color{cr1} Eulerian-Lagrangian} scheme with zero viscosity is a key point
in order to avoid the damping of Kelvin-Helmholtz instabilities.
Furthermore, Figure \ref{fig:WBT_sl_mu0}, which shows  the results obtained considering
spatial polynomial degrees $2$ and $4$, highlights the importance of employing
a high order scheme in order to capture the small secondary vortex structures. 

% % % % % % % % % % % % % % % % % % % % % % % % % % % % % %
\subsection{Warm bubble in three space dimensions}\label{sec:WBT3D}

We now extend the warm bubble test to a three-dimensional domain, following the setup described in \cite{TDg16,BLY17}. To this end, we define the computational domain  $\Omega=\left[ 0,1 \right]\times \left[ 0,1 \right]\times \left[ 0,1.5 \right]$ km. The warm bubble is initially located at $\mathbf{x}_{b}=(0.5,0.5,0.35)$ and the temperature is defined by \eqref{eq:WBT:_temperaturefun} with $r = \left\| \mathbf{x} - \mathbf{x}_b \right\|$. Furthermore, we consider $r_{b}=0.25$km, $\theta_{0}=303.15$K and zero viscosity, i.e. $\nu=0$.

\begin{figure}
	\vspace{-1.5cm}	
	\begin{center}
		\begin{tabular}{cc}
		\includegraphics[width=0.3\linewidth]{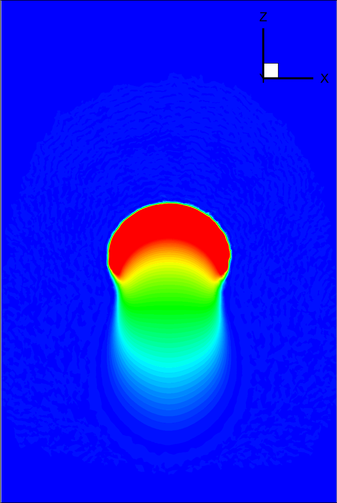} &
		\includegraphics[width=0.3\linewidth]{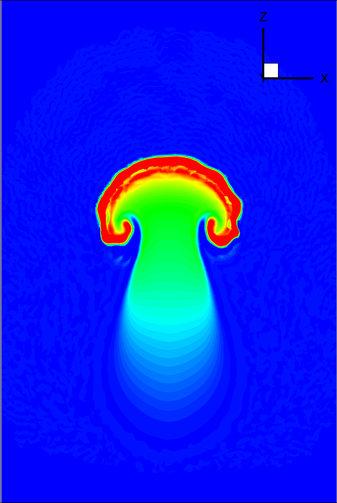} \\
		\includegraphics[width=0.3\linewidth]{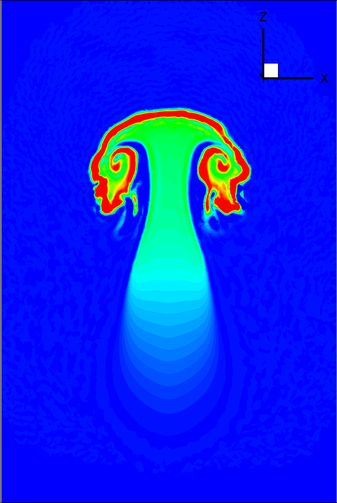} &
		\includegraphics[width=0.3\linewidth]{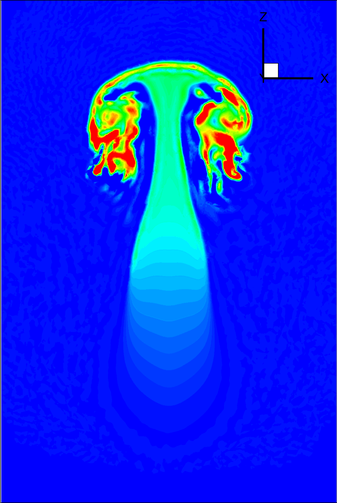} \\
		\includegraphics[width=0.3\linewidth]{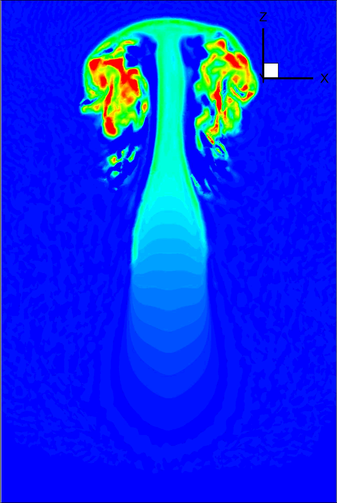} & 
		\includegraphics[width=0.3\linewidth]{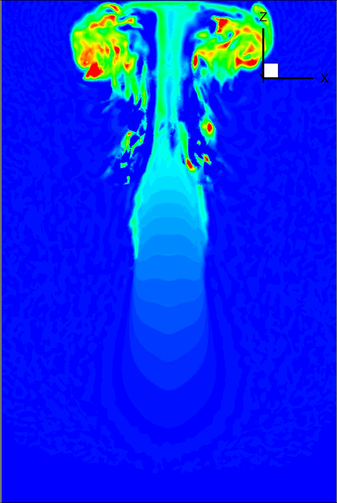}  
		\end{tabular} 
	\end{center}	
	\caption{Warm bubble test in 3D: temperature contours in the plane $x=0.5$ at times $t\in \left\lbrace 400,500,600,700,800,900 \right\rbrace$s.}
	\label{fig:WBT3D_temp}
\end{figure}

\begin{figure}
	\vspace{-1.5cm}	
	\begin{center}
		\includegraphics[width=0.34\linewidth]{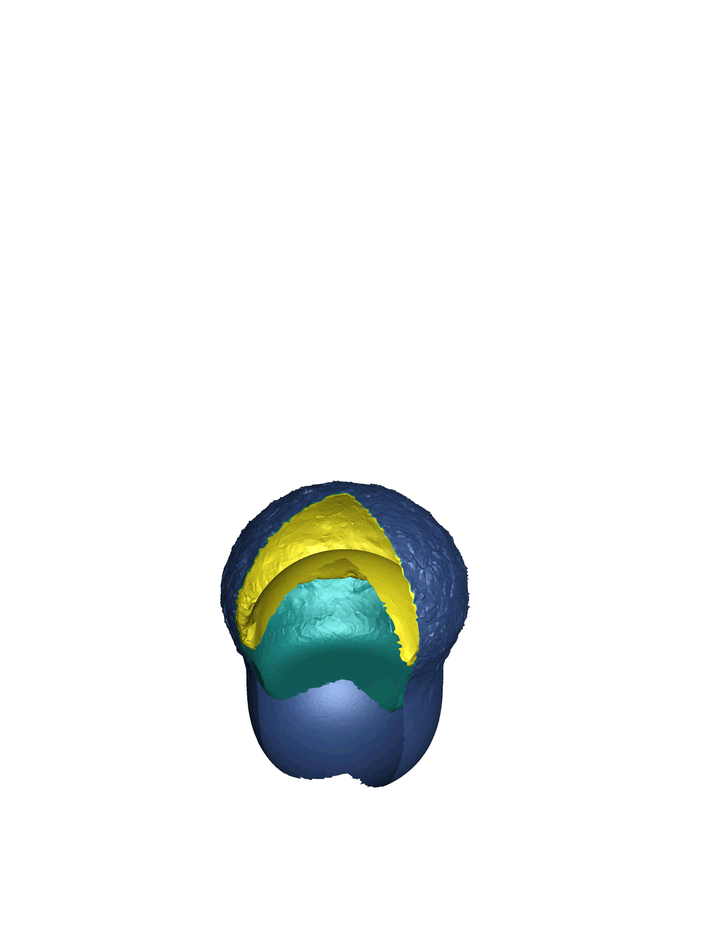} 
		\includegraphics[width=0.34\linewidth]{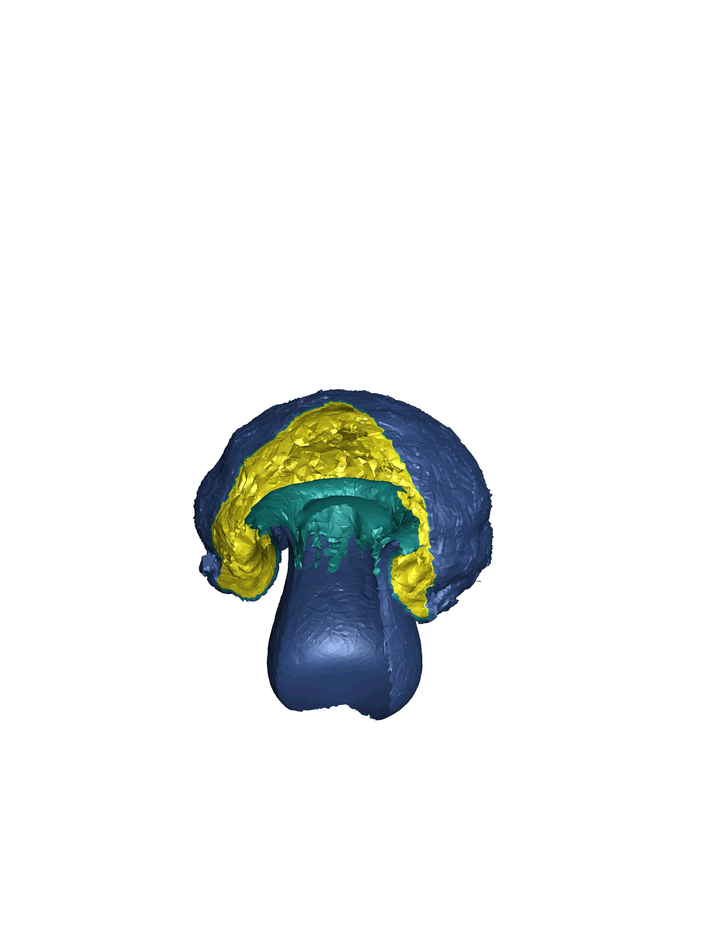} 
		\includegraphics[width=0.34\linewidth]{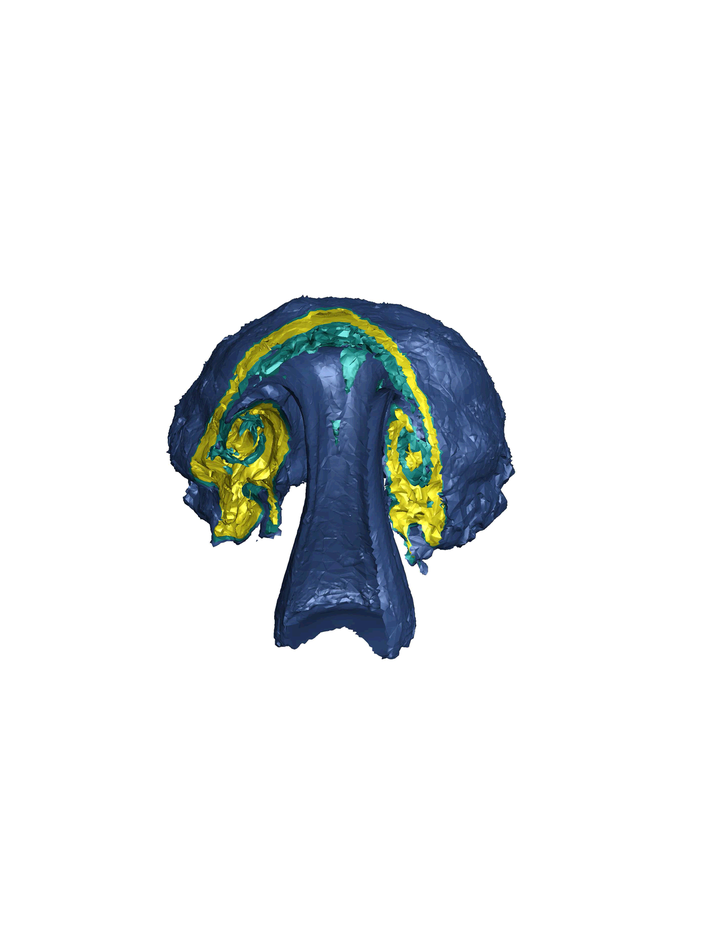} 
		\includegraphics[width=0.34\linewidth]{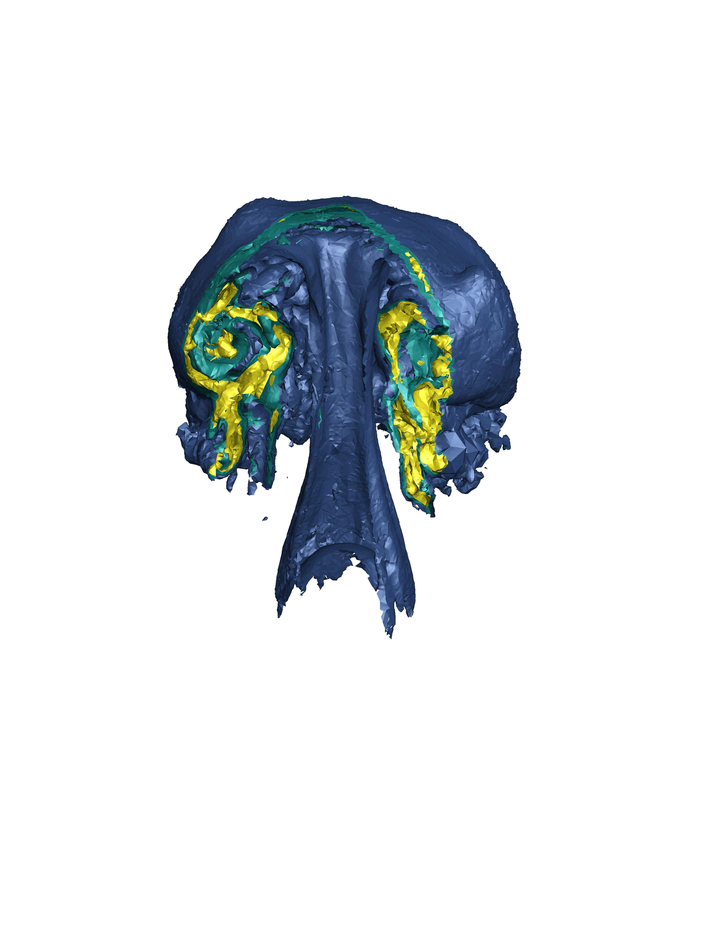}
		\includegraphics[width=0.34\linewidth]{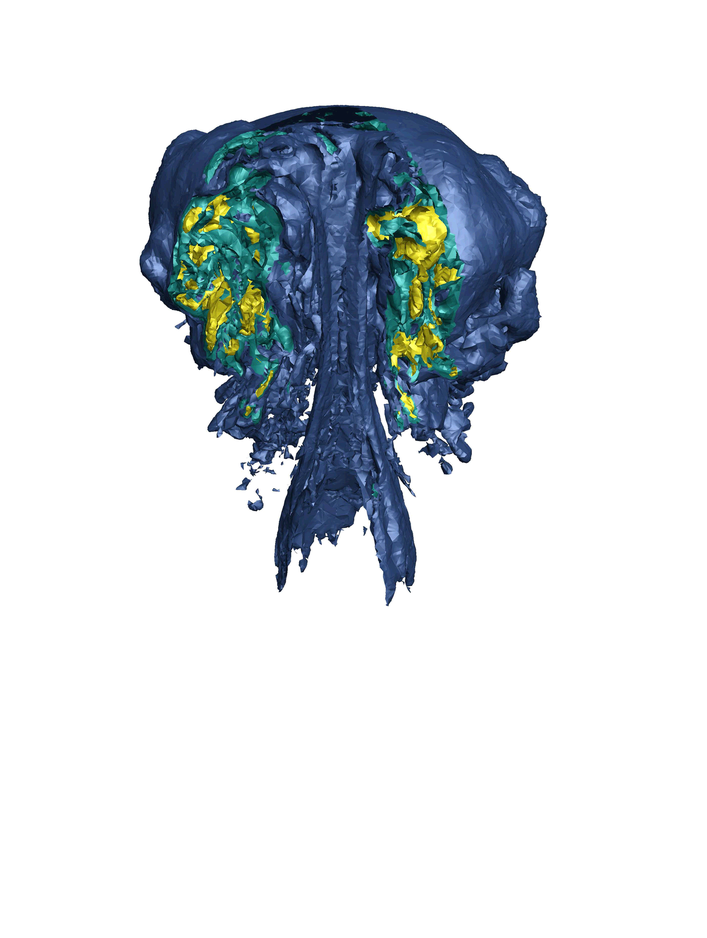}  
		\includegraphics[width=0.34\linewidth]{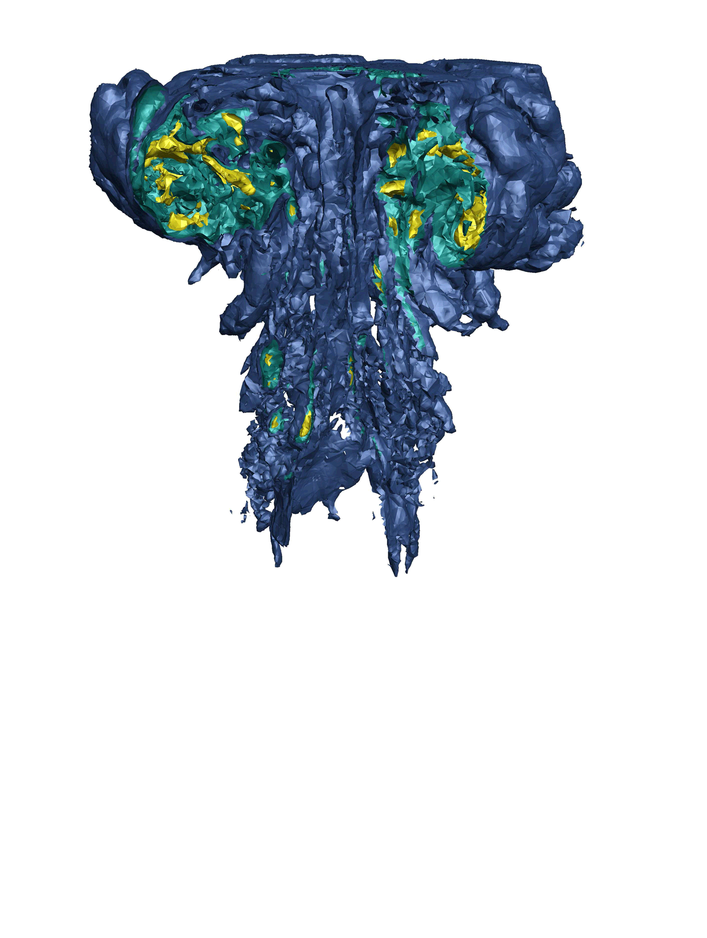}  
	\end{center}
	\caption{Warm bubble test in 3D: isosurfaces of $\theta$ at times $t\in \left\lbrace 400,500,600,700,800,900 \right\rbrace$s. }
	\label{fig:WBT3D_isot}
\end{figure}

The computational mesh counts a total number of $1,331,442$ primal tetrahedral elements and we use piecewise polynomials of degree $p=3$ for the representation of the discrete solution. This amounts to a total of $26,628,840$ degrees of freedom for the pressure. The computer code was parallelized using the MPI standard and the simulation was run on 960 CPU cores of the SuperMUC-NG supercomputer at the LRZ in Garching, Germany. Figures \ref{fig:WBT3D_temp}-\ref{fig:WBT3D_isot} depict the numerical results obtained at different time instants.
The three-dimensional results mimic the behavior of the warm bubble already observed in the two-dimensional setting. The differences in temperature induce a velocity field in the flow that will produce an upward movement of the bubble, which yields the development of the main mushroom-shaped structure for the temperature. 
However, we observe that the Kelvin-Helmholtz instabilities arising in the bi-dimensional test case do not appear in 3D. 
This may be due to the coarse grid resolution used here. As already noticed in Figure \ref{fig:WBT_sl_mu0}, an increase in the spatial order of accuracy of the scheme could {\color{cr1} improve} the final results.

% % % % % % % % % % % % % % % % % % % % % % % % % % % % % %
% % % % % % % % % % % % % % % % % % % % % % % % % % % % % %
\subsection{Comparison of incompressible and compressible flow solvers for a Gaussian bubble in 2D}\label{sec:GB2D}

\begin{figure}
	\begin{center}
		\includegraphics[width=0.07\linewidth]{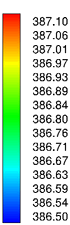}\hfill
		\includegraphics[width=0.29\linewidth]{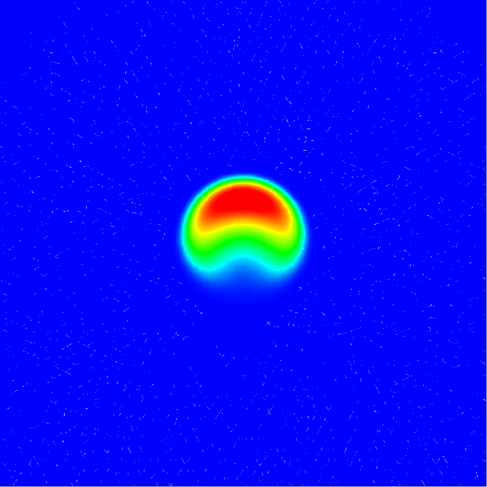}\hfill
		\includegraphics[width=0.29\linewidth]{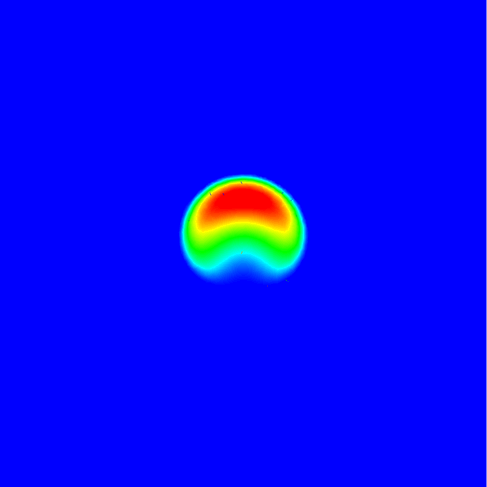}\hfill
		\includegraphics[trim= 0 10 10 0,clip,width=0.29\linewidth]{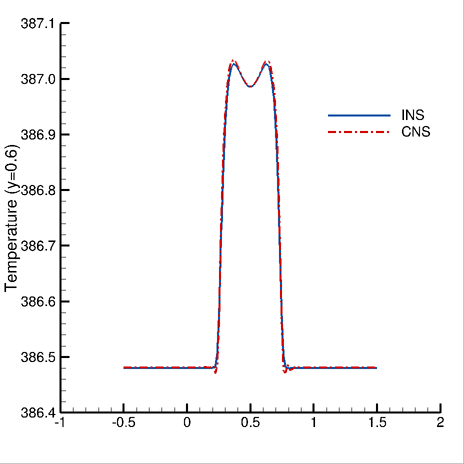}
		
		\vspace{0.2cm}	
		\includegraphics[width=0.06\linewidth]{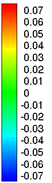}\hspace{0.01\linewidth}\hfill	
		\includegraphics[width=0.29\linewidth]{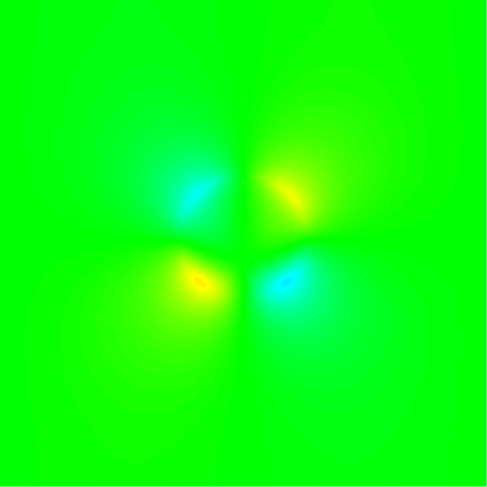}\hfill %-0.0672 0.069
		\includegraphics[width=0.29\linewidth]{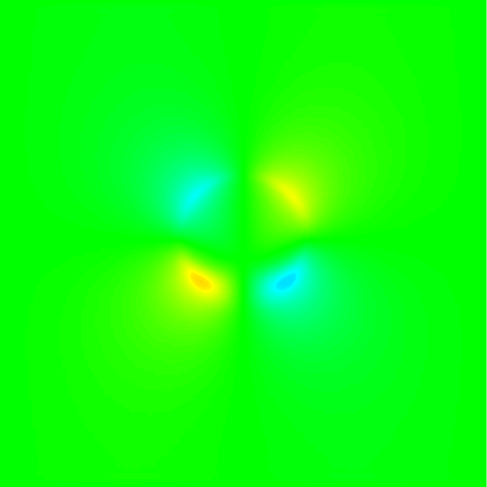}\hfill
		\includegraphics[trim= 0 10 10 0,clip,width=0.29\linewidth]{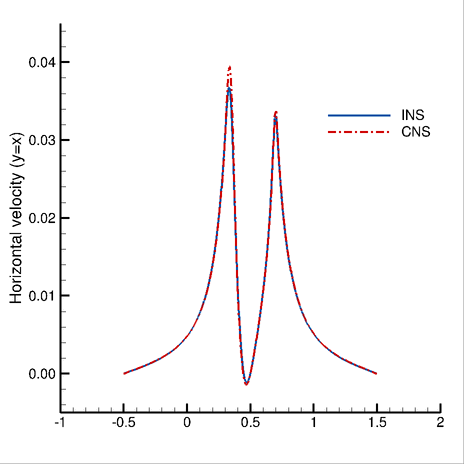} 
		
		\vspace{0.2cm}
		\includegraphics[width=0.06\linewidth]{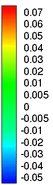}\hspace{0.01\linewidth}\hfill
		\includegraphics[width=0.29\linewidth]{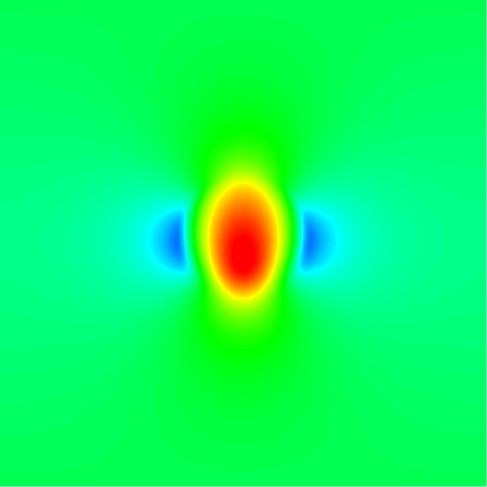}\hfill %-0.05 0.09
		\includegraphics[width=0.29\linewidth]{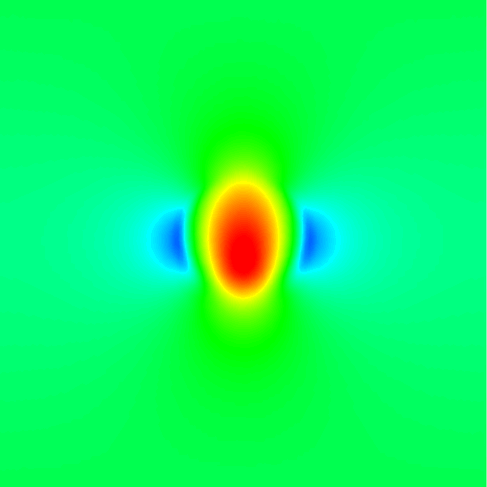}\hfill
		\includegraphics[trim= 0 10 10 0,clip,width=0.29\linewidth]{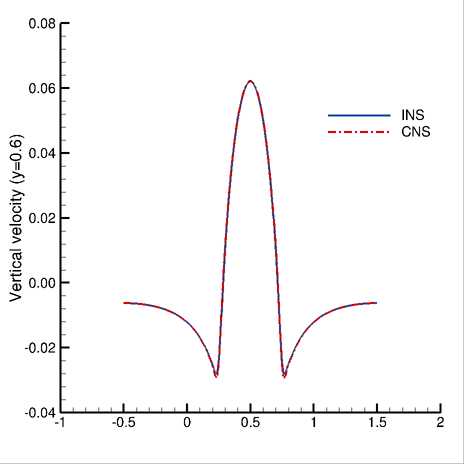}
	\end{center}
	\caption{Gaussian bubble test. From top to bottom: temperature, horizontal and vertical velocity, $t=10s$. Left: incompressible flow solver. Center: compressible flow solver. Right: profile obtained over a specific line.}\label{fig:GBTCNS_t10}
\end{figure}

\begin{figure}
	\begin{center}
		\includegraphics[width=0.07\linewidth]{./GBTINS_temperature_t15_legend.png}\hfill
		\includegraphics[width=0.29\linewidth]{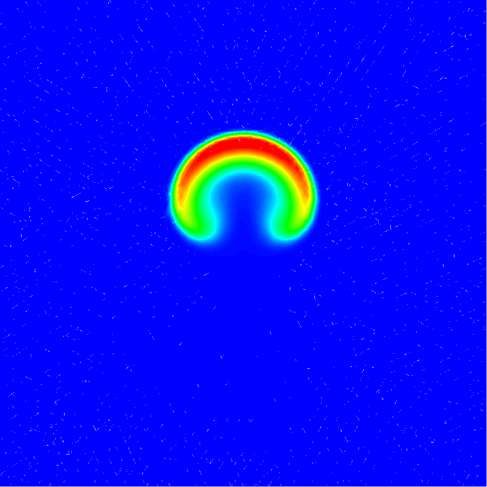}\hfill
		\includegraphics[width=0.29\linewidth]{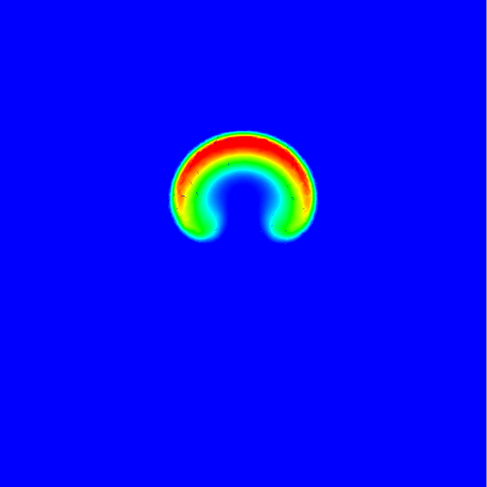}\hfill
		\includegraphics[trim= 0 10 10 0,clip,width=0.29\linewidth]{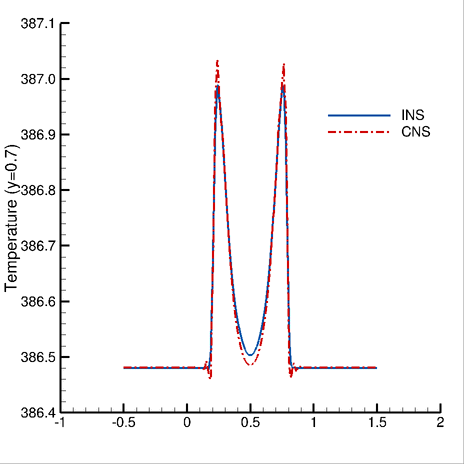}
		
		\vspace{0.2cm}		
		\includegraphics[width=0.06\linewidth]{./GBTINS_velx_t15_legend.png}\hspace{0.01\linewidth}\hfill
		\includegraphics[width=0.29\linewidth]{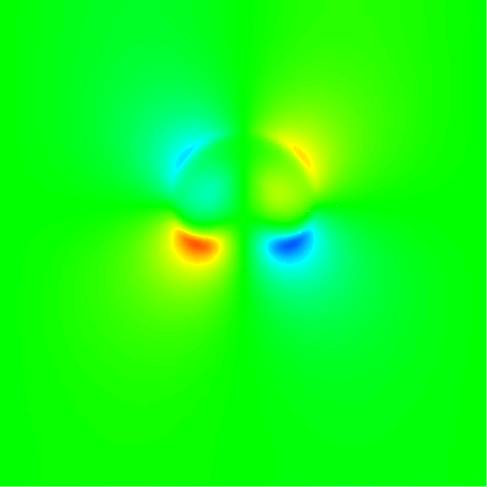}\hfill %-0.0672 0.069
		\includegraphics[width=0.29\linewidth]{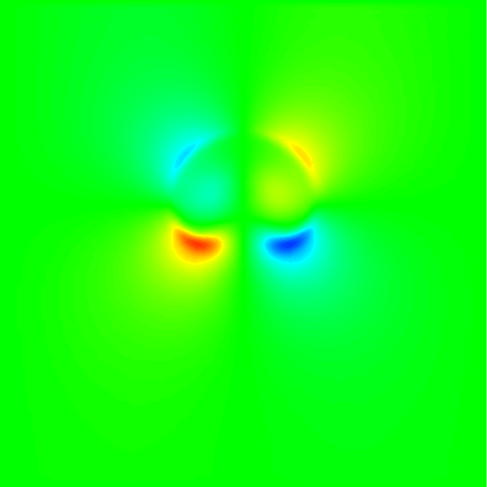}\hfill 
		\includegraphics[trim= 0 10 10 0,clip,width=0.29\linewidth]{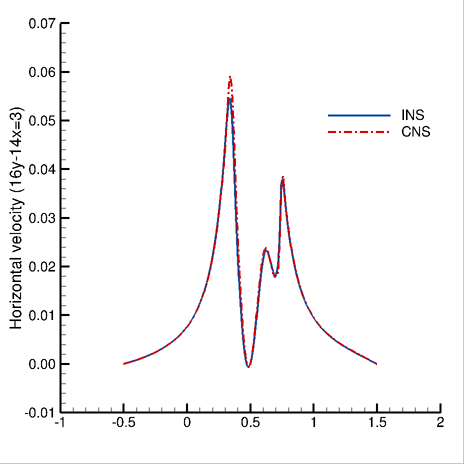}
		
		\vspace{0.2cm}
		\includegraphics[width=0.06\linewidth]{./GBTINS_vely_t15_legend.png}\hspace{0.01\linewidth}\hfill
		\includegraphics[width=0.29\linewidth]{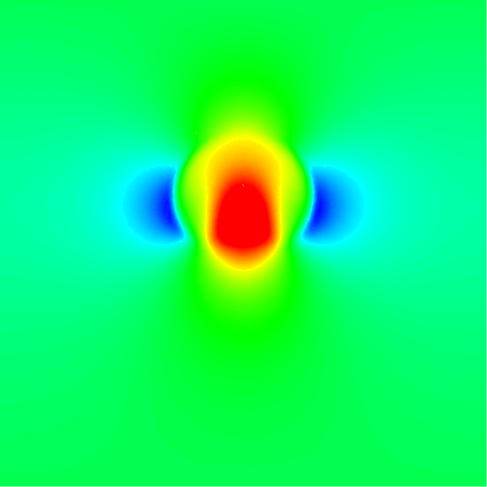}\hfill %-0.05 0.09
		\includegraphics[width=0.29\linewidth]{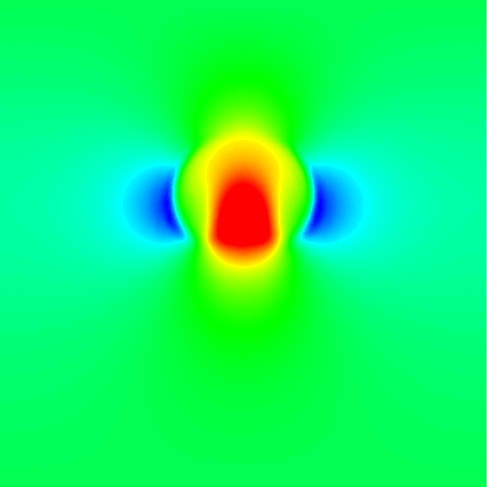}\hfill
		\includegraphics[trim= 0 10 10 0,clip,width=0.29\linewidth]{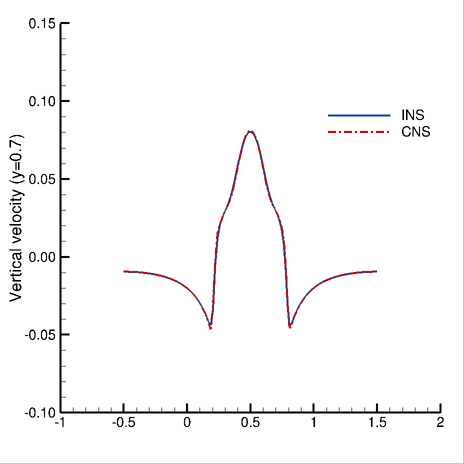}		
	\end{center}
	\caption{Gaussian bubble test. From top to bottom: temperature, horizontal and vertical velocity, $t=15s$. Left: incompressible flow solver. Center: compressible flow solver. Right: profile obtained over a specific line.}\label{fig:GBTCNS_t15}
\end{figure}

\begin{figure}
	\begin{center}
		\includegraphics[width=0.07\linewidth]{./GBTINS_temperature_t15_legend.png}\hfill
		\includegraphics[width=0.29\linewidth]{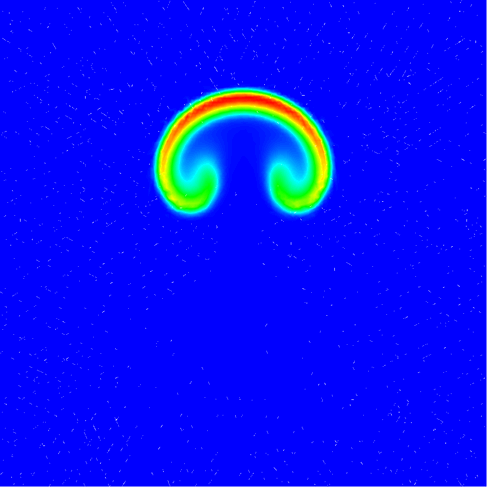}\hfill
		\includegraphics[width=0.29\linewidth]{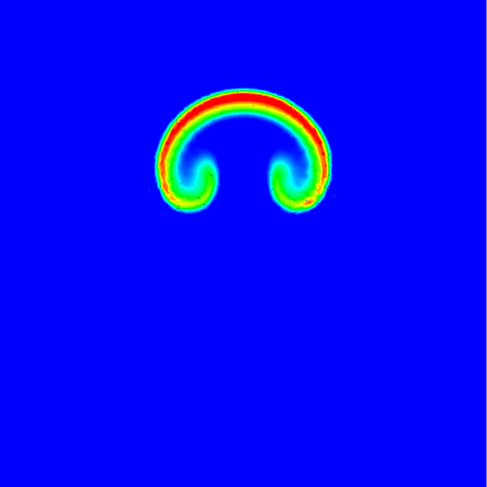}\hfill
		\includegraphics[trim= 0 10 10 0,clip,width=0.29\linewidth]{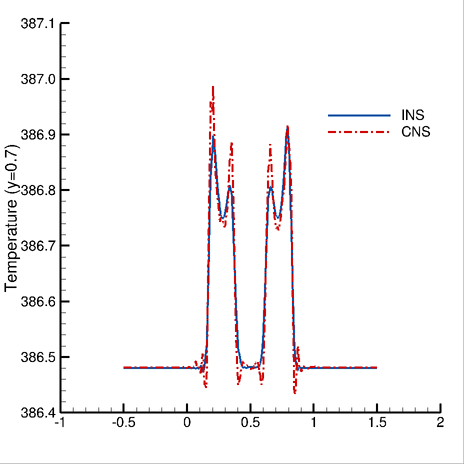}
		
		\vspace{0.2cm}		
		\includegraphics[width=0.06\linewidth]{./GBTINS_velx_t15_legend.png}\hspace{0.01\linewidth}\hfill
		\includegraphics[width=0.29\linewidth]{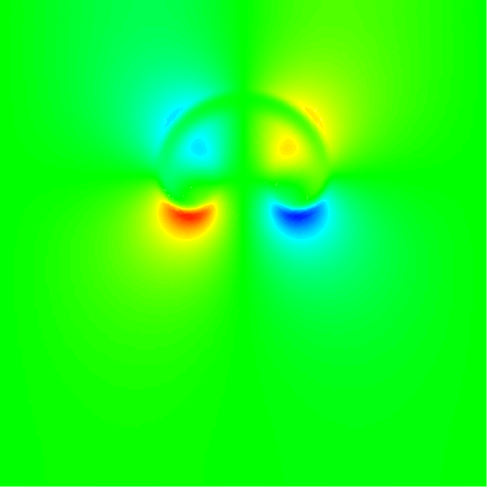}\hfill %-0.0672 0.069
		\includegraphics[width=0.29\linewidth]{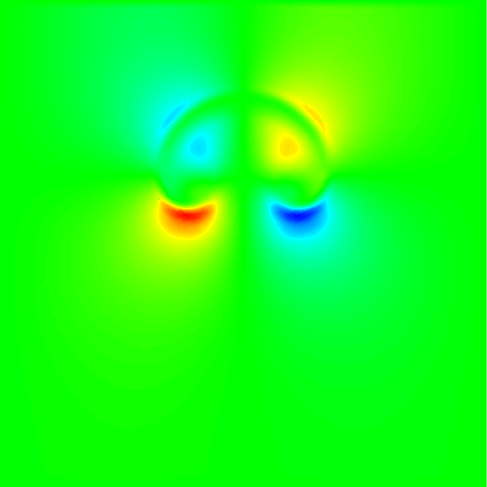}\hfill
		\includegraphics[trim= 0 10 10 0,clip,width=0.29\linewidth]{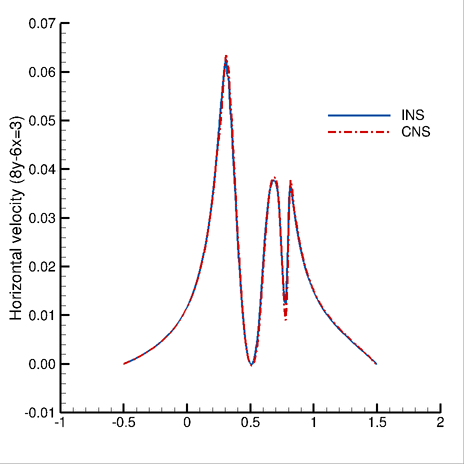}
		
		\vspace{0.2cm}
		\includegraphics[width=0.06\linewidth]{./GBTINS_vely_t15_legend.png}\hspace{0.01\linewidth}\hfill
		\includegraphics[width=0.29\linewidth]{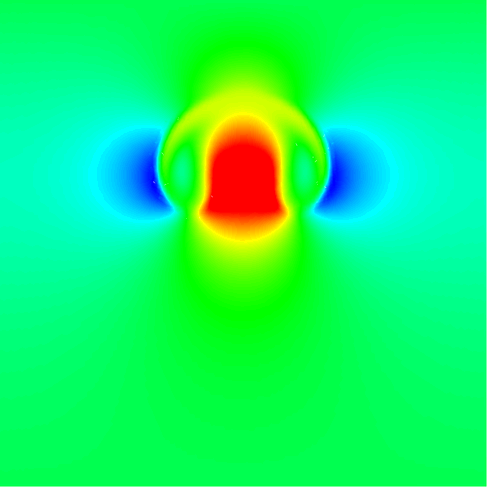}\hfill %-0.05 0.09
		\includegraphics[width=0.29\linewidth]{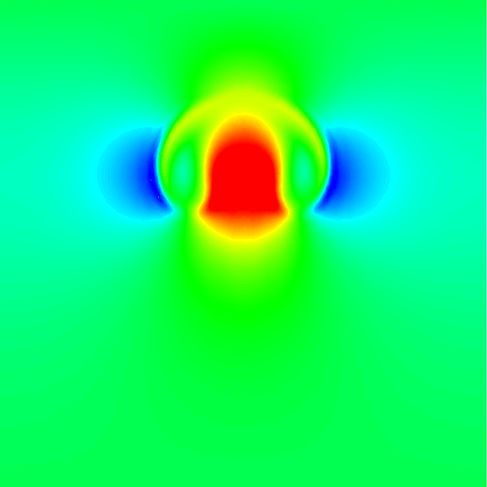}\hfill
		\includegraphics[trim= 0 10 10 0,clip,width=0.29\linewidth]{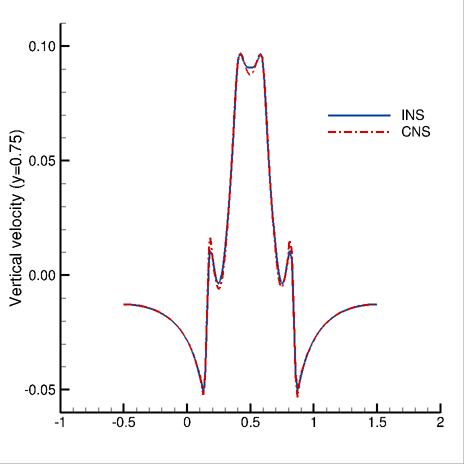}
	\end{center}
	\caption{Gaussian bubble test. From top to bottom: temperature, horizontal and vertical velocity, $t=20s$. Left: incompressible flow solver. Center: compressible flow solver. Right: profile obtained over a specific line.}\label{fig:GBTCNS_t20}
\end{figure}

In this section we compare the results obtained with both the incompressible and the compressible flow solver by considering a rising bubble test defined in the computational domain $\Omega= \left[-0.5,1.5\right]\times \left[-0.5,1.5\right]$ with periodic boundary conditions on the lateral boundaries. 
The initial temperature profile is the one of a Gaussian bubble and reads 
\begin{gather}
\theta(\mathbf{x},0) = \left\lbrace \begin{array}{lr}
386.48 = \theta_0, & \mathrm{if}\; r >r_{b},\\
\dfrac{p_0}{R\left(1.0 - 0.1 e^{-\frac{r^2}{\sigma^2}} \right) },  & \mathrm{if}\; r \leq r_{b},
\end{array}\right.
\end{gather}
where $r = \left\| \mathbf{x} - \mathbf{x}_b \right\|$ is the distance from the center of the bubble $\mathbf{x}_{b}=(0.5,0.35)$; $r_{b}=0.25$ is the bubble radius and $\sigma=2$ is a halfwidth; $p_{0} =10^{5}$ is the initial pressure; and $R=287$ denotes the specific gas constant. Moreover, we consider the following operating conditions:
\begin{gather*}
c_{\pi}=1004.5, \quad 
\gamma=1.4, \quad 
\mu= 1.68\cdot 10^{-5}, \quad
\alpha= 2.625\cdot 10^{-5}, \quad
\beta= 2.587\cdot 10^{-3}, \quad
\mathbf{g} = \left(0,-9.81\right)^{T}.
\end{gather*}
Consequently, the main dimensionless numbers are given by
\begin{gather*}
Re = 8.049\cdot 10^{2},\quad
Pr=0.71,\quad 
Gr= 7.603\cdot 10^{6}, \quad
Pe = 5.715\cdot 10^{2}, \quad
Ra = 5.399\cdot 10^{5}, \quad
\beta \Delta \theta=1.72\cdot 10^{-3}
, \quad M\sim 10^{-2}.
\end{gather*}
The simulations were run in the Eulerian framework taking  {\color{cr12} $p=3$, $\Theta=0.51$}. The  primal mesh consists of $3602$ control volumes. Figures \ref{fig:GBTCNS_t10}, \ref{fig:GBTCNS_t15} and \ref{fig:GBTCNS_t20} depict the numerical results at output times $t\in\left\lbrace 10,20,30\right\rbrace$. As expected, we notice that the temperature and velocity contours obtained with the incompressible (INS) and compressible (CNS) flow solver are really close to each other. This is confirmed by a more detailed analysis of the numerical solution computed along specific cuts along the $x-$axis in the computational domain, as we can observe in the plots included at the right hand side of Figures \ref{fig:GBTCNS_t10}-\ref{fig:GBTCNS_t20}. Therefore, for a flow regime verifying the hypothesis related to the Boussinesq approximation, we can use both sets of equations to solve natural convection problems obtaining similar results. One of the main advantages of the incompressible model is the lower computational cost compared to the compressible one, that is approximately seven times slower. Moreover, the algorithm for the incompressible model is much simpler, since we can neglect density variations, hence avoiding the non-linearity of the system presented for the compressible case (see Section \ref{sec:picard_cns}). On the other hand, the compressible model is more general. Since the algorithm has been developed to perform well at all Mach numbers, see \cite{TD17}, we are able to simulate a broader variety of problems including both small and large temperature fluctuations.


{\color{cr12} As second part of this test case, we compare the results obtained considering two different time discretizations. Figure \ref{fig:GBT_sp_t20} reports the numerical solution computed with the incompressible model, when using two sets of space-time basis functions, namely $(p,p_{\gamma})=(3,1)$ and $(p,p_{\gamma})=(3,2)$. We can observe that the theta method produces the same results of the second and third order accurate space-time DG schemes. 
\begin{figure}
	{\color{cr12}\begin{center}
		\includegraphics[width=0.07\linewidth]{./GBTINS_temperature_t15_legend.png}\hfill
		\includegraphics[width=0.29\linewidth]{./GBTINS_temperature_t20.png}\hfill
		\includegraphics[width=0.29\linewidth]{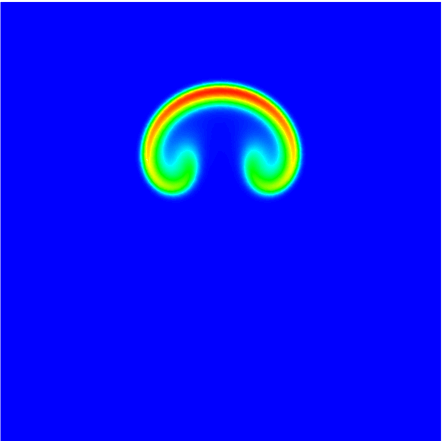}\hfill
		\includegraphics[width=0.29\linewidth]{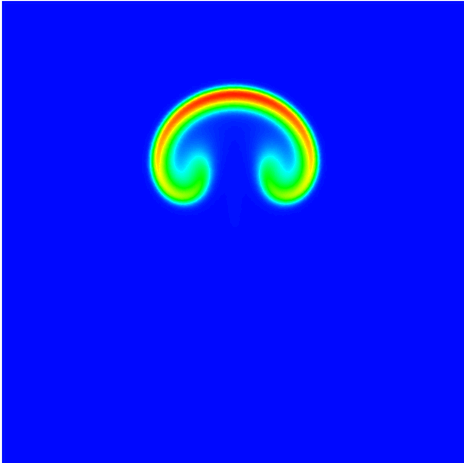}
		
		\vspace{0.2cm}		
		\includegraphics[width=0.06\linewidth]{./GBTINS_velx_t15_legend.png}\hspace{0.01\linewidth}\hfill
		\includegraphics[width=0.29\linewidth]{./GBTINS_velx_t20.png}\hfill %-0.0672 0.069
		\includegraphics[width=0.29\linewidth]{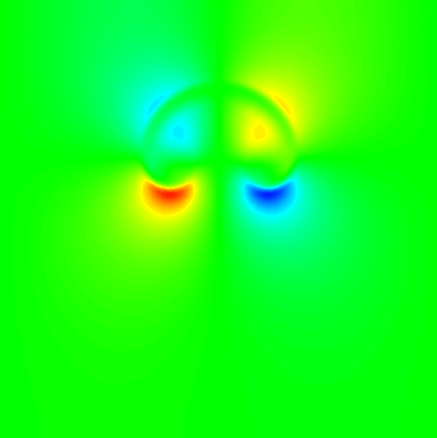}\hfill
		\includegraphics[width=0.29\linewidth]{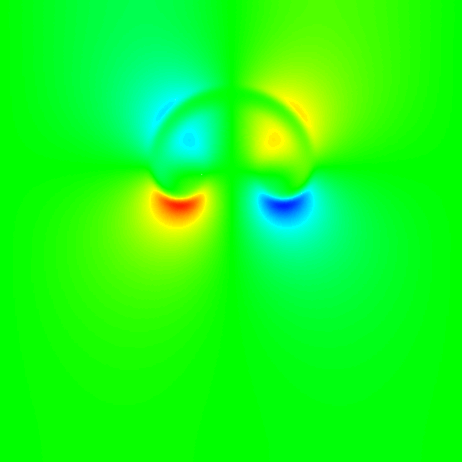}
		
		\vspace{0.2cm}
		\includegraphics[width=0.06\linewidth]{./GBTINS_vely_t15_legend.png}\hspace{0.01\linewidth}\hfill
		\includegraphics[width=0.29\linewidth]{./GBTINS_vely_t20.png}\hfill %-0.05 0.09
		\includegraphics[width=0.29\linewidth]{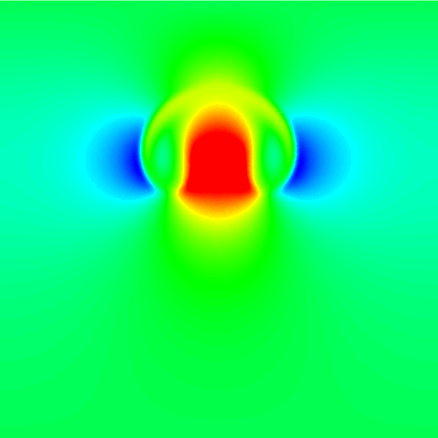}\hfill
		\includegraphics[width=0.29\linewidth]{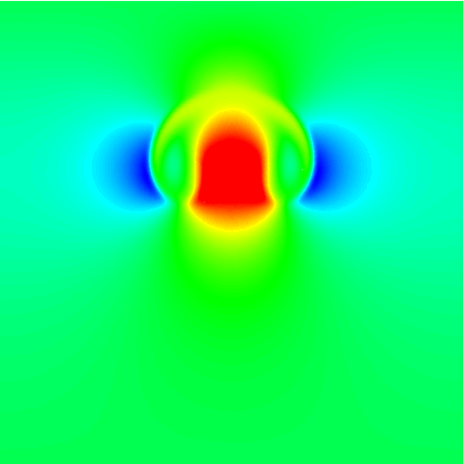}
	\end{center}
	\caption{Gaussian bubble test. From top to bottom: temperature, horizontal and vertical velocity, $t=20s$. Left: $p=3$, $\Theta=0.51$. Center: $(p,p_{\gamma})=(3,1)$. Right: $(p,p_{\gamma})=(3,2)$.}}\label{fig:GBT_sp_t20}
\end{figure}
}

%% % % % % % % % % % % % % % % % % % % % % % % % % % % % % %
%% % % % % % % % % % % % % % % % % % % % % % % % % % % % % %
\subsection{Locally heated cavity}\label{sec:LHC2D}

\begin{figure}
	\vspace{-0.5cm}
	\begin{center}
		\includegraphics[width=0.45\textwidth]{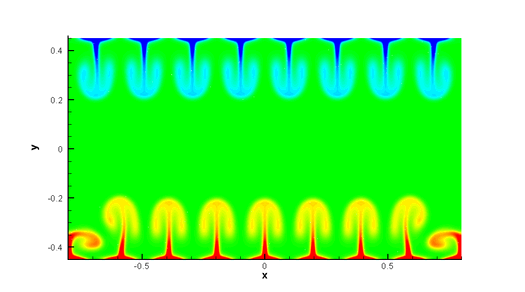}
		\includegraphics[width=0.45\textwidth]{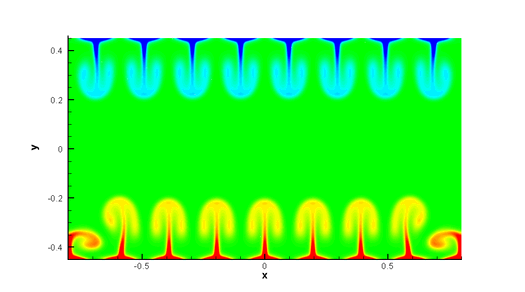} \\
		\includegraphics[width=0.45\textwidth]{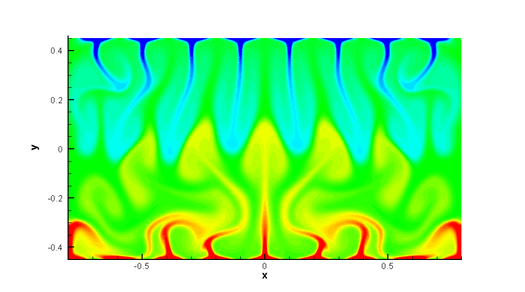}
		\includegraphics[width=0.45\textwidth]{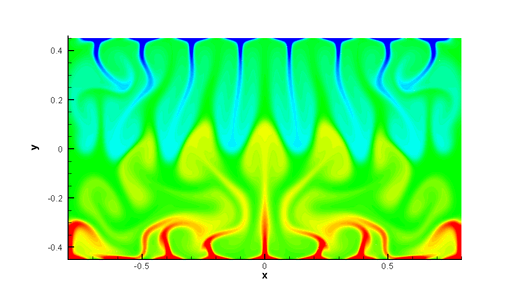} \\
		\includegraphics[width=0.45\textwidth]{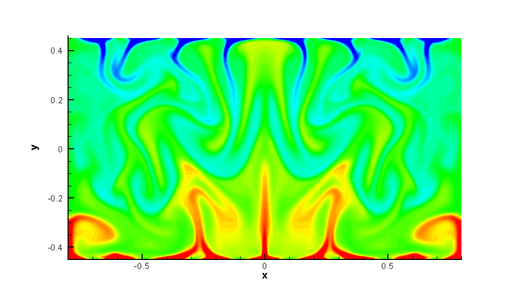}
		\includegraphics[width=0.45\textwidth]{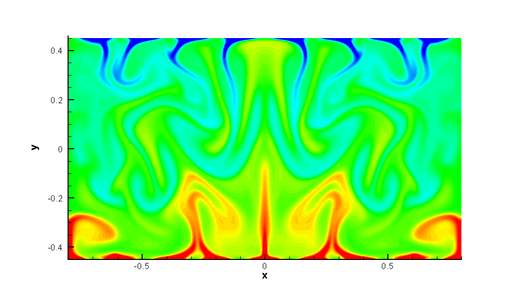} \\
		\includegraphics[width=0.45\textwidth]{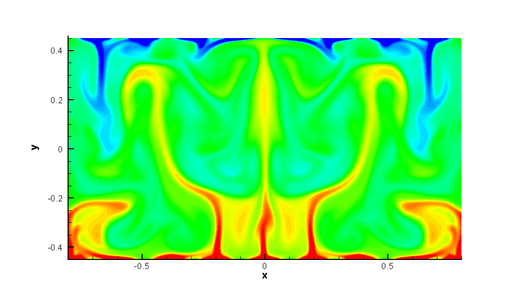}
		\includegraphics[width=0.45\textwidth]{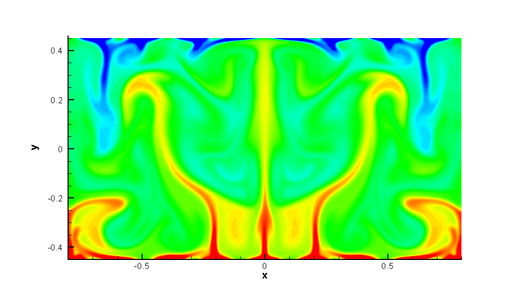} \\		
		\includegraphics[width=0.45\textwidth]{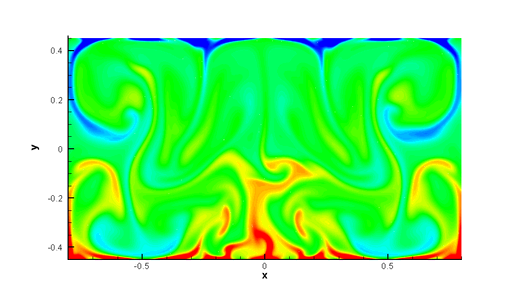}
		\includegraphics[width=0.45\textwidth]{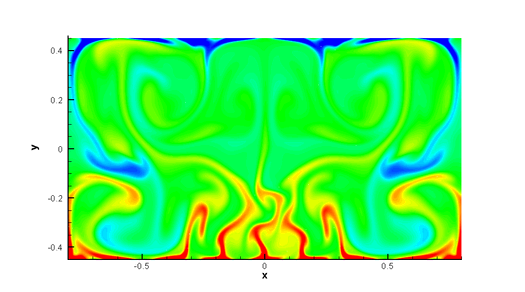} \\
		\includegraphics[width=0.4\textwidth]{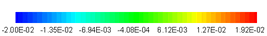}
	\end{center}
	\vspace{-0.4cm}
	\caption{Locally heated cavity. From top to bottom: temperature profile at times $t\in\left\lbrace5,10,12,14,17\right\rbrace$. Left: mesh refinement factor 2. Right: mesh refinement factor 3.}
	\label{fig.MC1}
\end{figure}

\begin{figure}
	\vspace{-0.5cm}
	\begin{center}
		\includegraphics[width=0.45\textwidth]{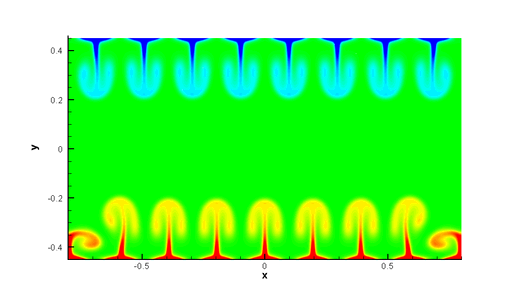}
		\includegraphics[width=0.45\textwidth]{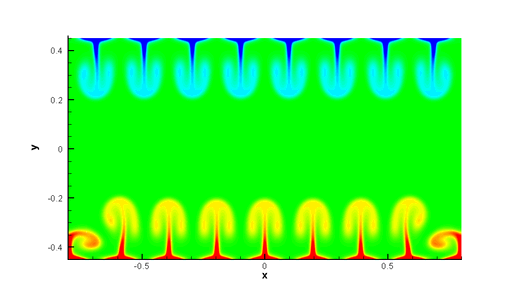} \\
		\includegraphics[width=0.45\textwidth]{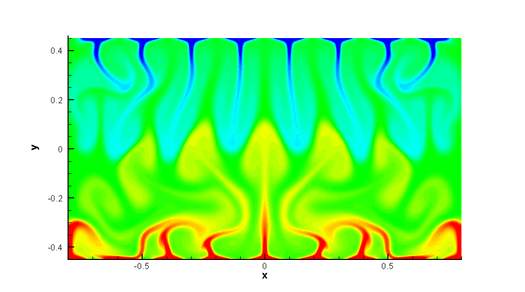}
		\includegraphics[width=0.45\textwidth]{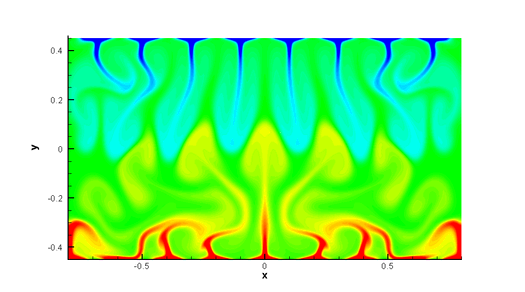} \\
		\includegraphics[width=0.45\textwidth]{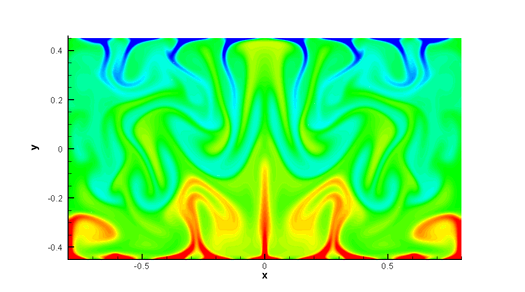}
		\includegraphics[width=0.45\textwidth]{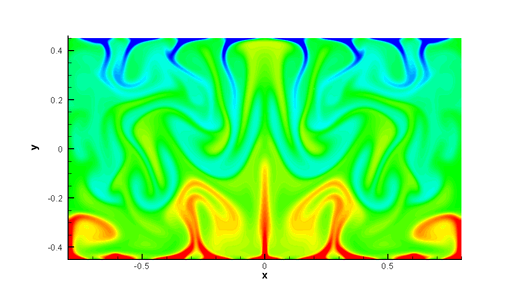} \\
		\includegraphics[width=0.45\textwidth]{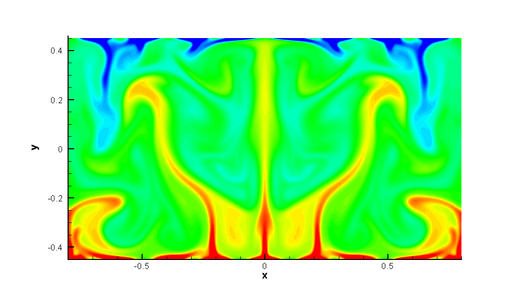}
		\includegraphics[width=0.45\textwidth]{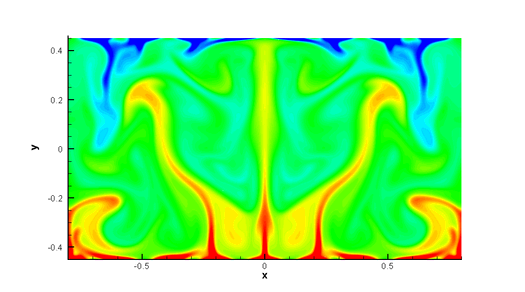} \\
		\includegraphics[width=0.45\textwidth]{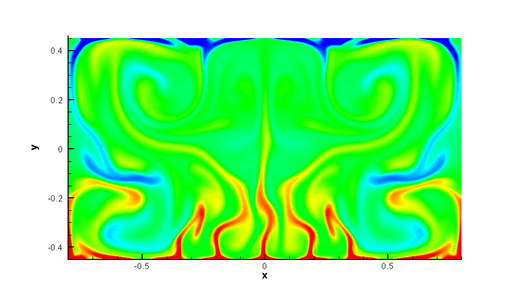}
		\includegraphics[width=0.45\textwidth]{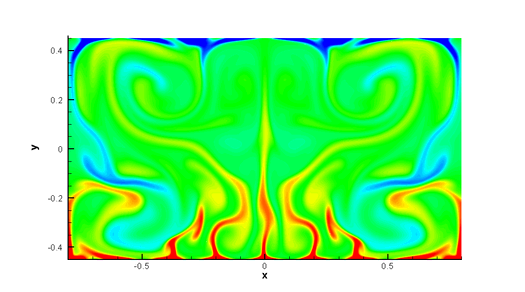} \\
		\includegraphics[width=0.4\textwidth]{./FF_legendc.png}
	\end{center}
	\vspace{-0.4cm}
	\caption{Locally heated cavity. From top to bottom: temperature profile at times $t\in\left\lbrace5,10,12,14,17\right\rbrace$.  Left: mesh refinement factor 4. Right: mesh refinement factor 5.}
	\label{fig.MC2}
\end{figure}

As last test case, we introduce a locally heated cavity test which looks similar to the 
classical one reported in Section \ref{sec:DHC}. Positive and negative heat 
sources on the bottom and upper wall, respectively, are imposed as boundary 
conditions. Due to the physical instability of the chosen boundary conditions, 
we drive the generation of the rising/falling structures by setting 
sinusoidal functions. The computational domain $\Omega=[-0.8,0.8]\times[-0.45,0.45]$, 
is initially discretized using $5172$ triangular elements. Then, we perform 
some refinements in order to show mesh convergence. We use adiabatic no-slip 
walls for the left and right boundaries and we prescribe
\begin{equation}
\mathbf{v}=\boldsymbol{0}, \quad \theta(x,y,t) = \left\lbrace \begin{array}{lr}
0.1-0.1 |\cos (16 x)|,  & \mathrm{if}\; y = -0.45,\\
-0.1+0.1 |\sin (16 x)|, & \mathrm{if}\; y = -0.45,
\end{array} \right.
\end{equation}
on the top and bottom boundaries. Finally, we set $\mathbf{g} = \left(0,-9.81\right)^{T}$, $\mu=10^{-4}$, $\alpha=10^{-3}$, and $p=3$.
The resulting time evolution of the temperature for mesh refinements 
by a factor of $2,3,4$ and $5$ in each space dimension is reported in Figures \ref{fig.MC1}-\ref{fig.MC2}. 
We can observe the generation of some mushroom-shape structures that arise from the flow that is only driven by the temperature boundary conditions. Later in time, they start to mix and to form 
complex substructures. Let us remark that, for refinement levels $2$ and $3$,
we obtain essentially the same results up to $t=12s$. However, at $t=14s$ we 
encounter some small differences that become evident at the final time, 
see Figure \ref{fig.MC1}. However, a good agreement is observed for long 
time solutions when increasing the mesh resolution, see Figure \ref{fig.MC2}.

% % % % % % % % % % % % % % % % % % % % % % % % % % % % % %
% % % % % % % % % % % % % % % % % % % % % % % % % % % % % %
%                  Conclusions                            %
% % % % % % % % % % % % % % % % % % % % % % % % % % % % % %
% % % % % % % % % % % % % % % % % % % % % % % % % % % % % %
\section{Conclusions}\label{sec:conclusions}
In this work, we have presented a new high order accurate staggered semi-implicit discontinuous Galerkin finite element scheme for the solution of natural convection problems. The algorithm is based on the work proposed in \cite{TD14,TD16,TD17}. 
A unified  framework for the discretization of incompressible and compressible Navier-Stokes 
equations with gravity terms has been introduced. The computational cost of the global 
algorithm has been reduced thanks to the development of a novel {\color{cr1} Eulerian-Lagrangian} approach 
for the treatment of the nonlinear convective terms, which leads to an unconditionally stable scheme 
for the incompressible Navier-Stokes system. Furthermore, the new methodology is 
able to handle flows with large temperature and velocity gradients. The final algorithms 
developed within this work have been assessed using several classical benchmarks, showing good agreement 
with the reference data. A detailed comparison between the fully Eulerian and the {\color{cr1} Eulerian-Lagrangian}
approaches for advection has been performed, highlighting the main advantages and drawbacks of both methodologies.   
Moreover, the numerical results obtained with the incompressible Navier-Stokes equations in combination with the Boussinesq approximation have also been validated against a numerical solution of the full compressible Navier-Stokes equations. We have shown computational results for a rising bubble in three space dimensions using more than 26 million spatial degrees of freedom, which clearly shows that the proposed computational approach can also be used on modern massively parallel distributed memory supercomputers. This test case has been set to demonstrate the \textit{potential} capability of the proposed high order staggered semi-implicit DG algorithm for direct numerical  simulations (DNS) on unstructured grids at moderate Reynolds numbers, i.e. the full resolution of all small scale structures present in the flow without the use of any turbulence models.  

Future research will involve the development of a conservative {\color{cr1} Eulerian-Lagrangian} approach 
for the treatment of convective and viscous terms as well as its extension to the 
compressible regime. Another research direction will be the use of subgrid scale turbulence models for 
large eddy simulations (LES) of gravity-driven flows at high Reynolds numbers.

% % % % % % % % % % % % % % % % % % % % % % % % % % % % % %
% % % % % % % % % % % % % % % % % % % % % % % % % % % % % %
%                Acknowledgment                          %
% % % % % % % % % % % % % % % % % % % % % % % % % % % % % %
% % % % % % % % % % % % % % % % % % % % % % % % % % % % % %
\section*{Acknowledgments}
This work was financially supported by {\color{cr1} INdAM} (\textit{Istituto Nazionale di Alta Matematica}, Italy) 
under a Post-doctoral grant of the research project \textit{Progetto premiale FOE 2014-SIES}; M.T. and M.D. acknowledge partial support 
of the European Union's Horizon 2020 Research and Innovation Programme under 
the project \textit{ExaHyPE}, grant no. 671698 (call FETHPC-1-2014); W.B. was partially financed by the GNCS group of INdAM and the program \textit{Young Researchers Funding 2018}.
The simulations were performed on the HazelHen supercomputer at the HLRS in Stuttgart, 
Germany and on the SuperMUC-NG supercomputer at the LRZ in Garching, Germany. The authors acknowledge funding from the Italian Ministry of Education, University 
and Research (MIUR) in the frame of the Departments of Excellence Initiative 2018--2022 
attributed to DICAM of the University of Trento (grant L. 232/2016) and in the frame of the 
PRIN 2017 project \textit{Innovative numerical methods for evolutionary partial differential equations and  applications}. Furthermore, M.D. has also received funding from the University of Trento via the Strategic 
Initiative \textit{Modeling and Simulation}.  

The authors would like to thank the two anonymous referees for their helpful comments and suggestions, which helped to improve the quality of this manuscript. 

% % % % % % % % % % % % % % % % % % % % % % % % % % % % % %
% % % % % % % % % % % % % % % % % % % % % % % % % % % % % %
%                  References                             %
\bibliographystyle{elsarticle-num}
\bibliography{./mibiblio}

\begin{thebibliography}{100}
\expandafter\ifx\csname url\endcsname\relax
  \def\url#1{\texttt{#1}}\fi
\expandafter\ifx\csname urlprefix\endcsname\relax\def\urlprefix{URL }\fi
\expandafter\ifx\csname href\endcsname\relax
  \def\href#1#2{#2} \def\path#1{#1}\fi

\bibitem{TD14}
M.~Tavelli, M.~Dumbser, A staggered semi-implicit discontinuous {Galerkin}
  method for the two dimensional incompressible {N}avier-{S}tokes equations,
  Appl. Math. Comput. 248 (2014) 70 -- 92.

\bibitem{TD15}
M.~Tavelli, M.~Dumbser, A staggered space-time discontinuous {G}alerkin method
  for the incompressible {Navier-Stokes} equations on two-dimensional
  triangular meshes, Comput. Fluids 119 (2015) 235 -- 249.

\bibitem{TD16}
M.~Tavelli, M.~Dumbser, A staggered space-time discontinuous {Galerkin} method
  for the three-dimensional incompressible {Navier-Stokes} equations on
  unstructured tetrahedral meshes, J. Comput. Phys. 319 (2016) 294 -- 323.

\bibitem{TD17}
M.~Tavelli, M.~Dumbser, A pressure-based semi-implicit space-time discontinuous
  {G}alerkin method on staggered unstructured meshes for the solution of the
  compressible {N}avier-{S}tokes equations at all {Mach} numbers, J. Comput.
  Phys. 341 (2017) 341 -- 376.

\bibitem{GR08}
F.~X. Giraldo, M.~Restelli, A study of spectral element and discontinuous
  {G}alerkin methods for the {N}avier-{S}tokes equations in nonhydrostatic
  mesoscale atmospheric modeling: Equation sets and test cases, J. Comput.
  Phys. 227~(8) (2008) 3849 -- 3877.

\bibitem{BB11}
M.~Ben\'itez, A.~Berm\'udez, A second order characteristics finite element
  scheme for natural convection problems, J. Comput. Appl. Math. 235~(11)
  (2011) 3270 -- 3284.

\bibitem{BZG14}
A.~Ba\"iri, E.~Zarco-Pernia, J.-M.~G. de~Mar\'ia, A review on natural
  convection in enclosures for engineering applications. the particular case of
  the parallelogrammic diode cavity, Appl. Therm. Eng. 63~(1) (2014) 304 --
  322.

\bibitem{DRB17}
D.~Das, M.~Roy, T.~Basak, Studies on natural convection within enclosures of
  various (non-square) shapes - a review, Int. J. Heat Mass Transfer 106 (2017)
  356 -- 406.

\bibitem{GCD18}
E.~Gaburro, M.~J. Castro, M.~Dumbser, Well-balanced
  {Arbitrary-Lagrangian-Eulerian} finite volume schemes on moving nonconforming
  meshes for the {E}uler equations of gas dynamics with gravity, Mon. Not. R.
  Astron. Soc. 477~(2) (2018) 2251--2275.

\bibitem{Yi18}
T.~H. Yi, Time integration of unsteady nonhydrostatic equations with dual time
  stepping and multigrid methods, J. Comput. Phys. 374 (2018) 873--892.

\bibitem{MNG15}
S.~Marras, M.~Nazarov, F.~X. Giraldo, Stabilized high-order galerkin methods
  based on a parameter-free dynamic sgs model for les, J. Comput. Phys. 301
  (2015) 77 -- 101.

\bibitem{MS18}
I.~V. Miroshnichenko, M.~A. Sheremet, Turbulent natural convection heat
  transfer in rectangular enclosures using experimental and numerical
  approaches: A review, Renewable Sustainable Energy Rev. 82 (2018) 40 -- 59.

\bibitem{HW65}
F.~Harlow, J.~Welch, Numerical calculation of time-dependent viscous
  incompressible flow of fluid with a free surface, Phys. Fluids 8 (1965)
  2182--2189.

\bibitem{PS72}
S.~V. Patankar, D.~B. Spalding, A calculation procedure for heat, mass and
  momentum transfer in three-dimensional parabolic flows, Int J Heat Mass
  Transfer 15~(10) (1972) 1787--1806.

\bibitem{Pat80}
V.~Patankar, Numerical {Heat} {Transfer} and {Fluid} {Flow}, Hemisphere
  Publishing Corporation, 1980.

\bibitem{Kan86}
J.~van Kan, {A second-order accurate pressure correction method for viscous
  incompressible flow}, SIAM Journal on Scientific and Statistical Computing 7
  (1986) 870--891.

\bibitem{TH73}
C.~Taylor, P.~Hood, {A numerical solution of the Navier-Stokes equations using
  the finite element technique}, Computers and Fluids 1 (1973) 73--100.

\bibitem{BH82}
A.~Brooks, T.~Hughes, {Stream-line upwind/Petrov Galerkin formulstion for
  convection dominated flows with particular emphasis on the incompressible
  Navier-Stokes equation}, Computer Methods in Applied Mechanics and
  Engineering 32 (1982) 199--259.

\bibitem{HMM86}
T.~Hughes, M.~Mallet, M.~Mizukami, {A new finite element formulation for
  computational fluid dynamics: II. Beyond SUPG}, Computer Methods in Applied
  Mechanics and Engineering 54 (1986) 341--355.

\bibitem{For81}
M.~Fortin, {Old and new finite elements for incompressible flows},
  International Journal for Numerical Methods in Fluids 1 (1981) 347--364.

\bibitem{Ver91}
R.~Verf\"urth, {Finite element approximation of incompressible Navier-Stokes
  equations with slip boundary condition II}, Numerische Mathematik 59 (1991)
  615--636.

\bibitem{HR82}
J.~G. Heywood, R.~Rannacher, {Finite element approximation of the nonstationary
  Navier-Stokes Problem. I. Regularity of solutions and second order error
  estimates for spatial discretization}, SIAM J. Numer. Anal. 19 (1982)
  275--311.

\bibitem{HR88}
J.~G. Heywood, R.~Rannacher, {Finite element approximation of the nonstationary
  Navier-Stokes Problem. III. Smoothing property and higher order error
  estimates for spatial discretization}, SIAM J. Numer. Anal. 25 (1988)
  489--512.

\bibitem{BR97}
F.~Bassi, S.~Rebay, A high-order accurate discontinuous finite element method
  for the numerical solution of the compressible {Navier}-{Stokes} equations,
  J. Comput. Phys. 131 (1997) 267--279.

\bibitem{BO99}
C.~Baumann, J.~Oden, A discontinuous hp finite element method for
  convection-diffusion problems, Comput. Methods Appl. Mech. Eng. 175~(3-4)
  (1999) 311--341.

\bibitem{BO99b}
C.~Baumann, J.~Oden, A discontinuous hp finite element method for the euler and
  navier-stokes equations, Int. J. Numer. Methods Fluids 31~(1) (1999) 79--95.

\bibitem{BCPR06}
F.~Bassi, A.~Crivellini, D.~D. Pietro, S.~Rebay, On a robust discontinuous
  galerkin technique for the solution of compressible flow, J. Comput. Phys.
  218 (2006) 208--221.

\bibitem{HH06}
R.~Hartmann, P.~Houston, Symmetric interior penalty {DG} methods for the
  compressible {N}avier--{S}tokes equations {I}: Method formulation, Int. J.
  Num. Anal. Model. 3 (2006) 1--20.

\bibitem{BCPR07}
F.~Bassi, A.~Crivellini, D.~D. Pietro, S.~Rebay, {An implicit high-order
  discontinuous Galerkin method for steady and unsteady incompressible flows},
  Computers and Fluids 36 (2007) 1529--1546.

\bibitem{Gas07}
G.~Gassner, F.~Lorcher, C.~D. Munz, A contribution to the construction of
  diffusion fluxes for finite volume and discontinuous {G}alerkin schemes, J.
  Comp. Phys. 224~(2) (2007) 1049 -- 1063.

\bibitem{SFE07}
K.~Shahbazi, P.~F. Fischer, C.~R. Ethier, A high-order discontinuous galerkin
  method for the unsteady incompressible navier-stokes equations, J. Comput.
  Phys. 222 (2007) 391--407.

\bibitem{HH08}
R.~Hartmann, P.~Houston, An optimal order interior penalty discontinuous
  galerkin discretization of the compressible {N}avier-{S}tokes equations, J.
  Comput. Phys. 227 (2008) 9670--9685.

\bibitem{Dum10}
M.~Dumbser, Arbitrary high order {PNPM} schemes on unstructured meshes for the
  compressible {N}avier-{S}tokes equations, Comput. Fluids 39~(1) (2010)
  60--76.

\bibitem{FW11}
E.~Ferrer, R.~Willden, A high order discontinuous galerkin finite element
  solver for the incompressible navier-stokes equations, Computer and Fluids 46
  (2011) 224--230.

\bibitem{NPC11}
N.~Nguyen, J.~Peraire, B.~Cockburn, An implicit high-order hybridizable
  discontinuous galerkin method for the incompressible navier-stokes equations,
  J. Comput. Phys. 230 (2011) 1147--1170.

\bibitem{RC12}
S.~Rhebergen, B.~Cockburn, {A space-time hybridizable discontinuous Galerkin
  method for incompressible flows on deforming domains}, J. Comput. Phys. 231
  (2012) 4185--4204.

\bibitem{RCV13}
S.~Rhebergen, B.~Cockburn, J.~J. van~der Vegt, {A space-time discontinuous
  Galerkin method for the incompressible Navier-Stokes equations}, J. Comput.
  Phys. 233 (2013) 339--358.

\bibitem{CAB13}
A.~Crivellini, V.~D'Alessandro, F.~Bassi, {High-order discontinuous Galerkin
  solutions of three-dimensional incompressible RANS equations}, Computers and
  Fluids 81 (2013) 122--133.

\bibitem{KKO13}
B.~Klein, F.~Kummer, M.~Oberlack, {A SIMPLE based discontinuous Galerkin solver
  for steady incompressible flows}, J. Comput. Phys. 237 (2013) 235--250.

\bibitem{TBR13}
G.~Tumolo, L.~Bonaventura, M.~Restelli, {A semi-implicit, semi-Lagrangian,
  p-adaptive discontinuous Galerkin method for the shallow water equations },
  J. Comput. Phys. 232 (2013) 46--67.

\bibitem{GR10}
F.~X. Giraldo, M.~Restelli, High-order semi-implicit time-integrators for a
  triangular discontinuous galerkin oceanic shallow water model, Int. J. Numer.
  Methods Fluids 63 (2010) 1077--1102.

\bibitem{Dol08}
V.~Dolejsi, Semi-implicit interior penalty discontinuous {G}alerkin methods for
  viscous compressible flows, Comm. Comput. Phys. 4 (2008) 231--274.

\bibitem{DF04}
V.~Dolejsi, M.~Feistauer, A semi-implicit discontinuous {G}alerkin finite
  element method for the numerical solution of inviscid compressible flow, J.
  Comput. Phys. 198 (2004) 727--746.

\bibitem{DFH07}
V.~Dolejsi, M.~Feistauer, J.~Hozman, Analysis of semi-implicit {DGFEM} for
  nonlinear convection-diffusion problems on nonconforming meshes, Comput.
  Methods Appl. Mech. Eng. 196 (2007) 2813--2827.

\bibitem{PM05}
J.~Park, C.-D. Munz, Multiple pressure variables methods for fluid flow at all
  {M}ach numbers, International journal for numerical methods in fluids 49~(8)
  (2005) 905--931.

\bibitem{TV12}
E.~Toro, M.~V\'azquez-Cend\'on, {Flux splitting schemes for the Euler
  equations}, Computers and Fluids 70 (2012) 1--12.

\bibitem{DC16}
M.~Dumbser, V.~Casulli, {A conservative, weakly nonlinear semi-implicit finite
  volume scheme for the compressible Navier-Stokes equations with general
  equation of state}, Applied Mathematics and Computation 272 (2016) 479--497.

\bibitem{CasulliCompressible}
V.~Casulli, D.~Greenspan, Pressure method for the numerical solution of
  transient, compressible fluid flows, International Journal for Numerical
  Methods in Fluids 4~(11) (1984) 1001--1012.

\bibitem{LSTZ07}
Y.~J. Liu, C.~W. Shu, E.~Tadmor, M.~Zhang, Central discontinuous {G}alerkin
  methods on overlapping cells with a non-oscillatory hierarchical
  reconstruction, SIAM J. Numer. Anal. 45 (2007) 2442--2467.

\bibitem{LSTZ08}
Y.~J. Liu, C.~W. Shu, E.~Tadmor, M.~Zhang, L2-stability analysis of the central
  discontinuous {G}alerkin method and a comparison between the central and
  regular discontinuous {G}alerkin methods, Mathematical Modeling and Numerical
  Analysis 42 (2008) 593--607.

\bibitem{CE06}
E.~Chung, B.~Engquist, {Optimal discontinuous Galerkin methods for wave
  propagation}, SIAM J. Numer. Anal. 44 (2006) 2131--2158.

\bibitem{CL12}
E.~T. Chung, C.~S. Lee, {A staggered discontinuous Galerkin method for the
  convection--diffusion equation}, Journal of Numerical Mathematics 20 (2012)
  1--31.

\bibitem{FambriDumbser}
F.~Fambri, M.~Dumbser, Spectral semi-implicit and space-time discontinuous
  {G}alerkin methods for the incompressible {N}avier-{S}tokes equations on
  staggered {C}artesian grids, Applied Numerical Mathematics 110 (2016) 41--74.

\bibitem{AMRDGSI}
F.~Fambri, M.~Dumbser, Semi-implicit discontinuous {G}alerkin methods for the
  incompressible {N}avier-{S}tokes equations on adaptive staggered {C}artesian
  grids, Computer Methods in Applied Mechanics and Engineering 324 (2017)
  170--203.

\bibitem{TB19}
M.~Tavelli, W.~Boscheri, A high‐order parallel eulerian‐lagrangian
  algorithm for advection‐diffusion problems on unstructured meshes, Int. J.
  Numer. Methods Fluids (2019).
\newblock \href {https://doi.org/https://doi.org/10.1002/fld.4756}
  {\path{doi:https://doi.org/10.1002/fld.4756}}.

\bibitem{toro4}
E.~Toro, V.~Titarev, Solution of the generalized {Riemann} problem for
  advection-reaction equations, Proc. Roy. Soc. London (2002) 271--281.

\bibitem{titarevtoro}
V.~A. Titarev, E.~F. Toro, {ADER} schemes for three-dimensional nonlinear
  hyperbolic systems, Journal of Computational Physics 204 (2005) 715--736.

\bibitem{Toro:2006a}
E.~F. Toro, V.~A. Titarev, {Derivative Riemann solvers for systems of
  conservation laws and ADER methods}, Journal of Computational Physics 212~(1)
  (2006) 150--165.

\bibitem{Wel55}
P.~Welander, {Studies on the general development of motion in a two-dimensional
  ideal fluid}, Tellus 17 (1955) 141--156.

\bibitem{Wii59}
A.~Wiin-Nielson, {On the application of trajectory methods in numerical
  forecasting}, Tellus 11 (1959) 180--186.

\bibitem{Rob81}
A.~Robert, {A stable numerical integration scheme for the primitive
  meteorological equations}, Atmos. Ocean 19 (1981) 35--46.

\bibitem{BD82}
J.~Bates, A.~McDonald, {Multiply-upstream, semi-Lagrangian advective schemes:
  analysis and application to a multi-level primitive equation model}, Mon.
  Wea. Rev. 110 (1982) 1831--1842.

\bibitem{Cas88}
V.~Casulli, {On Eulerian-Lagrangian methods for the Navier-Stokes equations at
  high Reynolds number}, International Journal for Numerical Methods in Fluids
  8 (1988) 1349--1360.

\bibitem{CS11}
V.~Casulli, G.~S. Stelling, Semi-implicit subgrid modelling of
  three-dimensional free-surface flows, Int. J. Numer. Methods Fluids 67 (2011)
  441--449.

\bibitem{BDR13}
W.~Boscheri, M.~Dumbser, M.~Righetti, {A semi-implicit scheme for 3D free
  surface flows with high-order velocity reconstruction on unstructured Voronoi
  meshes}, Int. J. Numer. Methods Fluids 72 (2013) 607--631.

\bibitem{BPR19}
W.~Boscheri, G.~Pisaturo, M.~Righetti, High order divergence-free velocity
  reconstruction for free surface flows on unstructured voronoi meshes, Int. J.
  Numer. Methods Fluids 90 (2019) 296--321.

\bibitem{Bon00}
L.~Bonaventura, {A Semi-implicit Semi-Lagrangian Scheme Using the Height
  Coordinate for a Nonhydrostatic and Fully Elastic Model of Atmospheric
  Flows}, J. Comput. Phys. 158 (2000) 186--213.

\bibitem{RBS06}
M.~Restelli, L.~Bonaventura, R.~Sacco, A semi-lagrangian discontinuous
  {G}alerkin method for scalar advection by incompressible flows, J. Comput.
  Phys. 216 (2006) 195--215.

\bibitem{BFR18}
L.~Bonaventura, R.~Ferretti, L.~Rocchi, {A fully semi-Lagrangian discretization
  for the 2D incompressible Navier-Stokes equations in the
  vorticity-streamfunction formulation}, Applied Mathematics and Computation
  323 (2018) 132--144.

\bibitem{BDDV98}
A.~Berm\'udez, A.~Dervieux, J.~A. Desideri, M.~E. V\'azquez-Cend\'on, Upwind
  schemes for the two-dimensional shallow water equations with variable depth
  using unstructured meshes, Comput. Methods Appl. Mech. Eng. 155~(1) (1998)
  49--72.

\bibitem{THD09}
E.~F. Toro, A.~Hidalgo, M.~Dumbser, {FORCE} schemes on unstructured meshes {I}:
  Conservative hyperbolic systems, J. Comput. Phys. 228~(9) (2009) 3368 --
  3389.

\bibitem{TD14sw}
M.~Tavelli, M.~Dumbser, {A high order semi-implicit discontinuous Galerkin
  method for the two dimensional shallow water equations on staggered
  unstructured meshes}, Appl. Math. Comput. 234 (2014) 623--644.

\bibitem{DHCPT10}
M.~Dumbser, A.~Hidalgo, M.~Castro, C.~Par\'es, E.~F. Toro, {FORCE} schemes on
  unstructured meshes {II}: Non-conservative hyperbolic systems, Comput.
  Methods Appl. Mech. Eng. 199 (2010) 625--647.

\bibitem{BFSV14}
A.~Berm\'udez, J.~L. Ferr\'in, L.~Saavedra, M.~E. V\'azquez-Cend\'on, A
  projection hybrid finite volume/element method for low-{M}ach number flows,
  J. Comp. Phys. 271 (2014) 360--378.

\bibitem{BFTVC17}
S.~Busto, J.~L. Ferr\'in, E.~F. Toro, M.~E. V\'azquez-Cend\'on, A projection
  hybrid high order finite volume/finite element method for incompressible
  turbulent flows, J. Comput. Phys. 353 (2018) 169--192.

\bibitem{CGP06}
M.~J. Castro, J.~M. Gallardo, C.~Par\'es, High-order finite volume schemes
  based on reconstruction of states for solving hyperbolic systems with
  nonconservative products. applications to shallow-water systems, Mathematics
  of Computations 75 (2006) 1103--1134.

\bibitem{Par06}
C.~Par\'es, Numerical methods for nonconservative hyperbolic systems: a
  theoretical framework, SIAM J. Numer. Anal. 44 (2006) 300--321.

\bibitem{CG84}
V.~Casulli, D.~Greenspan, Pressure method for the numerical solution of
  transient, compressible fluid flows, Int. J. Numer. Methods Fluids 4 (1984)
  1001--1012.

\bibitem{CC92}
V.~Casulli, R.~T. Cheng, Semi-implicit finite difference methods for
  three--dimensional shallow water flow, Int. J. Numer. Methods Fluids 15
  (1992) 629--648.

\bibitem{CC94}
V.~Casulli, E.~Cattani, Stability, accuracy and efficiency of a semi-implicit
  method for three-dimensional shallow water flow, Computers \& Mathematics
  with Applications 27 (1994) 99--112.

\bibitem{CW00}
V.~Casulli, R.~A. Walters, An unstructured grid, three--dimensional model based
  on the shallow water equations, Int. J. Numer. Methods Fluids 32 (2000)
  331--348.

\bibitem{Rus62}
V.~V. Rusanov, The calculation of the interaction of non-stationary shock waves
  and obstacles, USSR Computational Mathematics and Mathematical Physics 1
  (1962) 304--320.

\bibitem{Mcg93}
J.~L. McGregor, Economical determination of departure points for
  semi-lagrangian models, Monthly Weather Review 121~(1) (1993) 221--230.

\bibitem{KleinMach}
R.~Klein, Semi-implicit extension of a godunov-type scheme based on low mach
  number asymptotics {I}: one-dimensional flow, Journal of Computational
  Physics 121 (1995) 213--237.

\bibitem{Klein2001}
R.~Klein, N.~Botta, T.~Schneider, C.~Munz, S.Roller, A.~Meister, L.~Hoffmann,
  T.~Sonar, Asymptotic adaptive methods for multi-scale problems in fluid
  mechanics, Journal of Engineering Mathematics 39 (2001) 261--343.

\bibitem{Munz2003}
C.~Munz, S.~Roller, R.~Klein, K.~Geratz, The extension of incompressible flow
  solvers to the weakly compressible regime, Computers and Fluids 32 (2003)
  173--196.

\bibitem{RussoAllMach}
S.~Boscarino, G.~Russo, L.~Scandurra, {All Mach number second order
  semi-implicit scheme for the Euler equations of gasdynamics}, Journal of
  Scientific Computing 77 (2018) 850--884.

\bibitem{CordierDegond}
F.~Cordier, P.~Degond, A.~Kumbaro, {An Asymptotic-Preserving all-speed scheme
  for the Euler and Navier-Stokes equations}, Journal of Computational Physics
  231 (2012) 5685--5704.

\bibitem{BottaKlein}
N.~Botta, R.~Klein, S.~Langenberg, S.~L\"utzenkirchen, Well balanced finite
  volume methods for nearly hydrostatic flows, Journal of Computational Physics
  196 (2004) 539--565.

\bibitem{Kapelli2014}
R.~K{\"a}ppeli, S.~Mishra, Well-balanced schemes for the euler equations with
  gravitation, Journal of Computational Physics 259 (2014) 199--219.

\bibitem{Klingenberg2015}
P.~Chandrashekar, C.~Klingenberg, {A second order well-balanced finite volume
  scheme for Euler Equations with gravity}, {Journal on Scientific Computing }
  37 (2015) B382--B402.

\bibitem{KlingenbergPuppo}
C.~Klingenberg, G.~Puppo, M.~Semplice, {Arbitrary order finite volume
  well-balanced schemes for the Euler equations with gravity}, SIAM Journal on
  Scientific Computing 41 (2019) A695--A721.

\bibitem{DBTM08}
M.~Dumbser, D.~S. Balsara, E.~F. Toro, C.-D. Munz, A unified framework for the
  construction of one-step finite volume and discontinuous {G}alerkin schemes
  on unstructured meshes, J. Comput. Phys. 227~(18) (2008) 8209--8253.

\bibitem{DZ09}
M.~Dumbser, O.~Zanotti, Very high order pnpm schemes on unstructured meshes for
  the resistive relativistic mhd equations, J. Comput. Phys. 228~(18) (2009)
  6991 -- 7006.

\bibitem{CZ09}
V.~Casulli, P.~Zanolli, A nested newton-type algorithm for finite volume
  methods solving richards' equation in mixed form, SIAM J. Sci. Comput. 32~(4)
  (2010) 2255--2273.

\bibitem{VD83}
G.~de~Vahl~Davis, Natural convection of air in a square cavity: a benchmark
  numerical solution, International Journal for Numerical Methods in Fluids 3
  (1983) 249--264.

\bibitem{MNZ98}
N.~Massarotti, P.~Nithiarasu, O.~Zienkiewicz, Characteristic-based-split {CBS}
  algorithm for incompressible flow problems with heat transfer, Int. J. Numer.
  Methods Heat Fluid Flow 8~(8) (1998) 969--990.

\bibitem{MUC00}
D.~A. Mayne, A.~S. Usmani, M.~Crapper, h-adaptive finite element solution of
  high rayleigh number thermally driven cavity problem, Int. J. Numer. Methods
  Heat Fluid Flow 10~(6) (2000) 598--615.

\bibitem{WPW01}
D.~C. Wan, B.~S.~V. Patnaik, G.~W. Wei, A new benchmark quality solution for
  the buoyancy-driven cavity by discrete singular convolution, Numerical Heat
  Transfer, Part B: Fundamentals 40~(3) (2001) 199--228.

\bibitem{MD84}
E.~F. Moore, R.~W. Davis, Numerical solutions for steady natural convection in
  a square cavity, NASA STI/Recon Technical Report N 85 (Mar. 1984).

\bibitem{SX98}
C.~Shu, H.~Xue, Comparison of two approaches for implementing stream function
  boundary conditions in {DQ} simulation of natural convection in a square
  cavity, Int. J. Heat Fluid Flow 19~(1) (1998) 59--68.

\bibitem{Man99}
M.~T. Manzari, An explicit finite element algorithm for convection heat
  transfer problems, International Journal of Numerical Methods for Heat \&
  Fluid Flow 9~(8) (1999) 860--877.

\bibitem{DT02}
Q.-H. Deng, G.-F. Tang, Numerical visualization of mass and heat transport for
  conjugate natural convection/heat conduction by streamline and heatline, Int.
  J. Heat Mass Transfer 45~(11) (2002) 2373 -- 2385.

\bibitem{MBGW13}
A.~M\"uller, J.~Behrens, F.~X. Giraldo, V.~Wirth, Comparison between adaptive
  and uniform discontinuous galerkin simulations in dry 2{D} bubble
  experiments, J. Comput. Phys. 235 (2013) 371 -- 393.

\bibitem{YMLGW14}
L.~Yelash, A.~M\"uller, M.~Luk\'aov\'a-Medvid'ov\'a, F.~Giraldo, V.~Wirth,
  Adaptive discontinuous evolution {G}alerkin method for dry atmospheric flow,
  J. Comput. Phys. 268 (2014) 106 -- 133.

\bibitem{BLY17}
G.~Bispen, M.~Luk\'aov\'a-Medvid'ov\'a, L.~Yelash, Asymptotic preserving {IMEX}
  finite volume schemes for low {M}ach number {E}uler equations with
  gravitation, J. Comput. Phys. 335 (2017) 222 --248.

\bibitem{TDg16}
P.~Tsoutsanis, D.~Drikakis, Addressing the challenges of implementation of
  high-order finite-volume schemes for atmospheric dynamics on unstructured
  meshes, ECCOMAS Congress 2016 - Proceedings of the 7th European Congress on
  Computational Methods in Applied Sciences and Engineering, National Technical
  University of Athens, 2016, pp. 684--708.

\end{thebibliography}

%%%%%%%%%%%%%%%%%%%%%%%%%%%%%%%%%%%%%%%%%%%%%%%%%%%%%%%%
%%%%%%%%%%%%%%%%%%%%%%%%%%%%%%%%%%%%%%%%%%%%%%%%%%%%%%%%
\end{document}